 \documentclass[twocolumn]{aastex63}

\DeclareUnicodeCharacter{2212}{-}
\accepted{\today}
\shorttitle{Variability of 2M2208+2921}
\shortauthors{Manjavacas et al.}
\graphicspath{{./}{figures/}}
\begin{document}

\title{Revealing the Vertical Cloud Structure of a young low-mass Brown Dwarf, an analog to the $\beta$-Pictoris~b directly-imaged exoplanet, through Keck\,I/MOSFIRE spectro-photometric variability}

\author{Elena Manjavacas}
\affiliation{AURA for the European Space Agency (ESA), ESA Office, Space Telescope Science Institute, 3700 San Martin Drive, Baltimore, MD, 21218 USA}
\affiliation{W. M. Keck Observatory, 65-1120 Mamalahoa Hwy. Kamuela, HI, USA}

\author{Theodora Karalidi}
\affiliation{Department of Physics, University of Central Florida, 4000 Central Florida Blvd., Orlando, FL 32816, USA}

\author{Johanna M. Vos}
\affiliation{Department of Astrophysics, American Museum of Natural History, Central Park West at 79th Street, New York, NY 10024, USA}

\author{Beth A. Biller}
\affiliation{SUPA, Institute for Astronomy, University of Edinburgh, Blackford Hill, Edinburgh EH9 3HJ, UK}
\affiliation{Centre for Exoplanet Science, University of Edinburgh, Edinburgh, UK}

\author{Ben W. P. Lew}
\affiliation{Lunar and Planetary Laboratory, The University of Arizona, 1640 E. University Blvd., Tucson, AZ 85721, USA}
\affiliation{Department of Astronomy and Steward Observatory, The University of Arizona, 933 N. Cherry Ave., Tucson, AZ 85721, USA}

\correspondingauthor{Elena Manjavacas}
\email{emanjavacas@stsci.edu}



\begin{abstract}

Young brown dwarfs are analogs to giant exoplanets, as they share effective temperatures, near-infrared colors and surface gravities. Thus, the detailed characterization of young brown dwarfs might shed light on the study of giant exoplanets, that we are currently unable to observe with the sufficient signal-to-noise to allow a precise characterization of their atmospheres. 2MASS~J22081363+2921215 is a young L3  brown dwarf, member of the $\beta$-Pictoris young moving group (23$\pm$3~Myr), that shares its effective temperature and mass with the {$\beta$~Pictoris~b} giant exoplanet. We performed a $\sim$2.5~hr spectro-photometric $J$-band monitoring of 2MASS~J22081363+2921215 with the MOSFIRE multi-object spectrograph, installed at the Keck\,I telescope. We measured a minimum variability amplitude of 3.22$\pm$0.42\% for its $J$-band light curve. The ratio between the maximum and the minimum flux spectra of 2MASS~J22081363+2921215 shows a weak wavelength dependence, and a potential enhanced variability amplitude in the alkali lines. Further analysis suggests that the variability amplitude on the alkali lines is higher than the overall variability amplitude (4.5--11\%, depending on the lines). The  variability amplitudes in these lines are lower if we degrade the resolution of the original MOSFIRE spectra to R$\sim$100, which explains why this potential enhanced variability in the alkali lines has not been found previously in HST/WFC3 light curves. Using radiative-transfer models, we obtained the different cloud layers that might be introducing the spectro-photometric variability we observe for 2MASS~J22081363+2921215, that further support the measured enhanced variability amplitude in the alkali lines. We provide an artistic recreation of the vertical cloud structure of this {$\beta$-Pictoris b} analog.

\end{abstract}



\section{Introduction} \label{intro}

Brown dwarfs are substellar objects that are unable to sustain hydrogen fusion, contracting as they cool down over their lifetime. Thus, younger brown dwarfs have larger radii and lower surface gravity than their older counterparts. Young brown dwarfs and giant exoplanet atmospheres share similar colours, temperatures, and surface gravities \citep{Chauvin2004, Marois2008, Faherty2013}. Nevertheless,  young brown dwarfs, unlike giant exoplanets, are found in isolation, being technically easier to observe with  current instrumentation. Thus, the characterization of young free-floating brown dwarfs {might} improve our understanding of the atmospheres of imaged young giant exoplanets. Some examples of these class of objects are  2MASS~J00452143+1634446 (L2, $\sim$50~Myr, \citealt{Kendall2004}), PSO~318.5-22 (L7, 23$\pm$3~Myr, \citealt{Liu2013}), 2MASS~J00470038+6803543 (L7, 130$\pm$20~Myr, \citealt{Gizis2012}), 2MASS~J035523.37+113343.7 (L5, $\sim$120~Myr, \citealt{Reid2006}), and 2MASS~J22081363+2921215 (L3, 23$\pm$3~Myr, \citealt{Cruz2009}), among others. 

Photometric or spectro-photometric variability surveys  with ground  and space-based data have shown that the majority of brown dwarfs have signs of low-level variability across all spectral types,  most likely due to the existence of different layers of heterogeneous clouds in their atmospheres that evolve as they rotate \citep{Radigan2014, Buenzli2014, Metchev2015}. For example, \cite{Metchev2015} monitored 23 L-dwarfs, and 16 T-dwarfs using the Spitzer telescope, and concluded that $\sim$61\% of the L-dwarfs of their sample show photometric variability signs with amplitudes $>$0.2\%, and also, at least 31\% of the T-dwarfs showed signs of low-level variability. In addition, \cite{Metchev2015} suggested that variability amplitudes for low gravity brown dwarfs might be enhanced in comparison with the field brown dwarf population. 

Using the New Technology Telescope (NTT) and the United Kingdom Infrared Telescope (UKIRT), \cite{Vos2019} photometrically monitored a sample of 36 young brown dwarfs with spectral types between L0 and L8.5, finding that $\mathrm{30^{+16}_{-8}}$\% of the young brown dwarfs were variable. In contrast, \cite{Radigan2014} found that only $\mathrm{11^{+13}_{-4}}$\% of the field brown dwarfs with the same spectra types are variable using also ground-based data. These results suggest that the variability may be enhanced for low-gravity/low-mass exoplanet analogs. In fact, for free-floating young planetary mass objects like WISEP J004701+680352, VHS 1256-1267ABb and PSO J318.5-22,   very high variability amplitudes have been measured \citep{Lew2016,Biller2018, Zhou2020, Bowler2020}.

Finally, photometric and spectro-photometric variability  has been also predicted for giant exoplanets \citep[respectively]{Komacek2020, Tan2019}. 
%For hot Jupiters, this variability is most likely due to atmospheric circulation because of their large dayside-to-nightside flux contrast. The variability amplitudes predicted for hot Jupiters vary between 0.1\% and few percent. In fact, the search for photometric variability in hot Jupiters  have started to provide positive results in a few hot Jupiters, like Kepler-76b \citep{Jackson2019}, or at least to constrain the upper limits for others, like HD 189733b and HD 209458b \citep{Kilpatrick2020}, for which they constrain upper limits of variability amplitude of 5.6\% and 6.0\%, and 12\% and 1.6\%, respectively in the channels 1 and 2 of Spitzer. 
Their source of photometric variability is expected to be atmopsheric dynamics, as for brown dwarfs. Some attempts to measure photometric variability in giant exoplanets have been performed using Extreme Adaptive Optics instrumentation. For example,  \cite{Apai2016}, and \cite{Biller2021}, attempted to use VLT/SPHERE to measure the photometric variability of the HR~8799 system, but, due to the lack of a long data baseline, {just} upper limits for variability amplitude could be provided.  
In conclusion, detecting photometric and  spectro-photometric variability in  giant exoplanets is  challenging due to instrumental limitations. Thus, as young brown dwarfs and giant exoplanets share several physical characteristics, and given the  easier observability of young brown dwarfs, and the higher chances of finding detectable variability, spectro-photometric monitoring of these objects can provide insights on the heterogeneous cloud coverage of exoplanet atmospheres, and the vertical pressure levels at which those are found. 

This paper is structured as follows: in Section \ref{target}, we introduce the key properties of 2MASS~J22081363+2921215. In Section \ref{observations}, we describe the details of the Keck~I/MOSFIRE spectro-photometric monitoring we performed for 2MASS~J22081363+2921215. In Section \ref{reduction}, we describe the data reduction. In Section \ref{light_curves} we explain how the light curve production and correction was performed using the calibration stars. In Section \ref{systematics} we account for the potential influence of systematics in the target's light curve. In Section \ref{results} we present the results for photometric and spectro-photometric variability for 2MASS~J22081363+2921215. Finally, in Section \ref{interpretation}, we describe the interpretation of the spectro-photometric variability found for 2MASS~J22081363+2921215 using radiative-transfer models, and we provide a general picture of the cloud structure of the object might be given the spectro-photometric variability measured.

\section{2MASS~J22081363+2921215} \label{target}

2MASS~J22081363+2921215 (2M2208+2921), $\mathrm{M_{J}}$~=~15.8, was one of the first peculiar early L objects found \citep{Kirkpatrick2000},  because of its weak alkali lines. It was spectrally classified in the optical by \cite{Kirkpatrick2000}, as an L2  object.  Its peculiarity was later explained as an effect of low-surface gravity \citep{Kirkpatrick2008}. \cite{Cruz2009} classified it as an L3$\gamma$ in the near infrared. \cite{Allers2013} classified it as a very-low surface gravity object using spectral indices. Using the BT-Settl atmospheric models with solar metallicity, \cite{Manjavacas2014} estimated its effective temperature in 1800~K, and its surface gravity in log~g$\sim$4.0.

\cite{Zapatero_Osorio2014} provided a trigonometric parallax for 2M2208+2921 of $\pi$ = 21.2$\pm$0.7~mas, proper motions of $\mu_{\alpha} \cos \delta$ = 90.7$\pm$3.0~mas/yr, and $\mu_{\delta}$ = −16.2$\pm$3.7~mas/yr,  and a luminosity of $\mathrm{\log(L/L_{\odot}}$)~= -3.71$\pm$0.10. \cite{Gagne2014} found, with a modest probability of 53.8\%, that 2M2208+2921  belongs to the $\beta$-Pictoris young moving group (23$\pm$3~Myr, \citealt{Mamajek2014}). \cite{Dupuy2018} confirmed 2M2208+2921 to be a likely member of the $\beta$-Pictoris using also radial velocity measurements from \cite{Vos2017}. In this case, 2M2208+2921 would have an estimated mass between 9 and 11~$\mathrm{M_{Jup}}$, being an analog of the planet/brown dwarf companion {$\beta$~Pictoris~b \citep{Lagrange2009}.  $\beta$~Pictoris~b was one of the first directly-imaged planets detected. It is a companion to the  $\beta$-Pictoris star  at 8-14~AU, with a spectral type $\mathrm{L2\pm2}$ \citep{Bonnefoy2013}, and with a dynamical mass of $\mathrm{13^{+0.3}_{-0.4} M_{Jup}}$ \citep{Dupuy2019}. \cite{Dupuy2019} also showed that 2M2208+2921 and $\beta$-Pictoris~b share a similar position in the color-magnitude diagram, further confirming the similarity of the two objects}.

\cite{Metchev2015} measured a rotational period of  3.5$\pm$0.2~h for 2M2208+2921 using \textit{Spitzer} [3.6] and [4.5] bands, with variability amplitudes of 0.69$\pm$0.07\%, and 0.54$\pm$0.11\%, respectively. 
\cite{Miles-Paez2017_pol} measured low values of $J$-band polarization for the object. Finally, \cite{Vos2017} measured  an inclination of $i$ = 55$\pm$10~deg.

\section{Observations} \label{observations}

Performing spectro-photometric monitoring observations from the ground using single-slit spectrographs is technically challenging, since at least one calibration star is needed for spectral calibration, to account for telluric contamination, changes in the airmass, humidity and temperature variations in the atmosphere, etc, {that might potentially introduce spurious variability signals}. Normally, brown dwarfs are isolated, and no other object is found  {close enough to be observed simultaneously as spectro-photometric calibrator together with the target, in the few arseconds long of single-slits spectrographs. This is only possible in the case of well-resolved binary brown dwarfs like Luhman-16AB \citep{Kellogg2017}}. {Since brown dwarfs are in their majority single objects \citep{Bouy2003, Burgasser2003, Luhman2007}}, near-infrared multi-object spectrographs like MOSFIRE \citep{McLean2010, McLean2012} are needed to perform spectro-photometric monitoring of brown dwarfs from the ground. {MOSFIRE is installed at the Cassegrain focus of Keck\,I,} and it performs  simultaneous spectroscopy of up to 46 objects in a 6.1'x 6.1' field of view, using the Configurable Slit Unit (CSU),  a cryogenic robotic slit mask system that is reconfigurable electronically in less than 5~minutes without any thermal cycling of the instrument. A single photometric band is covered in each instrument setting ($Y$, $J$, $H$ or $K$).

\begin{figure}
    \centering
    \includegraphics[width=0.46\textwidth]{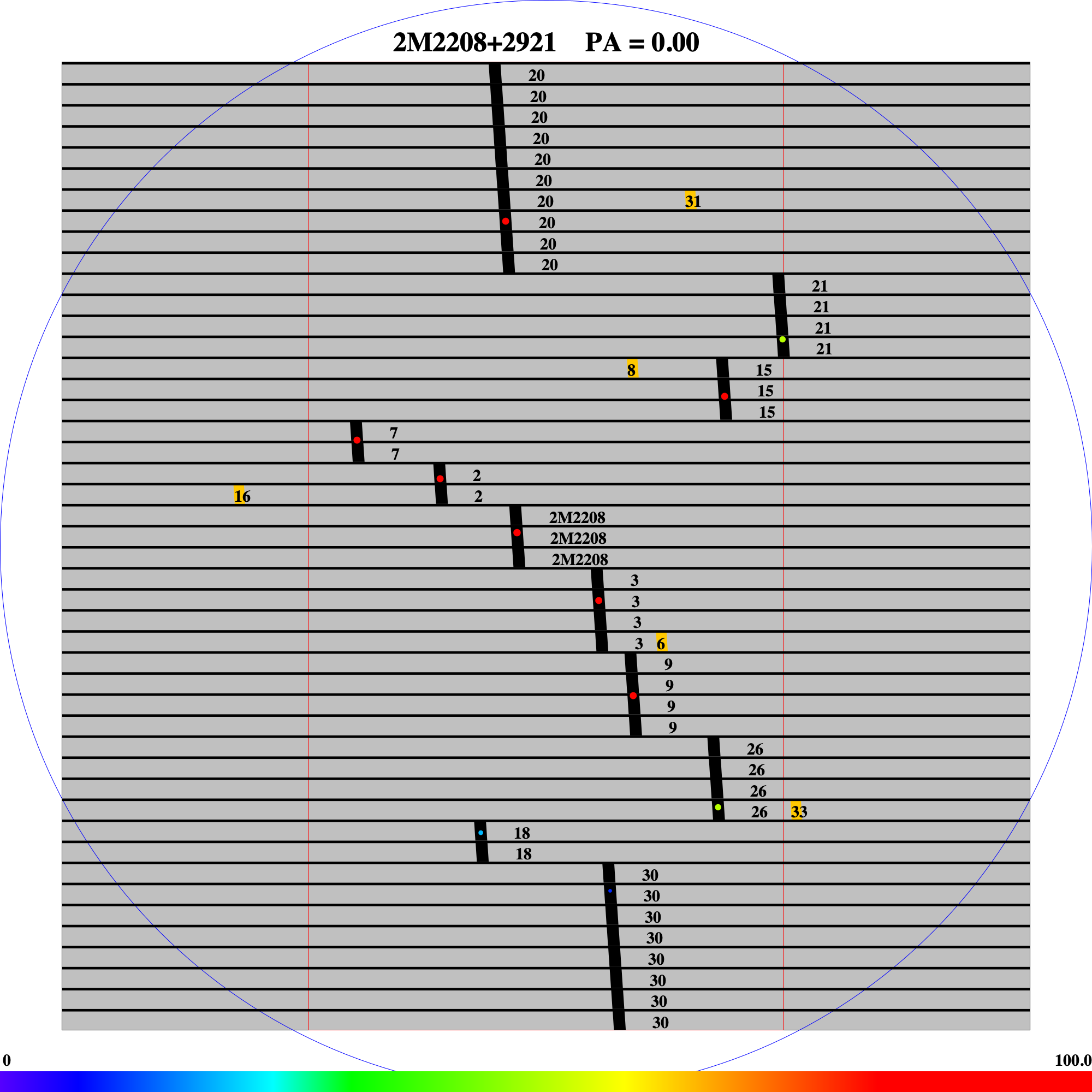}
    \caption{Illustration of the positioning of the CSU bars on MOSFIRE to obtain simultaneous multi-object spectroscopy of the field of 2M2208+2921 as produced by MAGMA, the MOSFIRE Automatic GUI-based Mask Application. Our target (named as 2M2208) is placed in the center of the field. The position of the comparison stars as shown in Table~\ref{calstars} are also marked. The round colored points show the expected positions of the target and calibration stars. The yellow squares show the position of the stars used for the alignment of the mask.}
    \label{mosfire_field}
\end{figure}

We observed 2M2208+2921 on UT 2019-10-13 with MOSFIRE at the Keck\,I telescope during half a night. We obtained in total 13 spectra of 2M2208+2921 in the $J$-band (1.153--1.352~$\mu$m) using an ABBA pattern during a total of $\sim$2.5~h of monitoring. We used wide slits of 4.5" to avoid slit losses for all 10 calibration stars and the target, obtaining a spectral resolution of R$\sim$1000. In Table~\ref{calstars} we show the list of objects used as calibrators, their coordinates, and their $J$-band magnitudes. In general, the calibration stars had similar magnitudes as the target. In Figure \ref{mosfire_field}, we show the configuration of the CSU mask, with the position of the target and the calibration stars. We used exposure times of 150~s for each nod position in the the first ABBA, and 180~s for each nod position of the rest of the ABBA patterns. We observed over a airmass range of 1.01 and 1.35.

For data reduction purposes, 13 dome flats of 11~s exposure were obtained. Due to challenges in producing a successful wavelength calibration using sky lines with 4.5" slits, we obtained on UT 2020-03-05 four $J$-band "sky" spectra  using the same configuration for the multiobject mask as for the observations, but using 1.0" slits to obtain higher resolution sky lines. The 1.0" slits provided spectra of the skylines with enough resolution to allow the pipeline to produce an accurate wavelength calibration.

\begin{table}
	\caption{Information about the calibration objects in the field of 2M2208+2921.}  
	\label{calstars}
	\centering
	\begin{center}
		\begin{tabular}{llllll}
        \hline
		\hline 
			
Num. mask & Num. obj. & RA & DEC & $\mathrm{M_{J}}$  \\   
\hline

20 & 1 & 22:08:13.962 & 29:23:19.62 & 16.14   \\
21 & 2  & 22:08:05.925 & 29:22:34.83 & 16.32   \\
15 & 3   & 22:08:05.925 & 29:22:34.83 & 15.61   \\
7  & 4  & 22:08:18.266 & 29:21:56.62 & 15.83   \\
2  & 5   & 22:08:15.857 & 29:21:41.9 & 15.86   \\
2M2208 & 2M2208 &  22:08:13.631 & 29:21:21.54 & 15.80  \\
3 & 6 &  22:08:11.258 & 29:20:55.81 & 15.88  \\
9 & 7 &  22:08:10.257 & 29:20:19.74 & 15.16     \\
26 & 8 &   22:08:07.798 & 29:19:37.43 & 16.43  \\
18 & 9 & 22:08:14.681 & 29:19:27.78 & 16.49   \\
30 & 10 &  22:08:10.930 & 29:19:05.76 & 16.04  \\

			\hline			
		\end{tabular}
	\end{center}

\end{table}

\section{Data Reduction} \label{reduction}

We used the version 1.0 of \textit{PypeIt\footnote{https://github.com/pypeit/PypeIt}} to reduce the multi-object spectroscopic data acquired with MOSFIRE in the $J$-band. PypeIt is a Python-based data reduction pipeline for spectroscopic data, applicable to a variety of spectrographs in different facilities \citep{Prochaska2019,Prochaska2020}. The pipeline corrected all the raw images from dark current, and a bad pixels mask is generated. The edges of the slits were traced using the dome flats. A master flat was also created. PypeIt produced a wavelength calibration for our data using the sky arc frames taken using the same multiobject mask we employed for our observations, but with narrower slits of 1.0", to obtain well-resolved skylines that would allow PypeIt to find a wavelength solution automatically. The wavelength calibration accounted for the spectral tilt across the slit. The calibrations were applied to our science frames, and the sky was subtracted using the A-B or B-A frames following \cite{Kelson2003}. The 1D science spectra were extracted from the 2D sky-corrected frames. Finally we coadded A-B and B-A the extracted 1D science spectra to obtain a signal-to-noise of $\sim$65 at 13000~{$\AA$} for our science target. The signal-to-noise achieved for each object on the field is summarized in Table~\ref{stats_calstars}. No telluric calibration was performed for these spectra, {since the spectral types of the calibration stars, necessary to perform this correction,  could not be determined}. Instead, for the upcoming analysis, we have used the wavelength range between 12200 and 13200~{$\AA$}, avoiding the most prominent telluric contamination.

\section{Production and Correction of Light Curves}\label{light_curves}

We produce a $J$-band light curve for each object in the field, restricting the wavelength range of the spectra between 12200 and 13200~{$\AA$}, to avoid the most prominent telluric contamination that might introduce spurious variability for the objects in the field. 

As these data were obtained from the ground, there might be other additional sources of non-astrophysical contamination affecting the shape of the light curve extracted for each object, such as varying atmospheric transparency, change in the water vapor content of the atmosphere, the seeing, variations in outside temperature during the $\sim$2.5~h of the observation, wind speed and direction variations, airmass variations, etc. Thus, the science target light curve needs to be corrected for those potential sources of contamination. 

To perform the light curve correction, we followed a similar approach to \cite{Radigan2014}, but with  more conservative criteria to select the best calibration stars. We corrected each light curve by dividing it by a calibration light curve produced median combining the relative-flux light curves of all the other objects in the field, beside the science target. First, we normalized the light curves of all objects to the median flux for each of them. For each reference star, a calibration light curve was created by median combining the light curves of all the other objects, beside the science target. Then, the raw light curve of each calibration star was divided by its corresponding calibration light curve to obtain a corrected light curve. Finally, we measured the standard deviation, $\sigma$, of each corrected light curve for each calibration star. In Table~\ref{stats_calstars} we show the standard deviation of all the calibration stars and our science target before and after correcting each light curve. To perform an optimal correction of the 2M2208+2921 light curve, we choose first which calibration stars are less likely to show intrinsic astrophysical variability, due to star spots and/or flares. To choose the more stable calibration stars, we selected those for which their standard deviation is at most half of the standard deviation of the target's light curve ($\sigma_{star} < \sigma_{target}/2$). Using this criteria, we selected five stable calibration stars (stars 1, 4, 5, 6, and 8), that coincide with the calibration stars showing a smaller degree of variability amplitude after they were corrected using the rest of the calibration stars in the field (see Table~\ref{stats_calstars}). We show the uncorrected and corrected light curves for the calibration stars in the Appendix, Figures \ref{non_corr_LCs} and \ref{corr_LCs}. The uncertainties for the data points in the light curves are  the formal instrumental uncertainties provided by the \textit{PypeIt} pipeline.

For a comparison, we also explored for our sample the best-selection criteria for the calibration stars used by \cite{Radigan2014}, for which they subtracted from the corrected light curved of each calibration star, a shifted version of itself, and divided it by $\sqrt{2}$ ($\sigma_{s}=[f_{cal}-f_{cal\_shifted}]/\sqrt{2}$). \cite{Radigan2014} then  identified  poor-quality calibration stars as those where $\sigma_{s}>1.5\times \sigma_{target}$. This criteria did not reject any of the calibration stars in our field, thus, we used  the more conservative method detailed above to choose the most stable calibration stars. 

\begin{figure*}
    \centering
    \includegraphics[width=0.48\textwidth]{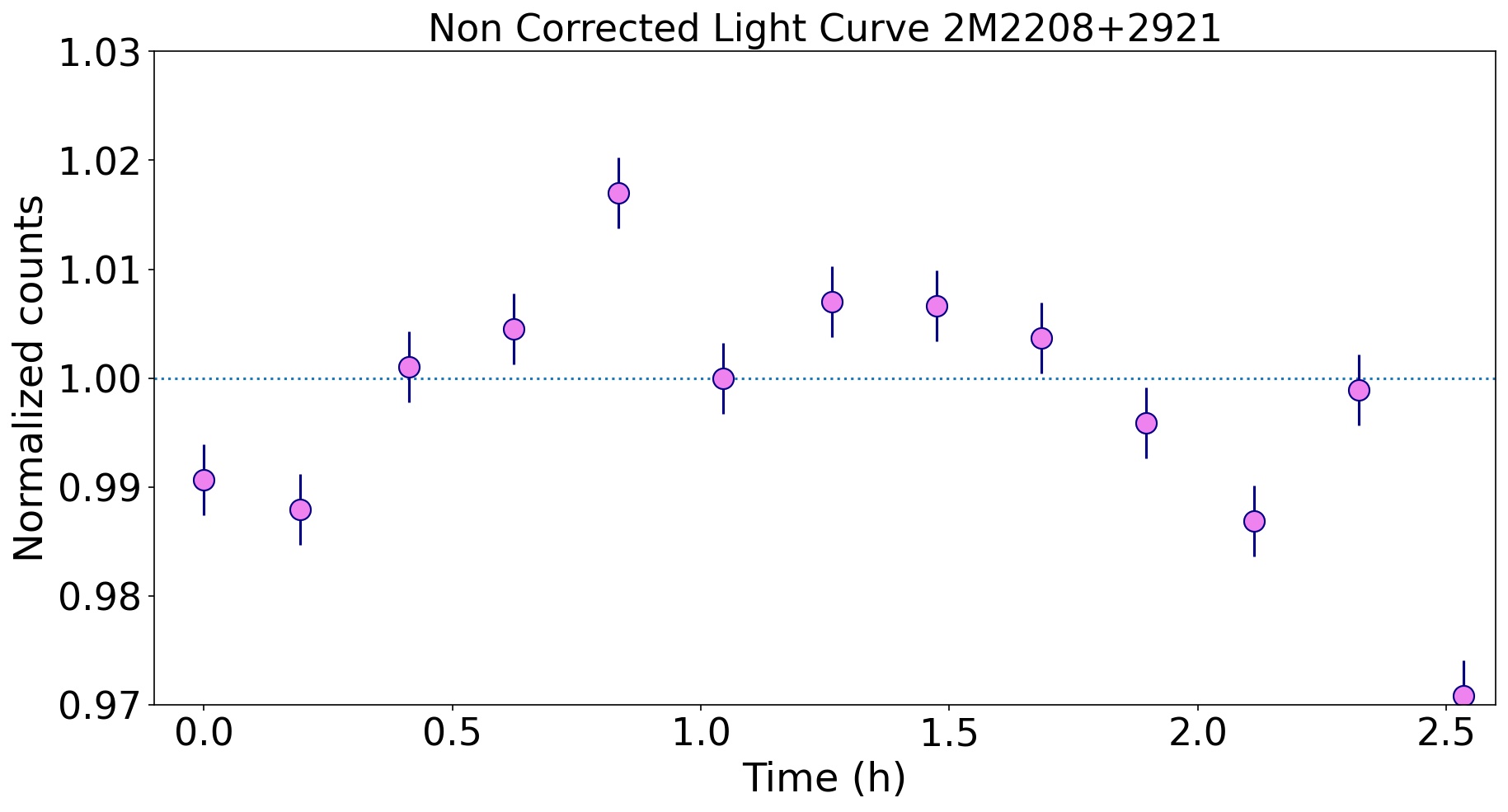}
    \includegraphics[width=0.48\textwidth]{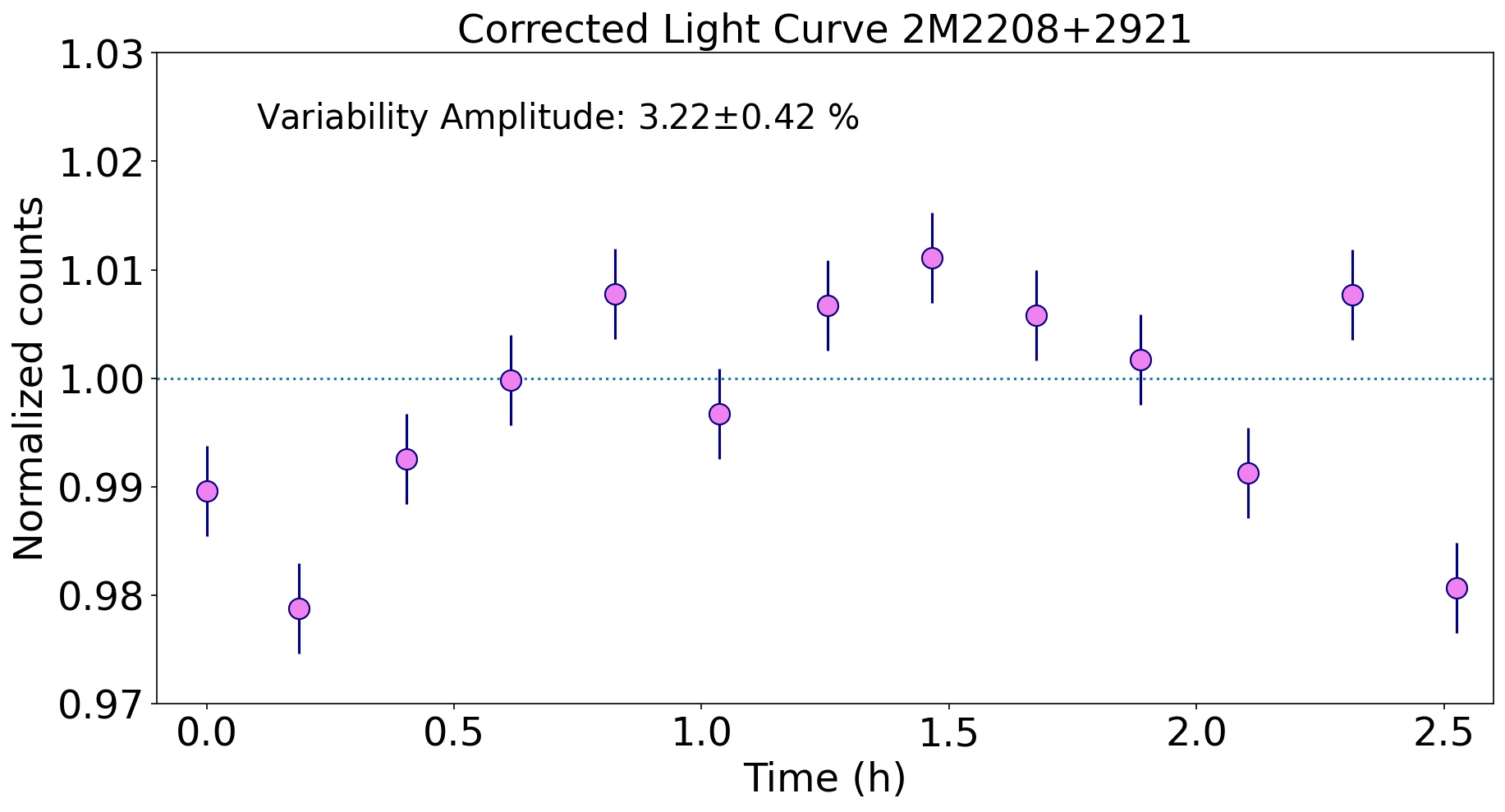}
    \caption{Normalized non-corrected (left) and corrected (right) light curves of  2M2208+2921.}
    \label{target_LCs}
\end{figure*}

The formal instrumental uncertainties provided by \textit{PypeIt} probably underestimate the uncertainties of 2M2208+2921 light curve, since it does not necessarily account for spurious variability introduced by changes in the Earth's atmosphere during the observation. Thus, we use a similar approach as \cite{Radigan2014} to estimate the uncertainties for each point of the light curve. We used the mean of the $\sigma_{s}$ calculated for the target and the selected calibration stars as the uncertainty for each point in the light curve of the target. This method  accounts for any residual uncorrected atmospheric contamination variability in the target's light curve. 
The non-corrected light curve of 2M2208+2921 is shown in Fig. \ref{target_LCs}, left, and the corrected light curve in Fig. \ref{target_LCs}, right.

\begin{table*}
	\caption{Statistics of the light curves of 2M2208+2921 and the calibration stars on its field. We highlight in bold face the best reference stars, selected as those with $\sigma_{calibration\_stars}$ $<$ $\sigma_{2M2208}/2$}  
	\label{stats_calstars}
	\centering
	\begin{center}
		\begin{tabular}{ccccc}
        \hline
		\hline 
			
Object Number & SNR at 13000~$\AA$ & $\sigma$ non-corrected light curve & $\sigma$ corrected light curve & Variability after correction \\   
\hline
 2M~2208+2921 &  64.3 & 1.12 x $10^{-2}$  & 9.70 x $10^{-3}$ & 3.22 \%  \\
{Object 1} & 48.5 & 5.30 x $10^{-3}$ & 4.17 x $10^{-3}$  & 1.32 \%\\
Object 2  & 36.2 & 1.07 x $10^{-2}$  & 5.15 x $10^{-3}$ &  2.00 \%  \\
Object 3  & 36.8 & 7.03 x $10^{-3}$ & 5.76 x $10^{-3}$ & 2.10 \% \\
{Object 4}  & 98.6 & 5.93 x $10^{-3}$ & 3.73 x $10^{-3}$  & 1.22 \% \\
{Object 5}   & 55.3 & 6.31 x $10^{-3}$ & 3.55 x $10^{-3}$   &  1.18 \% \\
{Object 6} &  53.0 & 5.59 x $10^{-3}$ & 3.40 x $10^{-3}$  & 1.39 \%  \\
{Object 8} &   76.4 &  8.31 x $10^{-3}$ &  3.72 x $10^{-3}$  & 1.26 \% \\
 Object 9 & 40.1 & 1.03 x $10^{-2}$ &  5.79 x $10^{-3}$  & 2.24 \% \\
 Object 10 &  43.4 & 1.02 x $10^{-2}$ & 6.93 x $10^{-3}$   & 2.21 \%  \\

			\hline			
		\end{tabular}
	\end{center}

\end{table*}

\subsection{BIC Test for Significant Variability} \label{BIC_test}

\begin{figure}
    \centering
    \includegraphics[width=0.50\textwidth]{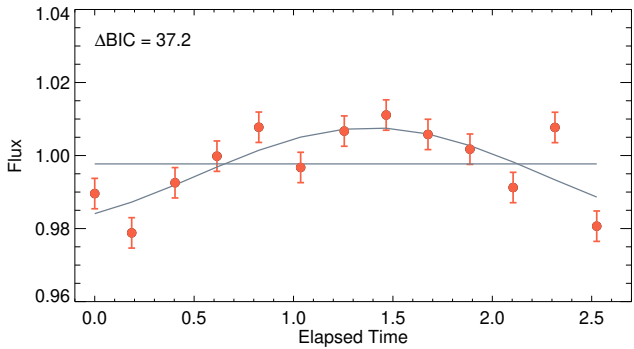}
    \caption{The light curve of 2M2208+2821 is shown by the orange points, with the best-fit non-variable (flat) and variable (sinusoidal) models are shown in grey. The BIC test shows that the variable model is strongly favored by the light curve.}
    \label{BIC_plot}
\end{figure}

To test the significance of the observed fluctuations in the light curve of 2M2208+2921, we use the Bayesian Information Criterion (BIC). The BIC is defined as 
\begin{equation}
    \mathrm{BIC}=-2~\mathrm{ln}~\mathcal{L}_\mathrm{max} + k ~\mathrm{ln}~ N
\end{equation}
where $\mathcal{L}_\mathrm{max}$ is the maximum likelihood achievable by the model, $k$ is the number of parameters in the model and $N$ is the number of data points used in the fit \citep{Schwarz1978}. In our case, we calculate $\Delta\mathrm{BIC} = \mathrm{BIC}_{flat} - \mathrm{BIC}_{\mathrm{sin}}$ to assess whether a variable sinusoidal or non-variable flat model is favored by the data. This method has previously been used for identifying brown dwarf variability by \citet{Naud2017, Vos2020}. The BIC penalizes the sinusoidal model for having additional parameters compared with the flat model. The sinusoidal and flat model are shown in Fig.~\ref{BIC_plot}. A $\Delta\mathrm{BIC}$ value of 37 implies that the variable model (sinusoidal) is very strongly preferred over a flat  model. 

\section{Systematics Corrections} \label{systematics}

\subsection{Comparison of Variability between Blue and Red Half of Spectrum}\label{LC_half}

Telluric contamination in the $J$-band  spectra asymmetrically affects the blue and the red edges, and also some intermediate wavelengths \citep{Rudolf2016}, that could potentially influence the variability amplitude we measure for 2M2208+2921.
To test the potential influence of telluric contamination in the light curve of 2M2208+2921, we produced two different light curves using only the first and the second half of the wavelength range of spectra. The first half spectra light curve was produced using the spectra between 12200 and 12700~$\AA$, and the second half light curve was produced using the range between 12700 and 13200~$\AA$ (see both light curves in the Appendix, Fig.~\ref{Correlation_LC_halves}). Both light curves looked visually  similar, but to quantitatively test that both light curves are  similar, and thus, the telluric contamination is not affecting significantly the spectra in the blue and red ends, we run a Mann-Whitney U test, which is a non-parametric test that checks the similarity on the distribution of two samples \citep{Neuhauser2011}. The Mann-Whitney U test does not assume a normal distribution in the data. The null hypothesis asserts that the medians of the two samples are identical. We calculate the value, U, and compared it to a tabulated $\mathrm{U_{critical}}$  given by  the number of points in the sample. If U $>$ $\mathrm{U_{critical}}$, the null hypothesis $H_{0}$ (samples are similar), is accepted. For the case of our target, the calculated U = 94.5 $>$ $\mathrm{U_{critical}}$ = 45, for a sample of 13 points, as in the case of 2M2208+2921 light curve. {We calculate the Kendall $\tau$ non-parametric correlation test \citep{Puka2011} between the target's light curve done with the first half and the second half wavelength range. We chose the Kendall $\tau$ correlation test since it is a non-parametric test to measure correlation of data, and more robust that other parametric correlation test like the Spearman $\rho$ test \citep{Langfelder2012}.} We obtained a Kendall $\tau$ coefficient of 0.85, (significance = 5.2$\times$$\mathrm{10^{-6}}$), indicating a strong correlation between both light curves,  supporting the U-test result.

\subsection{Correlation between Stars and Target Light Curves}\label{corr_LC}

To evaluate the effects of potential contamination on the target's light curve due to atmospheric effects, and potential thin clouds, we investigate the correlation between the non-corrected light curve of the target, and the comparison stars.  The Kendall's $\tau$ coefficients  suggests a weak to a moderate correlation between the light curves, depending on the "good" comparison star. The Kendall $\tau$ correlation coefficients vary between 0.18 (significance = 0.43) and 0.46 (significance = 0.03). In Fig. \ref{corr_LC_stars} in the Appendix, we show the correlation plots between the target and each of the stars that we use for calibration.

After correcting the 2M2208+2921 light curve using the method explained in Section \ref{light_curves}, we run the Kendall $\tau$ non-parametric correlation test again, finding correlation coefficients that range between 0.05 (significance = 0.85) and -0.33 (significance = 0.12), suggesting from non to a weak anticorrelation for some of the "good" comparison stars (see Fig. \ref{corr_LC_stars_corrected} in the Appendix).

\subsection{Correlation with Full Width Half Maximum of the Spectra}\label{corr_fwhm}

We obtained spectra following an ABBA pattern, thus, the slit losses might vary slightly at A and B positions of the pattern, potentially influencing the measured variability of the target. Thus, we investigated a potential relationship between the variability found for 2M2208+2921, and the Full Width Half Maximum (FWHM) of the target spectra taken during the 2.5~hr of monitoring with MOSFIRE. We measured the FWHM at three different positions of the coadded ABBA spectra in the spectral direction (x direction): across pixel x = 683, across pixel x = 1042, and pixel x = 1365, and then we calculated the mean of the FWHM at those three positions for each ABBA coadd. We obtained a median FWHM of 0.84$\pm$0.15 arcsec during the 2.5~hr of monitoring (see Fig, \ref{evol_FWHM}, left, in the Appendix). We calculated the Kendall $\tau$ correlation of the mean FWHM for each spectrum and the evolution flux of the spectra with time, obtaining a very weak negative correlation between both quantities ($\tau$ = -0.077, significance = 0.76, Fig. \ref{evol_FWHM}, right).

\subsection{Correlation with Atmospherical Parameters}\label{meteo}

The evolution of atmospheric conditions during the observation might influence the amount of flux collected by MOSFIRE, affecting simultaneously the target and the calibration stars. Namely, the most relevant factors that might potentially affect our observations are: the humidity content, the external temperature, and the airmass (Fig. \ref{corr_RH_temp_airm} in the Appendix). The evolution of these parameters are registered in the header, and/or in the Mauna Kea Weather Center webpage (\url{http://mkwc.ifa.hawaii.edu}). We calculated the $\tau$ Kendall correlation coefficient between the non-corrected light curve for 2M2208+2921, and each of the atmospheric parameters mentioned above. We found no correlation between the target's light curve and the humidity content ($\tau$ = -0.08, significance = 0.72), a weak correlation with the external temperature (0.35, significance = 0.09), and a weak anti-correlation with the airmass ($\tau$ = -0.20, significance = 0.37).

{Since these correlations are very small or not statistically significant, we conclude that there is no correlation between the external conditions and the target's light curve}.

\section{Results} \label{results}

\subsection{Photometric variability}\label{phot_variability}

As we did not cover the entire known rotational period of the target (3.5$\pm$0.2~hr, \citealt{Metchev2015}), with our MOSFIRE spectro-photometric observations,  we are just able to provide a minimum variability amplitude for this light curve in the $J$-band, that we found to be 3.22$\pm$0.42\%. As expected, this minimum variability amplitude is higher than the variability amplitude measured by Spitzer in the [3.6] and [4.5] channels \citep{Metchev2015}, which were 0.69$\pm$0.07\%, and 0.54$\pm$0.11\%, respectively. The $J$-band is tracing deeper layers of the atmosphere of 2M2208+2921 than the [3.6] and [4.5] bands, and thus, a higher variability amplitude is expected, assuming that the variability amplitudes measured with Spitzer have not changed significantly between epochs \citep{Yang2016}.

{We do not have enough time coverage to observe a full rotational period of the target (3.5$\pm$0.2~hr, \citealt{Metchev2015}), but still we  searched for other periods on the $J$-band light curve using a Lomb-Scargle periodogram \citep{Lomb1976, Scargle1982, Horne_Baliunas1986}, and a Bayesian Generalized Lomb–Scargle (BGLS) Periodogram \citep{Mortier2015} which did not find any periodicity in the $J$-band light curve}.

\subsection{Spectral variability} \label{spec_variability}

\begin{figure*}
\vspace{0.7cm}
    \centering
    \includegraphics[width=0.45\textwidth]{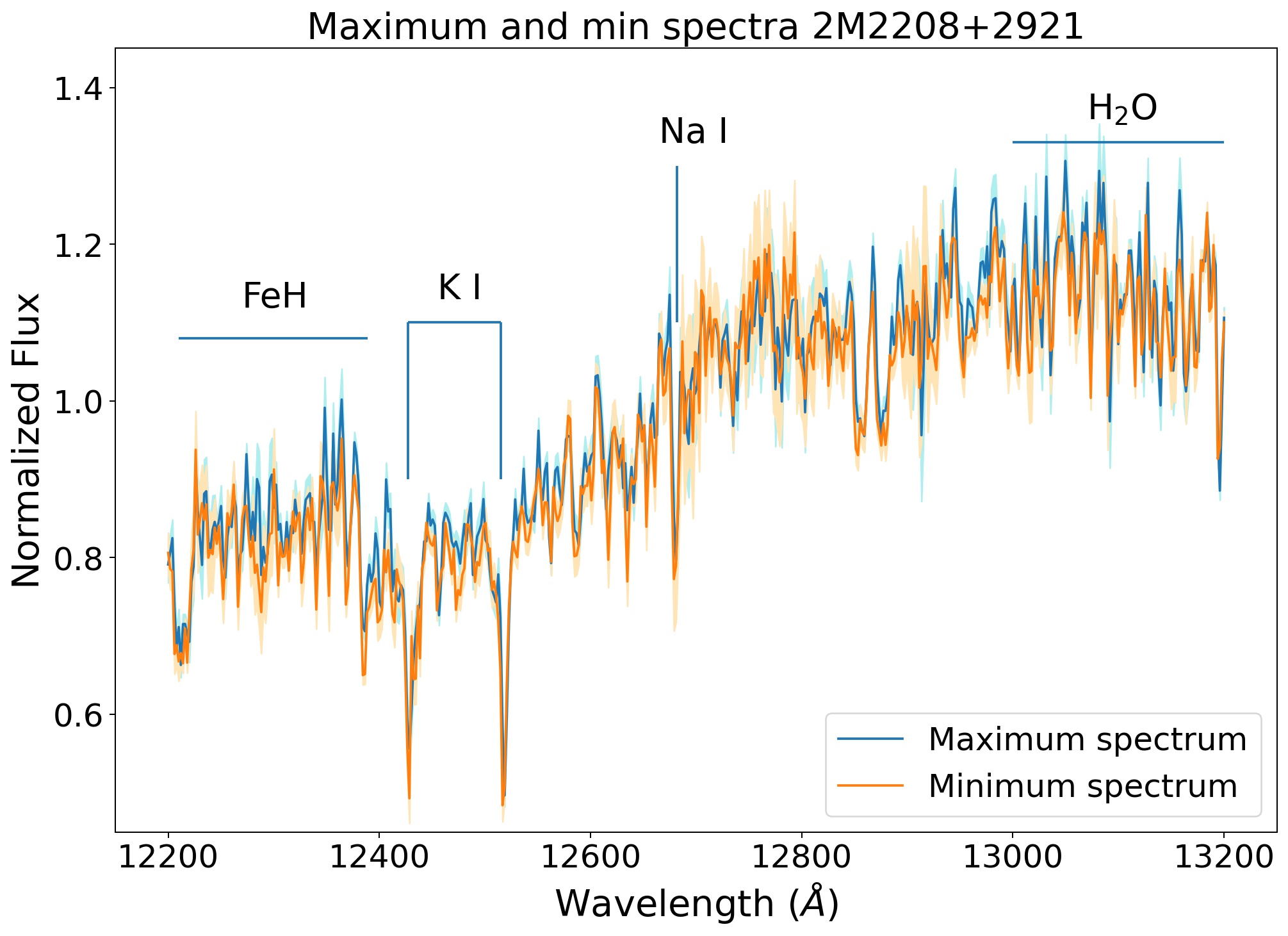}
    \includegraphics[width=0.45\textwidth]{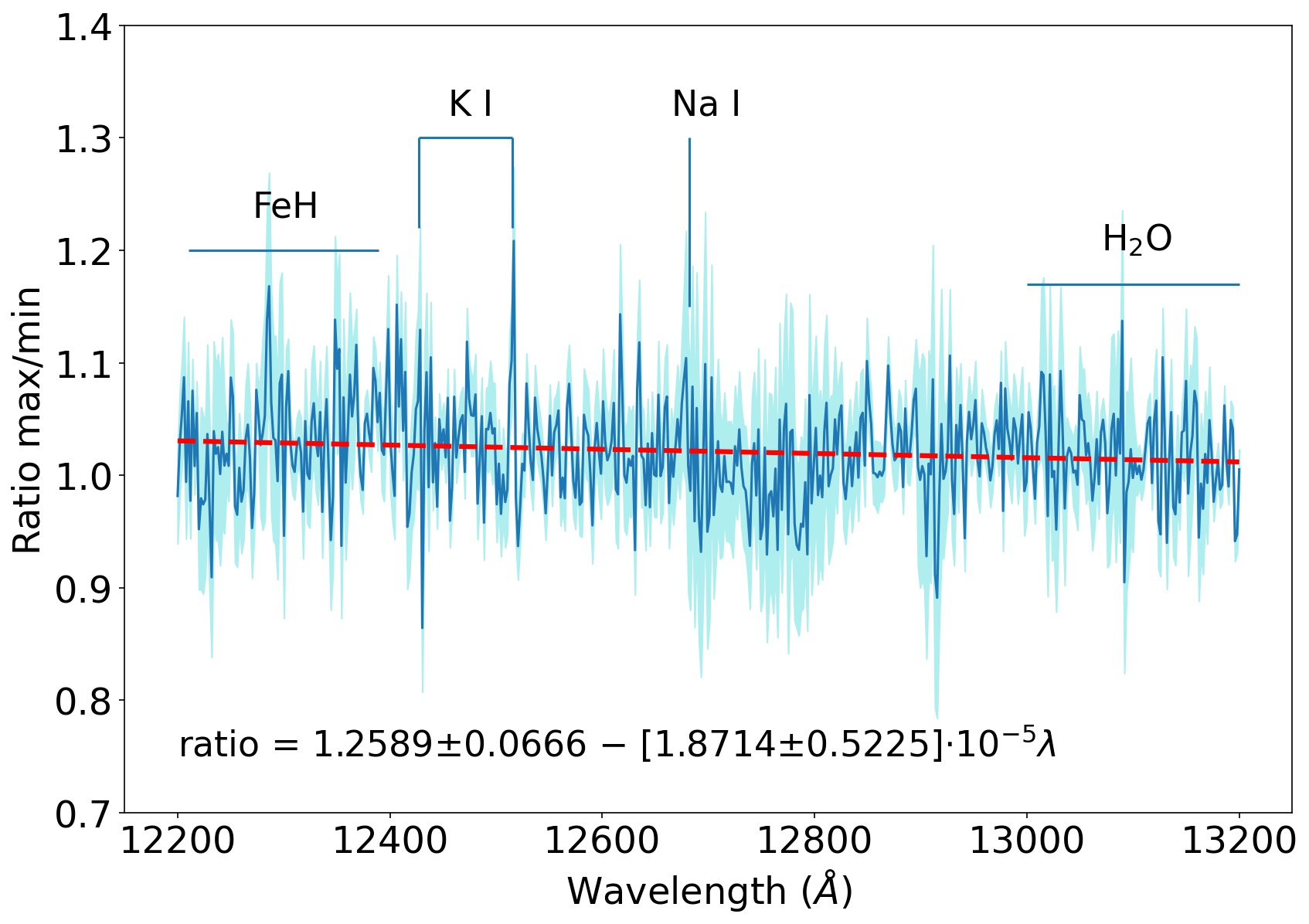}
    \caption{Left: We show the spectrum corresponding to the maximum flux obtained in the 2M2208+2921 light curve in blue, and the spectrum of the minimum spectrum in orange. Right: Ratio between the maximum and minimum flux spectrum of 2M2208+2921 (blue). The uncertainties of the ratio are in light blue. The fitted slope to the ratio is shown in red.}
    \label{ratio}
\end{figure*}

We explored the amplitude of the variability as a function of the wavelength by comparing the maximum and the minimum flux spectra among the 13 spectra obtained. In Figure \ref{ratio}, left, we show the brightest and faintest  spectrum, indicating the molecular and atomic absorption features for 2M2208+2921.

In Figure \ref{ratio}, right we show the ratio between the maximum and the minimum flux spectra, i.e. the relative amplitude across the spectral wavelength range, with its uncertainties, and  indicating as well the molecular and atomic absorption features for our target. We fit a line to the ratio of the maximum and minimum flux spectra using the \textit{numpy.polyfit} \textit{Python} library, obtaining a {negative slope to the ratio ($ratio = 1.2589\pm0.0666-[1.8714\pm0.5225]\times 10^{-5}\lambda$, see Fig. \ref{ratio}, right)}, suggesting that the variability amplitude decreases monotonically from 12200 to 13200~$\AA$, as it has been found for other {L-dwarfs like WISE0047+6803 ($ratio = 1.19\pm0.01 - [0.7\pm0.1]\times10^{-5} \lambda$), or LP261-75B ($ratio = 1.05\pm0.01 - [0.27\pm0.05]\times 10^{-5} \lambda]$)}. As proposed for WISE0047+6803 in \cite{Lew2016}, the variability amplitude, and wavelength dependence for 2M2208+2921 could be explained by the existence of hazes and dust particles in the atmosphere of the object. \cite{Hiranaka2016} proposed the existence of sub-micron sized particles in the atmospheres of L0-L6 brown dwarfs. {For L3 dwarfs \cite{Hiranaka2016} finds an effective radius of $\sim$0.27~$\mu$m, slightly smaller particles than for WISE0047+6803 atmosphere (0.3-0.4~$\mu$m, \citealt{Lew2016})}. For the same number of particles, smaller particles imply smaller variability amplitude, and a stronger wavelength dependence of the variability \citep{Hiranaka2016}, {which is what we find for 2M2208+2921 when compared to WISE0047+6803. WISE0047+6803 has a variability amplitude of $\sim$8\%, higher than the 3.22$\pm$0.42\% for 2M2208+2921, and a less strong wavelength dependence than 2M2208+2921}.

\subsection{Potential enhanced variability in the alkali lines}\label{alkali_var}

In Figure \ref{ratio} we found potentially prominent peaks at the wavelengths where the K\,I, and Na\,I alkali lines are located, suggesting a potential enhanced variability amplitude around those wavelengths. In the following, we investigate in depth the potential enhanced variability amplitude on those wavelengths. {For this purpose, in the following sections we measure the amplitude of variability inside the  K\,I doublet and the Na\,I alkali lines, and the variability amplitude of the blue and the red continuum of those lines. Finally we compare those variability amplitudes between them and with the overall $J$-band variability, and conclude if they are significantly different}.

\subsubsection{Variability of flux inside the K\,I and Na\,I lines} \label{Na_KI_line_variability}

%To investigate the potential enhanced variability amplitude of the K\,I and Na\,I alkali lines, we  measured the variability amplitude of pseudo equivalent widths (pEW) of those lines using different continuum wavelength ranges. We found that the actual measurement of the pEW for those lines  strongly depends on the continuum chosen. Thus, to find a minimum continuum range that provides a robust measurement of the pEW, we measured the value of the pEW with different continuum lengths from 0 to 100~$\AA$, in increments of 10~$\AA$, at both sides of the line. In Figure \ref{pEW_var_continuum} we show the variability of the pEW depending on the width of the continuum used. We observe that the value of the variability  has a tendency to stabilize for values higher than a determined width of the continuum. For both lines of the K\,I doublet and the Na\,I line, this value is approximately 40~$\AA$, as observed in Figure \ref{pEW_var_continuum}, where the stabilization width is marked with a black dashed line. For broader continuum widths, we obtained a similar measurement of the pEW variability, but higher uncertainties on its measurements, due most likely to higher noise added when we increment the width of the continuum used to measure the pEW.   Thus, we used 40~$\AA$  to obtain the most accurate value of the variability for the pEW for the K\,I doublet and the Na\,I line. 

We  investigated the variability of the flux inside the K\,I doublet and the Na\,I line themselves, {creating light curves using the flux inside these lines}. We  used the range between 12400--12463~$\AA$ for the K\,I line at 12430~$\AA$. The range between 12495--12540~$\AA$ for the K\,I line at 12525~$\AA$, and the range between 12675--12683~$\AA$ for the Na\,I line at 12682~$\AA$. To correct the light curves for the K\,I doublet and Na\,I lines from potential non-astrophysical contamination, we follow the same approach to correct the light curve as for the $J$-band light curve (see Section \ref{light_curves}), but using only the wavelength ranges of the calibration stars spectra corresponding to the K\,I doublet and Na\,I  wavelengths specified above (see correction light curves in Appendix, Fig. \ref{corr_LCs_obj1}, \ref{corr_LCs_obj4}, \ref{corr_LCs_obj5}, \ref{corr_LCs_obj6}, and \ref{corr_LCs_obj8}). This correction  accounts for potential telluric contamination at those specific wavelengths. This correction is particularly important for the Na\,I continuum and line, since there is a $\mathrm{O_{2}}$ telluric absorption between 12600~$\AA$ and 12750~$\AA$ \citep{Vernet2008, Sameshima2018}. In spite of our efforts to correct for telluric contamination at those wavelengths, we acknowledge that some contamination might remain uncorrected. 

{In Figure \ref{line_continuum_evol}, right panel, we show the corrected light curves corresponding to alkali lines}. We find that the variability of the flux for the K\,I lines at 12430~$\AA$, and 12525~$\AA$ is 2-3$\sigma$ higher than the  variability found for the $J$-band of 2M2208+2921 (4.60$\pm$0.54\%, and 4.48$\pm$0.54\%, respectively). Finally, the variability for the Na\,I line is about 2$\sigma$ higher than for the overall $J$-band, and also than the variability of the continuum, 10.93$\pm$3.17\%.

\subsubsection{Variability of alkali lines continuum flux} \label{Na_KI_continuum_variability}

We measured the variability of the continuum on the blue, and the red end of each line, expanding 40~$\AA$ in both ends. The wavelengths used as continuum for the K\,I at 12430~$\AA$  is 12360--12400~$\AA$ in the blue end, and 12463--12503~$\AA$ in the red end. For the K\,I line at 12525~$\AA$, we have used as blue side continuum the wavelength range between 12455--12495~$\AA$, and as red side continuum 12540--12580~$\AA$. Finally, for the Na\,I line at 12682~$\AA$, we have used as blue end continuum the wavelength between 12635--12675~$\AA$, and as red end continuum the range between 12720--12760~$\AA$.  We corrected the K\,I doublet and Na\,I continuum light curves as explained in Section \ref{light_curves} (see correction light curves in Appendix, Fig. \ref{corr_LCs_obj1}, \ref{corr_LCs_obj4}, \ref{corr_LCs_obj5}, \ref{corr_LCs_obj6}, and \ref{corr_LCs_obj8}). 

\begin{figure*}
    \centering
    \includegraphics[width=1.02\textwidth]{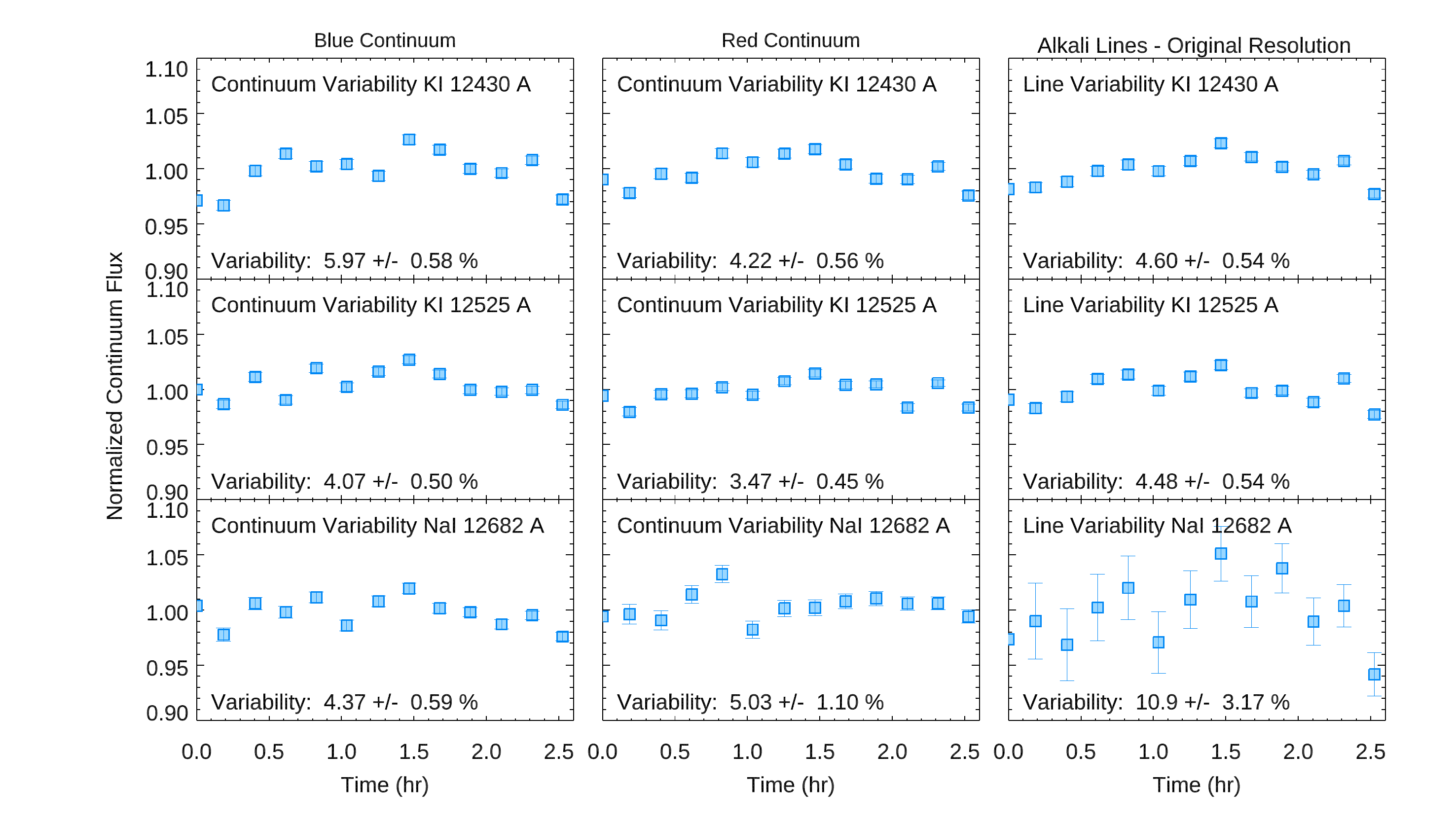}
    \caption{{Variability of the K\,I doublet, and the Na\,I lines and their blue and red continuum}. The continuum width used is 40~$\AA$ in both ends.}
    \label{line_continuum_evol}
\end{figure*}

In {Figure~\ref{line_continuum_evol}, left and middle panel}, we show the normalized continuum flux variability. As we observe in Fig.~\ref{line_continuum_evol}, {the variability amplitude of the continuum of the alkali lines, and the variability inside the alkali lines themselves is similar within the uncertainties for the K\,I lines. For the Na\,I line the variability of the line is  1--2$\sigma$ higher than the variability of the continuum. In any case,  the variability amplitude found for the continuum of the K\,I doublet and Na\,I alkali lines is slightly higher (1-2 $\sigma$)} than the  overall variability found in the $J$-band for 2M2208+2921 ({3.22$\pm$0.42\%})

\subsubsection{Comparison to low resolution spectro-photometric data}

Although some spectro-photometric data for other brown dwarfs of similar spectral types to 2M2208+2921  have already been collected using the \textit{Hubble Space Telescope} (HST), and its \textit{Wide Field Camera 3} (WFC3) with the G141 grism (R$\sim$100)  (e.g. 2MASS J17502484-0016151, a L4.5 brown dwarf from \citealt{Buenzli2014}, and 2MASS J18212815+1414010, a L5.0 from \citealt{Yang2015}), no enhanced variability amplitude has been found for the alkali lines in the $J$-band for those objects. Thus, we investigate if the enhanced variability inside those lines is washed out when the spectral resolution of the MOSFIRE/Keck~I spectra is degraded to the resolution of the HST/WFC3 + G141 grism spectra. For this purpose, we degraded the MOSFIRE/Keck~I spectra resolution (R$\sim$1000) to the resolution of HST/WFC3 + G141 (R$\sim$100) using a gaussian convolution. We reproduced the plots \ref{ratio}, and \ref{line_continuum_evol},  using the R$\sim$100 resolution spectra, after correcting the light curves following the same procedure than in Section \ref{light_curves} (see correction light curves in Appendix, Fig. \ref{corr_LCs_obj1_R100}, \ref{corr_LCs_obj4_R100}, \ref{corr_LCs_obj5_R100}, \ref{corr_LCs_obj6_R100}, and \ref{corr_LCs_obj8_R100}), and we compare the variability amplitude found for the continuum and inside the K\,I doublet and Na\,I alkali line.

\begin{figure*}
\vspace{0.7cm}
    \centering
    \includegraphics[width=0.42\textwidth]{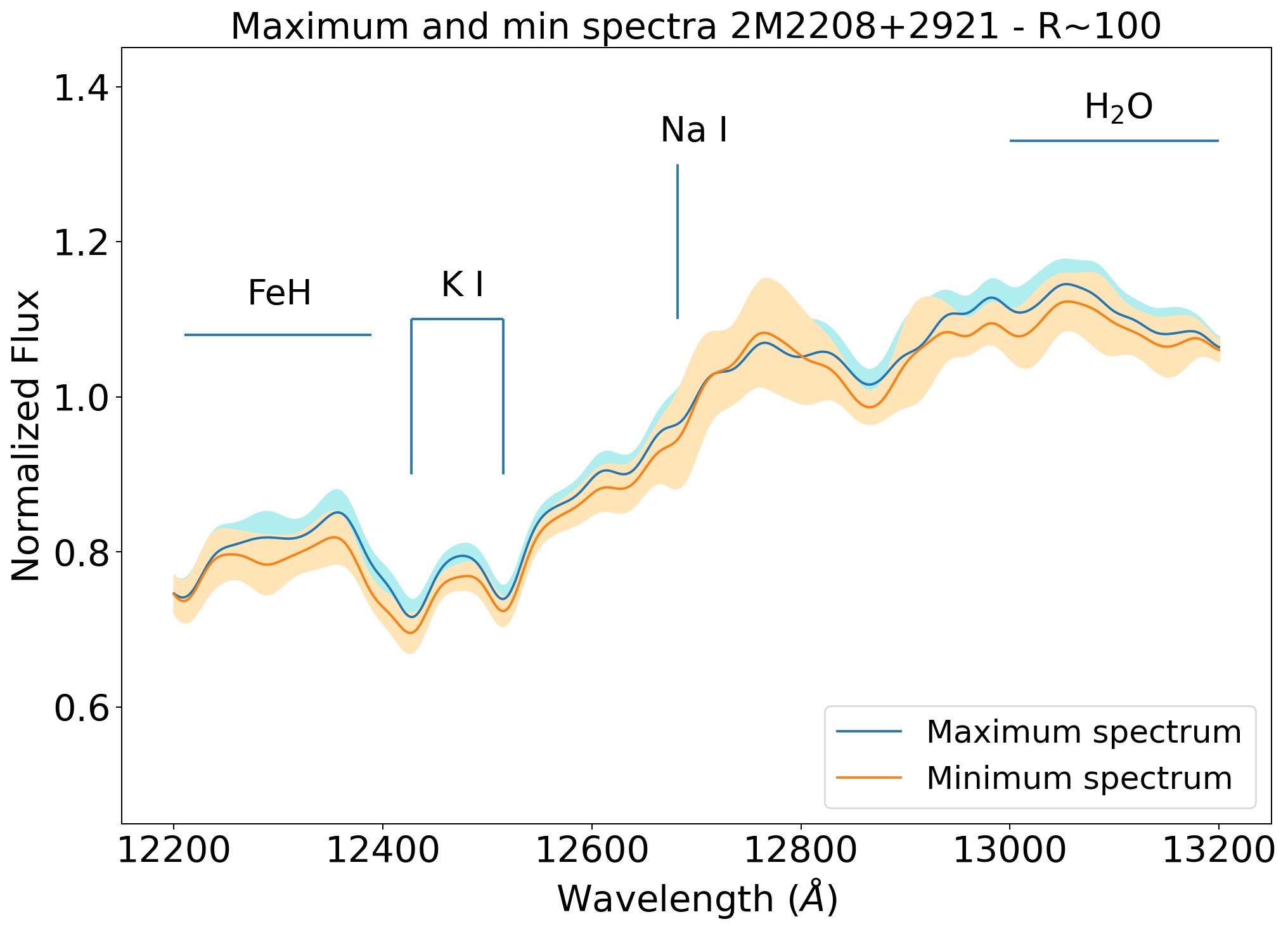}
    \includegraphics[width=0.42\textwidth]{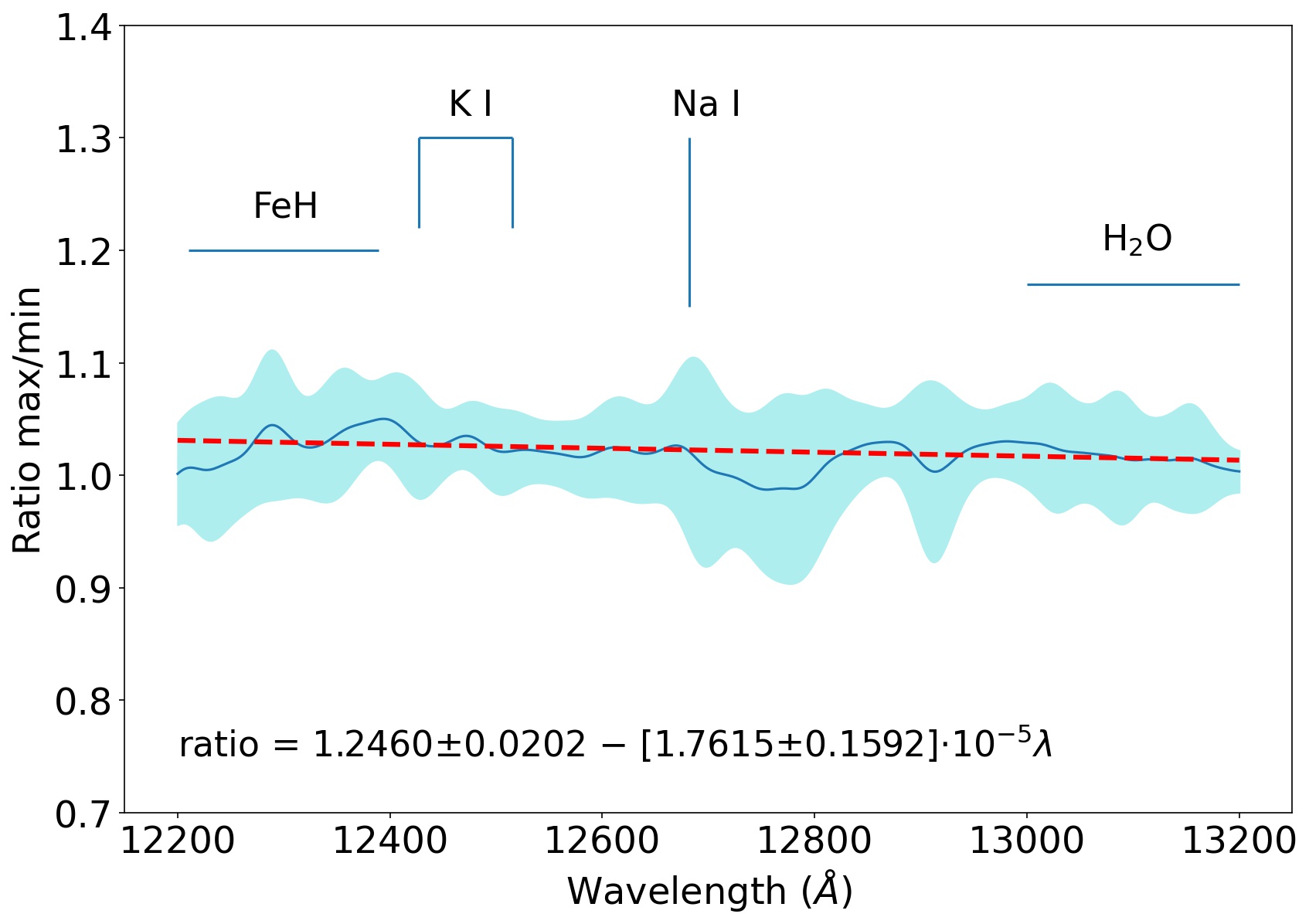}
    \caption{Same as Fig. \ref{ratio} for R$\sim$100 spectra similar to HST/WFC3 + G141 grism. Left: We show the spectrum corresponding to the maximum flux obtained in the 2M2208+2921 light curve in blue, and the spectrum of the minimum spectrum in orange for R$\sim$100. Right: Ratio between the maximum and minimum flux spectrum of 2M2208+2921 for R$\sim$100.}
    \label{ratio100}
\end{figure*}

In Figure \ref{ratio100}, similar as Figure \ref{ratio}, we show the comparison, and the ratio between the maximum and the minimum flux spectra in the 2M2208+2921 light curve. As in Fig. \ref{ratio}, we mark the atomic and molecular features for the $J$-band spectrum. We show the minimum spectrum in orange (corresponding to the second point the $J$-band light curve in Fig.~\ref{target_LCs}), and the maximum spectrum in blue (corresponding to the 8th point in the $J$-band light curve in Fig.~\ref{target_LCs}). In Fig. \ref{ratio100} right, we observe that within the uncertainties, the maximum and minimum spectra overlap, and in Fig. \ref{ratio100}, left, we observe that there is a wavelength dependent slope, as in Fig. \ref{ratio}, but we observe no remarkable peaks indicating potential enhanced variability amplitude in some wavelengths. Nevertheless, the overall maximum and/or minimum in the spectral lines does not necessarily coincide with the maximum and/or minimum of the $J$-band light curve.  The values of the linear fit to the ratio between the maximum and the minimum spectra are consistent with those in Fig. \ref{ratio} (right).

\begin{figure*}
    \centering
    \includegraphics[width=1.02\textwidth]{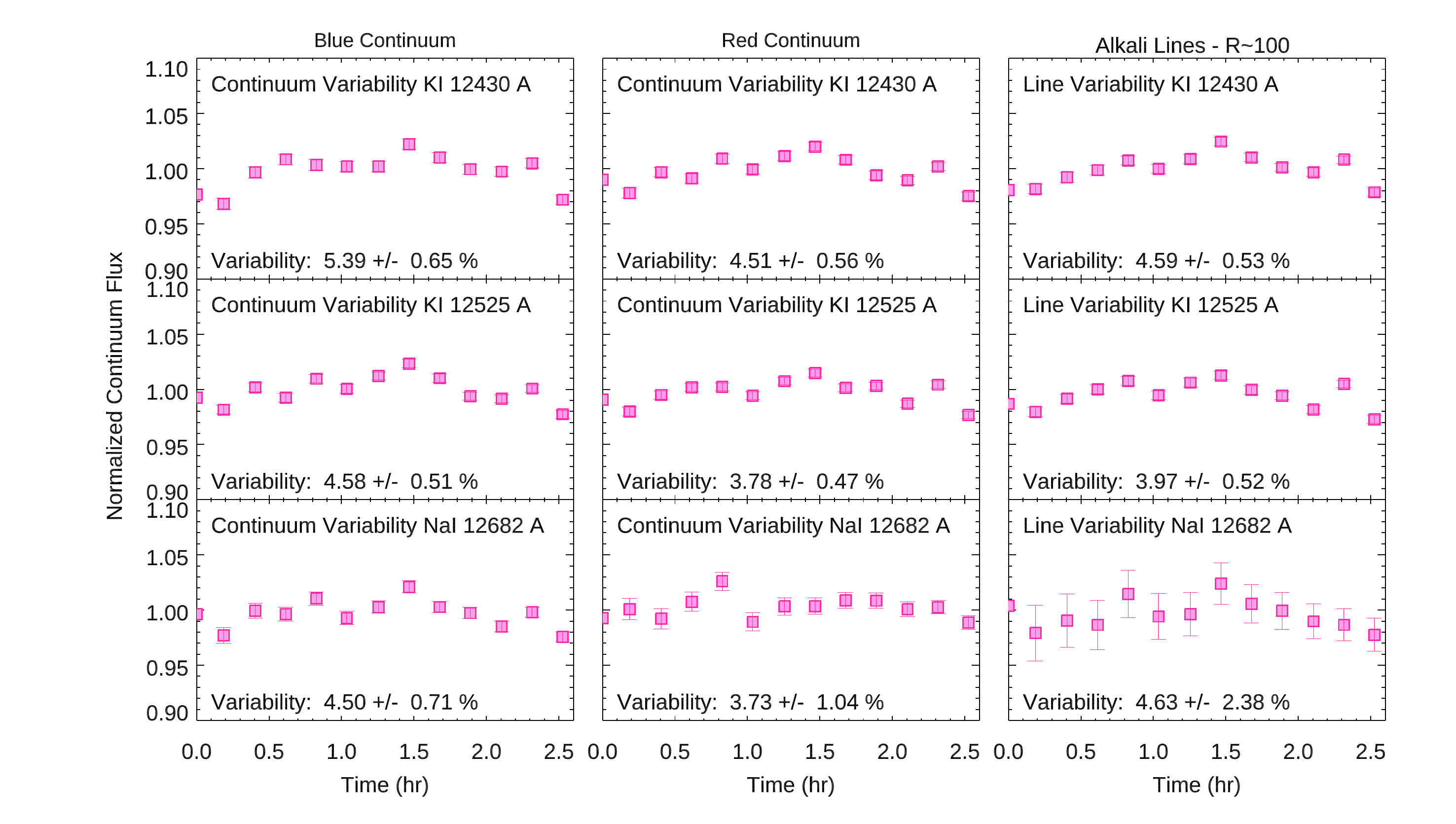}
    \caption{Same as Fig. \ref{line_continuum_evol} for R$\sim$100 spectra similar to HST/WFC3 + G141 grism spectra. {Variability of the K\,I doublet, and the Na\,I lines and their blue and red continuum for R$\sim$100}. The continuum width used is 40~$\AA$ in both ends.}
    \label{line_continuum_evol100}
\end{figure*}

In Figure \ref{line_continuum_evol100}, similar to Fig. \ref{line_continuum_evol}, we show the variability inside the K\,I doublet lines and the Na\,I alkali line measured as for the original resolution MOSFIRE/Keck I spectra, but on the degraded MOSFIRE/Keck I spectra to a resolution similar to the HST/WFC3 + G141 spectra. For the case of the K\,I doublet lines, the variability amplitude inside the lines for the original resolution spectra and the degraded spectra is similar within the uncertainties. For K\,I line at 12430~$\AA$ the variability amplitude inside the line for the original resolution is 3.95$\pm$0.54\%, and for R$\sim$100 is 3.90$\pm$0.53. For the K\,I line at 12525~$\AA$ the variability amplitude is 4.80$\pm$0.54\% for the original resolution spectra, and 4.27$\pm$0.53 for the R$\sim$100 spectra. 

Finally, for the Na\,I at 12682~$\AA$ line, the variability amplitude differs if it is measured at the original resolution spectra, or in the degraded resolution spectra. For the original resolution spectra, the variability of the Na\,I line is 10.93$\pm$3.17\%, and measured on the R$\sim$100 resolution spectra is 4.63$\pm$2.38\%, which is consistent with the variability amplitude measured for the overall $J$-band. Therefore, this result suggests that the enhanced variability inside the Na\,I line is partially washed out when the resolution of the spectra is low, and the individual alkali lines cannot be resolved, as it happens in the case of the HST/WFC3 + G141 grism spectra. Thus, this would explain why enhanced variability in the Na\,I line has not been found in HST/WFC3 + G141 grism spectra for brown dwarfs of a similar spectral type.

In Figure \ref{line_continuum_evol100}, similar to Fig. \ref{line_continuum_evol}, we show the variability of the continuum measured 40~$\AA$ around the alkali lines as done previously, but for the  MOSFIRE/Keck 1 spectra smoothed to R$\sim$100.  In Fig. \ref{line_continuum_evol100}, we observe that for both blue and red sides of the continuum for the K\,I doublet and the Na\,I lines the variability amplitudes are consistent with the variability amplitudes found for the continuum for the original resolution of the MOSFIRE/Keck I spectra within the uncertainties. Thus, degrading the resolution of the spectra does not significantly influence the measured variability amplitude for the continuum around the K\,I doublet, and the Na\,I line.

\section{Interpretation}\label{interpretation}

\subsection{Description of radiative-transfer models}\label{models}

The emergent flux at diverse wavelengths of the $J$-band MOSFIRE spectrum traces different pressure levels of the atmosphere of 2M2208+2921, {providing information about the cloud coverage at different levels of the atmosphere of the object}. Spectro-photometric variability at those wavelengths can be used to trace the various cloud layers in the atmosphere of the target. {We used a state-of-the-art radiative transfer to calculate the flux contribution of the different modeled pressure levels. We used the effective temperature and surface gravity estimated for 2M2208+2921 in \cite{Manjavacas2014}, with a VLT/ISAAC spectrum that covers the $J$, $H$, and $K$-bands. \cite{Manjavacas2014} used the BT-Settl models \citep{Allard2001, Allard2003, Allard2012a, Allard2012b} in two different released versions (2010 and 2013) to estimate effective temperatures and surface gravities for 2M2208+2921. The adopted atmospheric parameters were $\mathrm{T_{eff}}$ = 1800$\pm$100~K, and log~g~=~4.0$\pm$0.5. Further details on how the spectral fitting was performed can be found in \cite{Manjavacas2014}. }

To obtain the contribution functions for 2M2208+2921, we followed a similar approach to \cite{Yang2016}, using standard radiative-convective equilibrium atmosphere thermal structure models following the approach of \cite{Saumon_Marley2008}. Then, a temperature perturbation was applied at different pressure levels of the atmosphere of the object consecutively, and each time, a new temperature profile was generated, and a new emergent spectrum. The ratio between each  emission spectrum generated for each perturbation at each pressure level, and the spectrum to the baseline case, provides the sensitivity of each wavelength range to temperature perturbations at different pressure levels.

As in \cite{Yang2016}, this procedure was repeated at different pressure levels between 1.8~$\times \mathrm{10^{-4}}$ bars to $\sim$23 bars, obtaining the flux contributions for the wavelengths covered by the MOSFIRE $J$-band, after applying the MOSFIRE $J$-band bandpass, and also for the K\,I and the Na\,I alkali lines, that trace slightly different, and narrower pressure levels. As in \cite{Yang2016}, the results strictly apply only to variations in atmospheric temperature, but they reflect the atmospheric region to which the spectra at a given wavelength are most sensitive.

\subsection{Cloud Layers probed by alkali lines and $J$-band flux}\label{cloud_layers}

\begin{figure}
    \centering
    \includegraphics[width=0.46\textwidth]{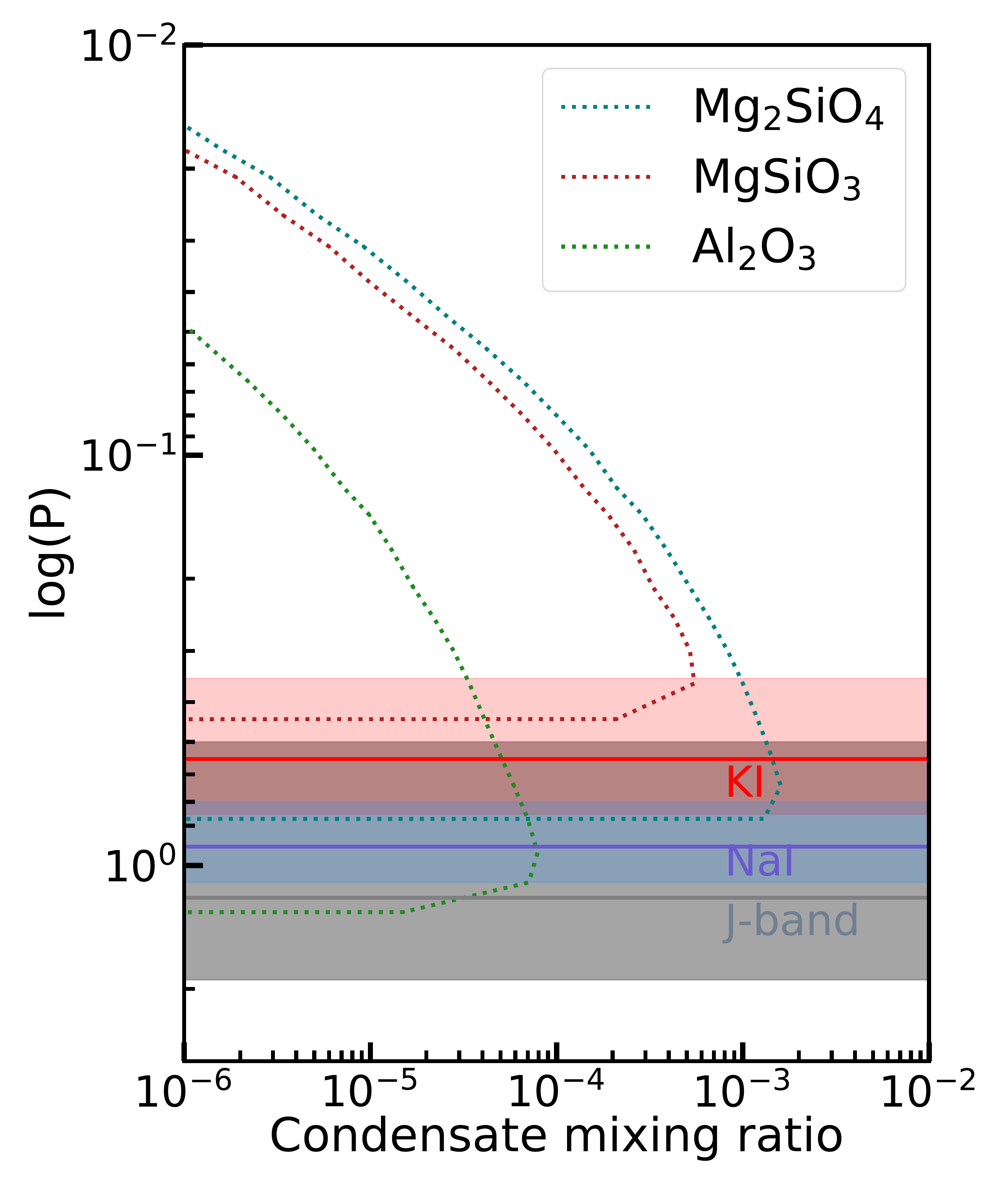}
    \caption{{Condensate mixing ratio (mole fraction) for different silicate clouds ($\mathrm{Mg_{2}SiO_{4}}$, blue dotted line, $\mathrm{MgSiO_{3}}$, red dotted line, and $\mathrm{Al_{2}O_{3}}$, green dotted line) versus vertical pressure in the atmosphere of a model comparable to 2M2208+2921. The grey band indicates the pressure levels that the $J$-band traces, the blue band indicates the pressure levels traced by the Na\,I line, and the red band indicates the pressures levels traced by the K\,I doublet.}}
    \label{vertical_structure}
\end{figure}

In Figure~\ref{vertical_structure}, we show the result of the radiative transfer model for the different atmospheric pressure levels traced by the MOSFIRE $J$-band spectrum, and the K\,I and the Na\,I alkali lines. We also include an uncertainty for the pressures probed by assigning an error-bar equal to the average pressure difference probed between the core and edge of the wings of the lines for the K\,I and the Na\,I alkali lines. For the $J$-band we use half the average pressure range probed in the band. 
We overplot the predicted condensate mixing ratio (mole fraction) for three different types of silicate clouds: $\mathrm{Mg_{2}SiO_{4}}$, $\mathrm{MgSiO_{3}}$, and $\mathrm{Al_{2}O_{3}}$. The pressure levels where the condensate mixing ratio reaches a maximum indicate the bottom of the that type of silicate cloud. Above that pressure level, the condensate mixing ration decreases as the pressure level decreases. The bottom of the $\mathrm{Mg_{2}SiO_{4}}$ cloud is around the 1.0~bars. For the $\mathrm{MgSiO_{3}}$ cloud is around 0.58~bar, and for the $\mathrm{Al_{2}O_{3}}$ is around 1.7~bar.

%As observed in Figure~\ref{vertical_structure}, the radiative transfer models predict that the Na\,I alkali lines traces slightly deeper atmospheric pressure levels than the K\,I doublet. The integrated flux over the $J$-band traces a much wider pressure level range in the atmosphere of 2M2208+2129, thus, so we lose precision on the vertical pressure levels at which the different layers of heterogeneous clouds introduce the spectro-photometric variability we measure. 

{As observed in Figure~\ref{vertical_structure}, the radiative transfer models predict that the K\,I lines trace around the 0.55~bars pressure level and above, Na\,I line traces the pressure level around 0.9~bars and above, and the $J$-band traces the pressure levels around the 1.5 bars, and above. Thus, with the integrated $J$-band light curve, we are observing the blended cloud maps of the three silicate clouds of layers ($\mathrm{Mg_{2}SiO_{4}}$, $\mathrm{MgSiO_{3}}$, and $\mathrm{Al_{2}O_{3}}$). With the integrated flux over the Na\,I line, we are sensitive to the top two layers of clouds ($\mathrm{Mg_{2}SiO_{4}}$, and $\mathrm{MgSiO_{3}}$). Finally, with the integrated flux over the K\,I doublet, we are tracing the uppermost layer ($\mathrm{MgSiO_{3}}$) of the atmosphere of 2M2208+2921.}

\subsection{Modeling the amplitudes and wavelength-dependence of spectral variability}\label{wavelength_dep}

\begin{figure}
    \centering
    \includegraphics[width=0.46\textwidth]{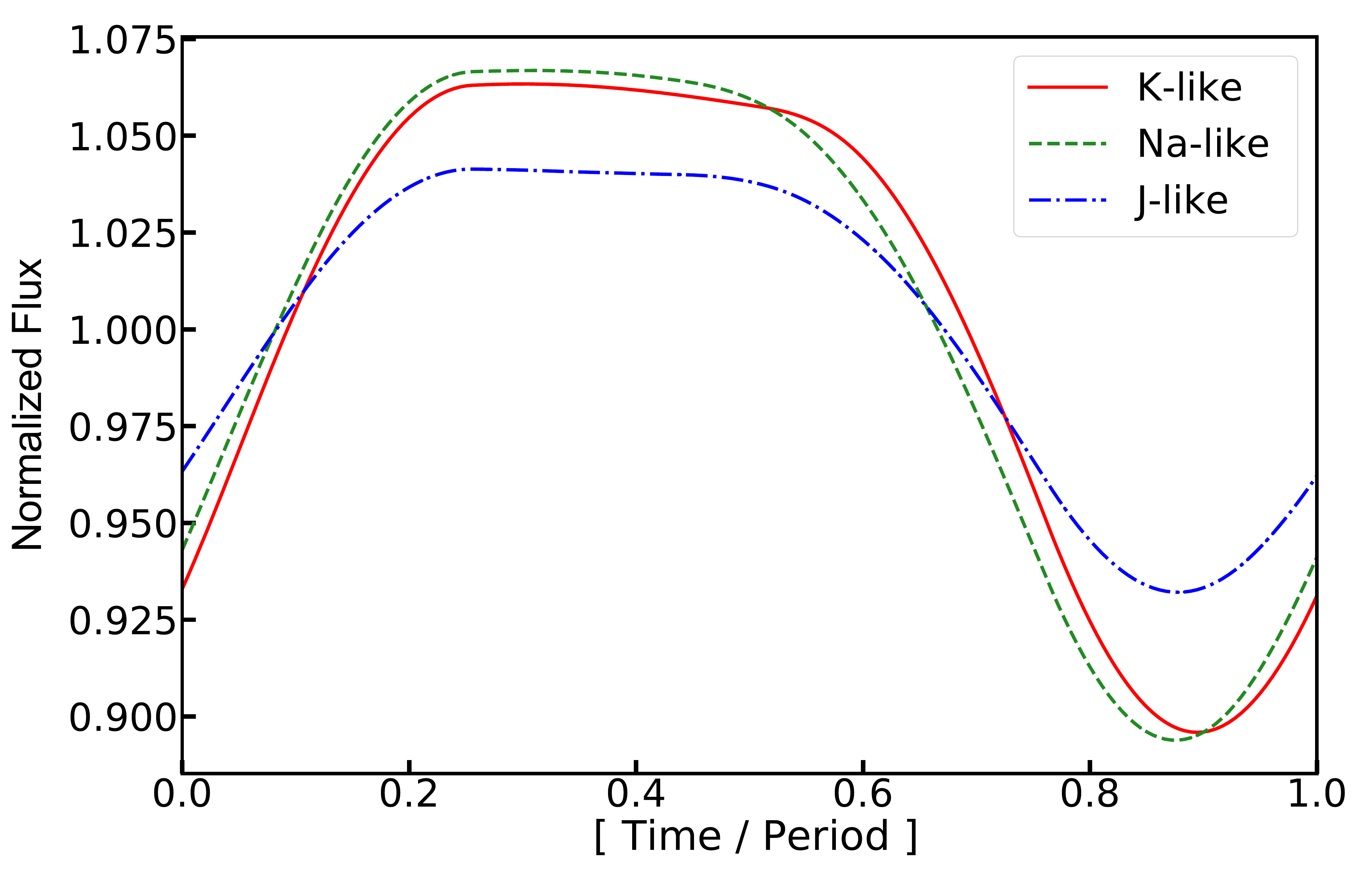}
    \caption{{Simulated} light curves ``observed'' at three different pressure layers (i.e., different wavelength bands) for a toy-model atmosphere with three cloud layers. We assumed random maps for each cloud layer and used information from  the contribution function of 2M2208+2129 to create the ``observed'' light curve at the top of the atmosphere for each band. }
    \label{vertical_structure_lc_amplitude}
\end{figure}

{The smaller amplitude variability measured in our MOSFIRE spectra for the $J$-band in comparison to the alkali lines can be due to a more homogeneous cloud-deck in the lower Al$_2$O$_3$ cloud, which would reduce the observed variability. The larger number of cloud layers probed, which added produce a more ``homogeneous'' cloud coverage, can also affect the observed amplitude of the $J$-band. To test the assumption that the different number of cloud layers probed could affect the observed variability in the $J$-band versus the alkali lines, we modeled the $J$-band, Na\,I and K\,I light curves produced from cloud maps at these three different pressure layers. To produce the light curves we used pixelated maps  \citep[similar to ][]{2015ApJ...814...65K} and compared their disk-integrated light curve shapes and variability amplitudes.  Fig.~\ref{vertical_structure_lc_amplitude} shows the light curves produced at the top of the atmosphere by blending three random, independent maps for three clouds layers of our model atmosphere.
%We made the maps using the mapping heart of \emph{Aeolus} \citep{2015ApJ...814...65K} by 
We randomly assigned two to four spots in each cloud layer and placed them in different, random locations on the map.
To calculate the contrast ratio of the cloud features to the background atmospheric layer, we used information from the temperature--pressure profile of our model atmosphere ($\mathrm{T_{eff}}$ = 1800~K, and $\log g$~= 4.0).%, and a contrast ratio of 150~K. 
We then calculated the average light curve we would observe at the top of the atmosphere by blending the individual light curves using the contribution function information as a weight for each one.} The relative shape of all light curves appears the same, in agreement with our {MOSFIRE K\,I, Na\,I and $J$-band light curves}. The light curve that would correspond to the $J$-band observation has the smallest peak-to-trough amplitude as the chances of a peak of one layer's light curve coinciding with a trough of another (i.e., a cloud clearing of one coinciding with a cloud-decked area of another layer) are larger.  This prediction actually agrees with the spectro-photometric variability amplitudes detected in the MOSFIRE data, as described previously in Section \ref{results}. 

\begin{figure}
    \centering
    \includegraphics[width=0.46\textwidth]{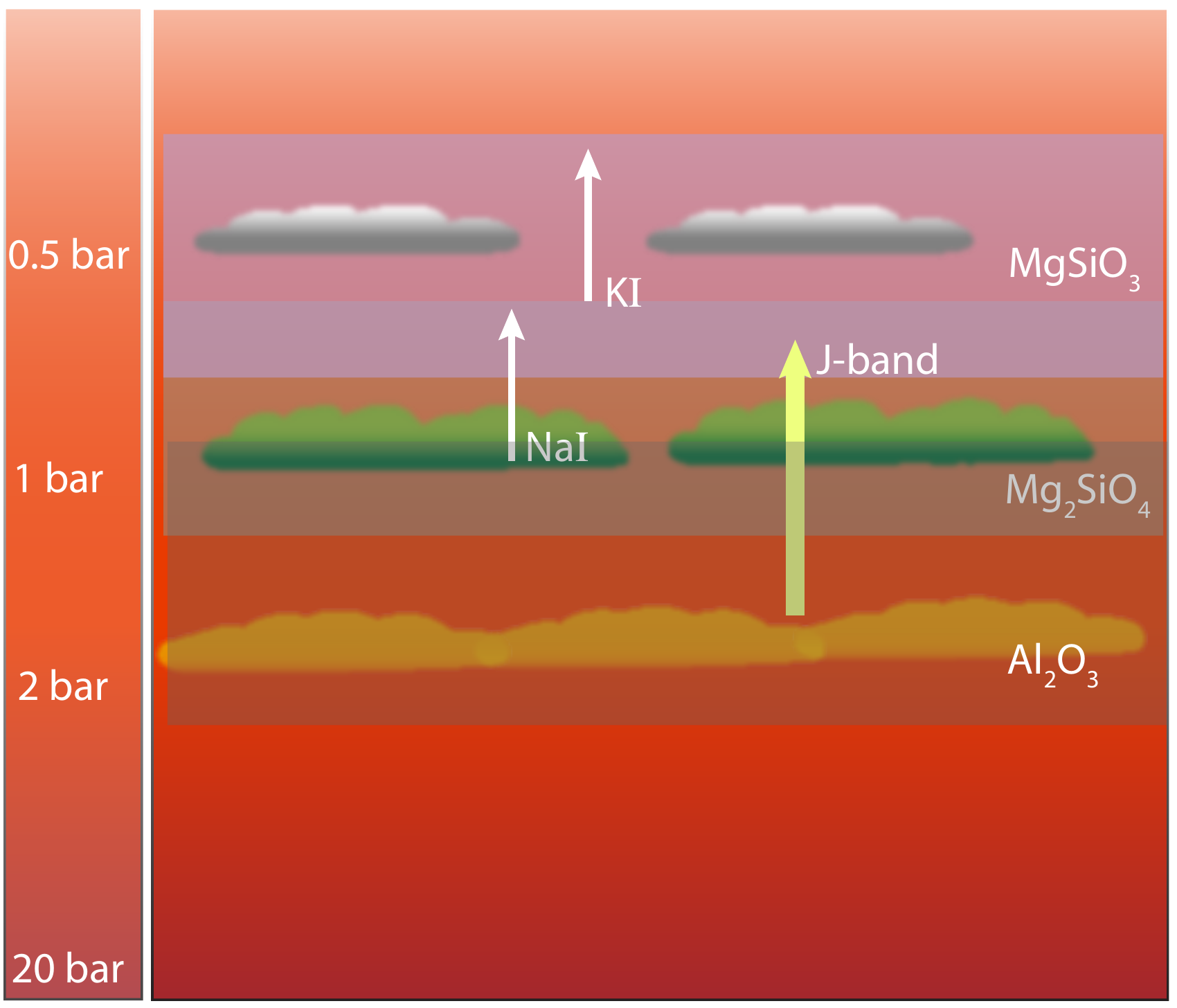}
    \caption{Vertical cloud structure of the atmosphere of 2M2208+2129 with the different heterogeneous cloud layers we can find at different vertical pressures. We include the pressures that the $J$-band, the K\,I doublet, and the Na\,I line trace. The arrows indicate the maximum pressures of the atmosphere each spectral characteristic trace.}
    \label{vertical_structure_picture}
\end{figure}

In Fig. \ref{vertical_structure_picture} we show an illustrative representation of the vertical structure of the atmosphere of 2M2208+2921, using the outcome of the radiative-transfer models, that indicate at which pressure levels the different silicate clouds condensate. In addition, we include the pressure levels that our light curves for the K\,I doublet, the Na\,I line and the entire $J$-band trace.

%add txt about probing different layers both in wav (e.g. NIRCAM) as well as in res (SOSS?) using cores of lines to probe different Ps than broadband

%In addition, we measure the spectro-photometric variability of the modeled spectrum of 2M2208+2921 at each phase of its light curve to compare with the observed MOSFIRE $J$-band spectra

\begin{figure}
    \centering
    \includegraphics[width=0.46\textwidth]{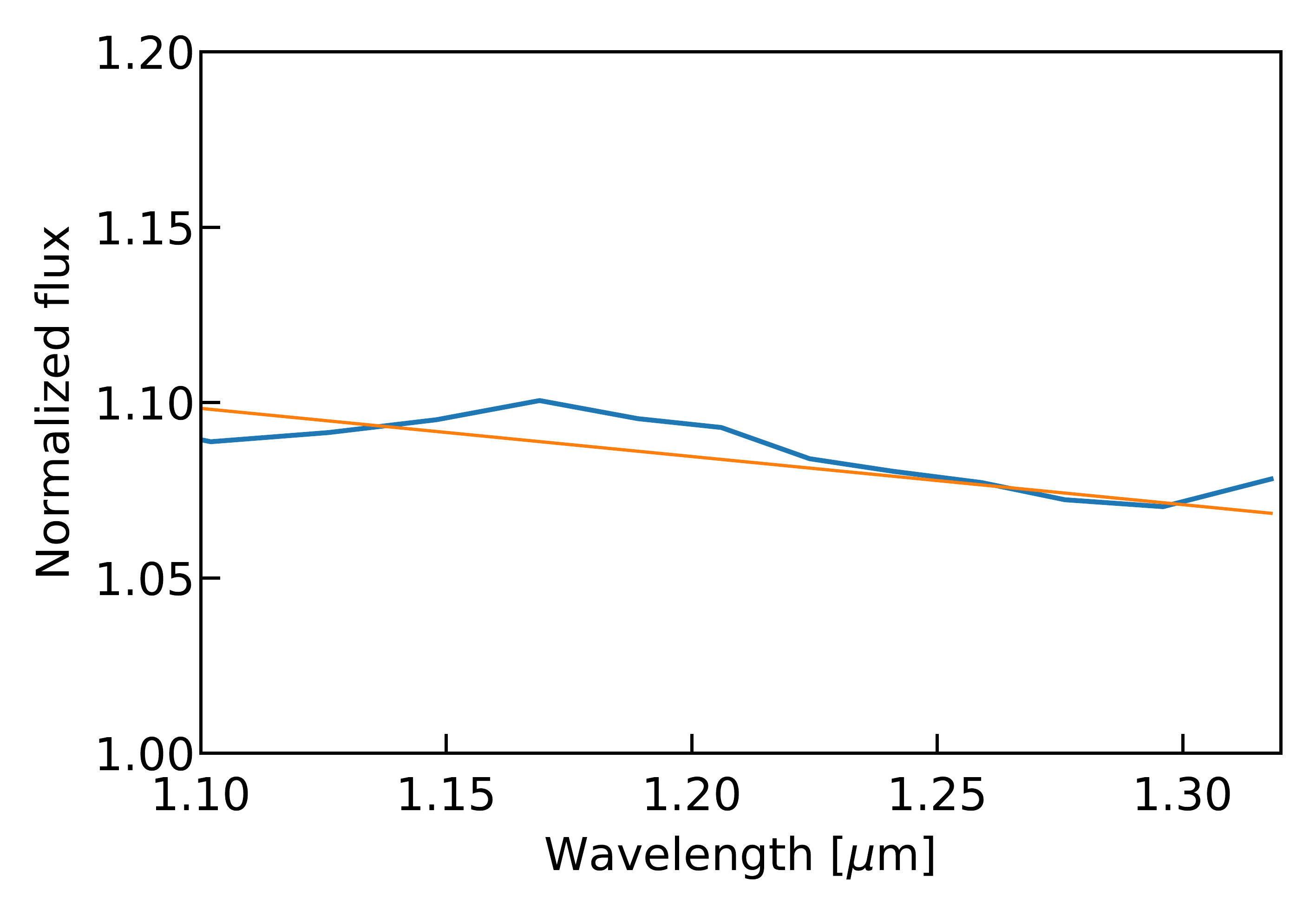}
    \caption{{{Best-fit model to the ratio between the maximum and minimum flux spectrum of 2M2208+2921 (Figure \ref{ratio}). We show the best-fit model ratio (blue line) and best-fit slope (orange line).}}}
    \label{model_slope}
\end{figure}

{We modeled the wavelength dependence of the ratio between the maximum and the minimum spectra of 2M2208+2921 in low-resolution (similar to Fig. \ref{ratio100}, right). We modeled the low-resolution ratio, since the slope is not affected by the resolution of the spectra, but the radiative-transfer models converge faster to a best fit. We used a grid of cloudy and truncated cloud models similar to \cite{2014ApJ...789L..14M}, and \cite{2020ApJ...903...15L}. We found that the best-fit model to the ratio of the maximum and the minimum 2M2208+2921 spectra is a combination of $T_\mathrm{eff}=1800$ K and $T_\mathrm{eff}=1650$ K models with a coverage fraction, $\delta A$, of 0.22. This means that 22\% of the atmosphere has $T_\mathrm{eff}=1650$ K and 78\% of the atmosphere has $T_\mathrm{eff}=1800$ K. In Fig.~\ref{model_slope} we show the best-fit model to the ratio of the maximum divided by the minimum spectrum (same as in Fig. \ref{ratio}, blue line), and best-fit to the slope (similar as in Fig. \ref{ratio}, orange line) plotted between 1.10 and 1.32~$\mu$m for plot clarity. The linear fit of the best-fit model is $1.2483-1.366*10^{-5} \lambda$, which within the error-bars agrees with our MOSFIRE observations slope in Fig.~\ref{ratio}, right panel.}

\begin{figure}
    \centering
    \includegraphics[width=0.49\textwidth]{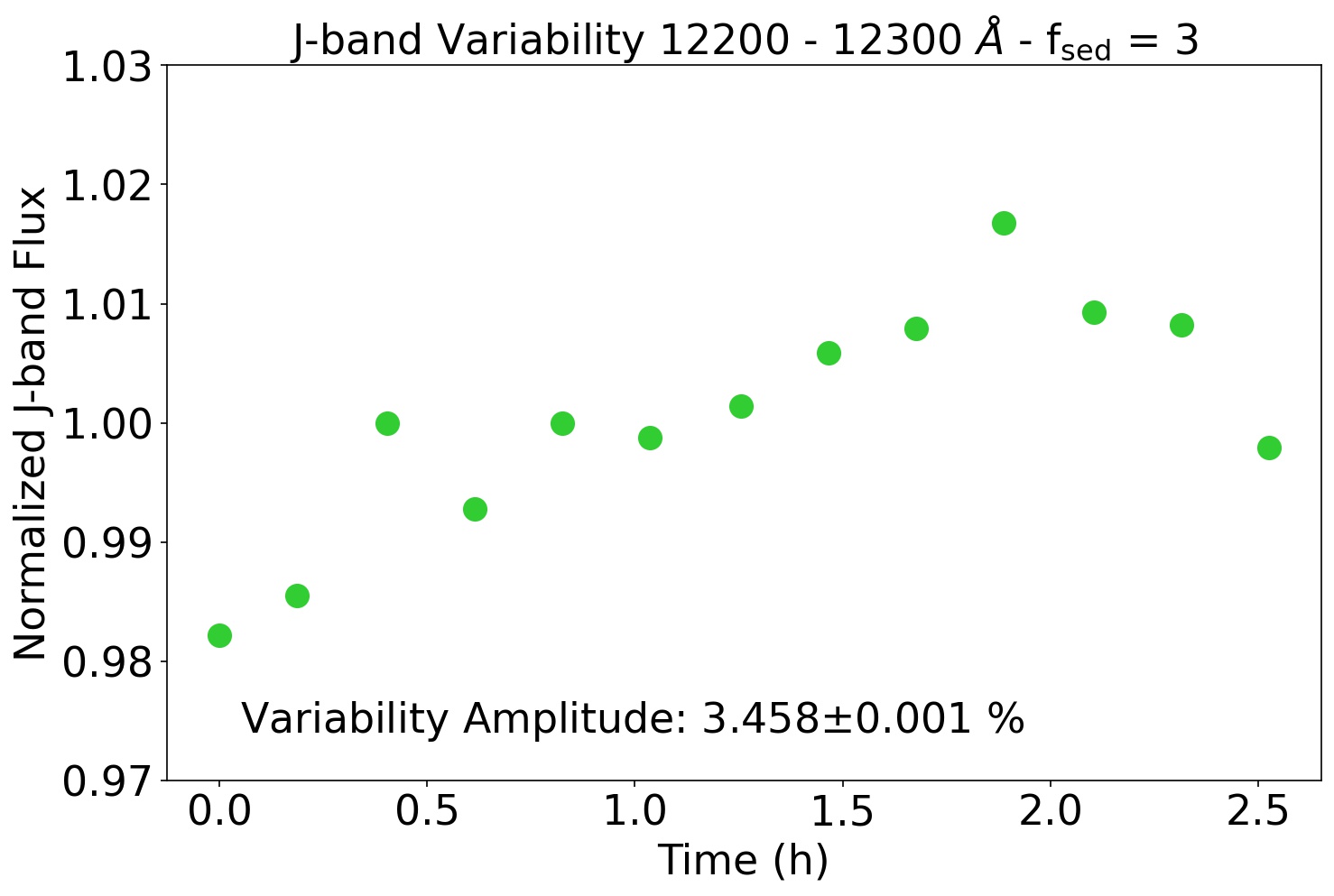}
    \caption{{Simulated $J$-band light curve using radiative-transfer models with nearly 3\% of variability amplitude, similar to our MOSFIRE $J$-band light curve. The best-fit model combination to reproduce a light curve with $\sim$3.5\% variability amplitude consisted of a 1800~K and a 1650~K with clouds with $\mathrm{f_{sed}}$ = 3.}}
    \label{var_model_J}
\end{figure}

%FIX WITH BETTER WORDING ALL RED STUFF
To test our approach to retrieve the spectro-photometric variability of 2M2208+2921 we then modeled a heterogeneous atmosphere that produces a light curve with a comparable amplitude to that of 2M2208+2921. We note that our aim was to test the validity of our method and not to map the atmosphere of 2M2208+2921, so we did not aim to find the best-fit phase-resolved combination of models that reproduces the observed {MOSFIRE $J$-band} light curve, but just a light curve with a comparable amplitude. {Our best-fit model combination consisted of a $T_\mathrm{eff}=1800$~K and a  $T_\mathrm{eff}=1600$~K with clouds with $\mathrm{f_{sed}=1}$ and 3 respectively with $\delta A$=0.13. Note that this model combination is slightly different from our best-fit model combination for the spectral slope mentioned before (1800~K and a 1650~K). The linear fit of this model is $1.0292-1.5879*10^{-5} \lambda$, which is a better fit than our best-fit model combination for the spectral slope in Fig. \ref{ratio} and \ref{ratio100}, right panels}. We blended the models in 13 time steps to create time-resolved simulated ``observations'' that create a sinusoidal-like light curve with a variability amplitude of $\sim$3\%, i.e., comparable to that of our MOSFIRE observations (see Figure \ref{var_model_J}). Each of the 13 model spectra was assigned a random poissonian noise  to mimic {their corresponding uncertainties}. We then used the same method we did for our MOSFIRE observations to obtain the modeled ``observed'' variability in the K\,I doublet, and the Na\,I alkali lines. 

\begin{figure*}
    \centering
    \includegraphics[width=1.02\textwidth]{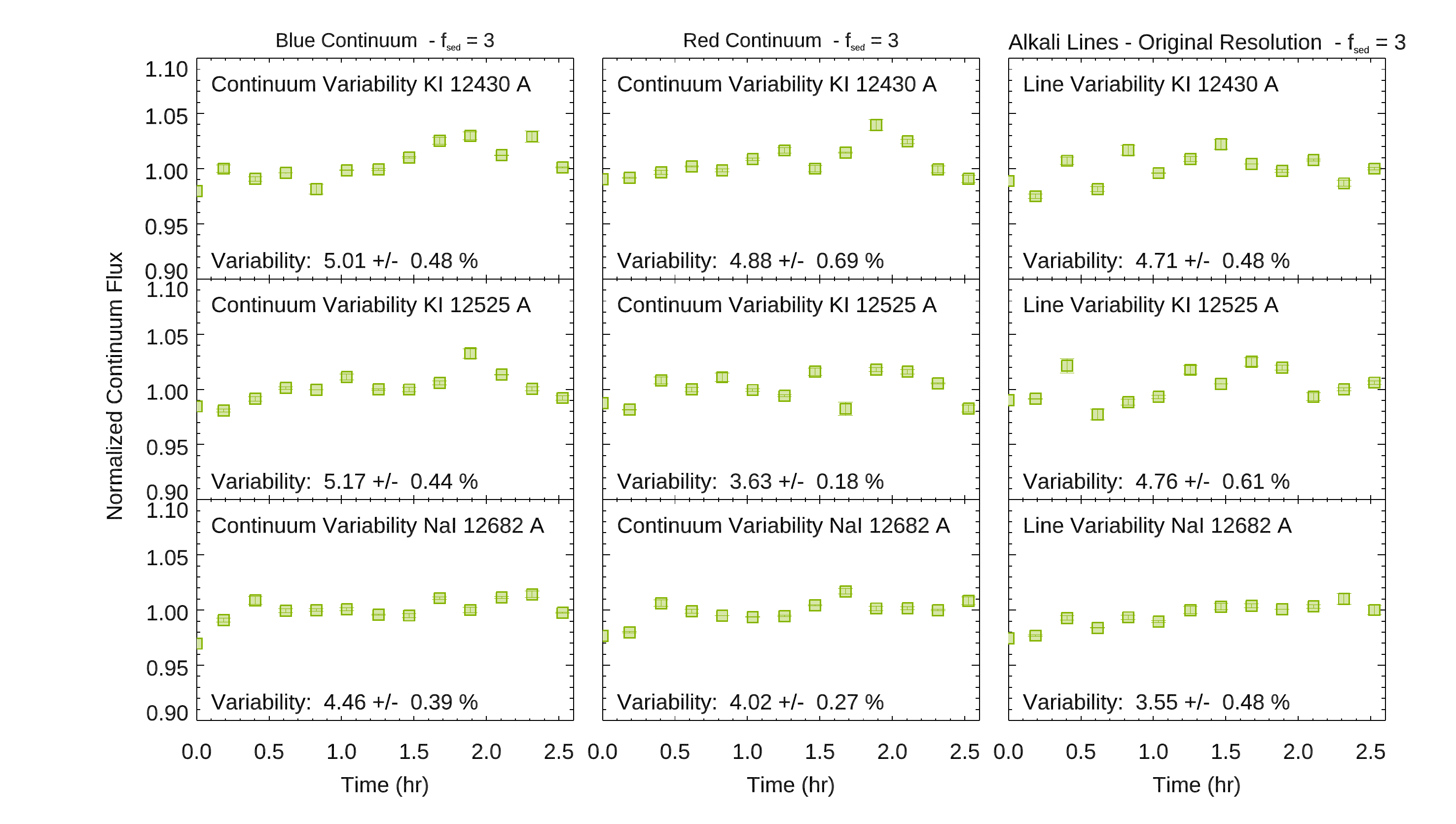}
    \caption{{Variability of the K\,I and Na\,I lines and their blue and red continuum as measured the modeled spectra for $f_{sed}$ = 3.}}
    \label{var_model_line_continuum}
\end{figure*}

Figure \ref{var_model_line_continuum} shows the variability of the K\,I doublet and the Na\,I alkali lines, and their respective blue and red continuum, measured in the modeled spectra following the same methodology as for our observed MOSFIRE spectra in Sections~\ref{Na_KI_line_variability}, and \ref{Na_KI_continuum_variability}. In Figure~\ref{var_model_line_continuum} we observe that the variability amplitudes of the {alkali lines, and their} blue and red continuums is between {3.6-5.1\%, in general inconsistent} with the variability amplitude of $\sim$3\% in the simulated $J$-band light curve. The enhanced variability amplitude predicted by the modeled spectra for the K\,I doublet is consistent, in amplitude value, with the enhanced variability amplitude measured in Sections~\ref{Na_KI_line_variability} and \ref{Na_KI_continuum_variability} for the observed MOSFIRE spectra at their original resolution. For the Na\,I line, we measured a variability amplitude of 10.93$\pm$3.17\% in the observed MOSFIRE spectra. Since such enhanced variability amplitude is not predicted  by the models, we suspect that there might be uncorrected telluric contamination remaining in the Na\,I light curve, even after the correction performed using the other calibration stars in the field. Nevertheless, qualitative, the radiative-transfer models still {predict} that the variability amplitude of the Na\,I is enhanced. 

\begin{figure*}
    \centering
    \includegraphics[width=1.02\textwidth]{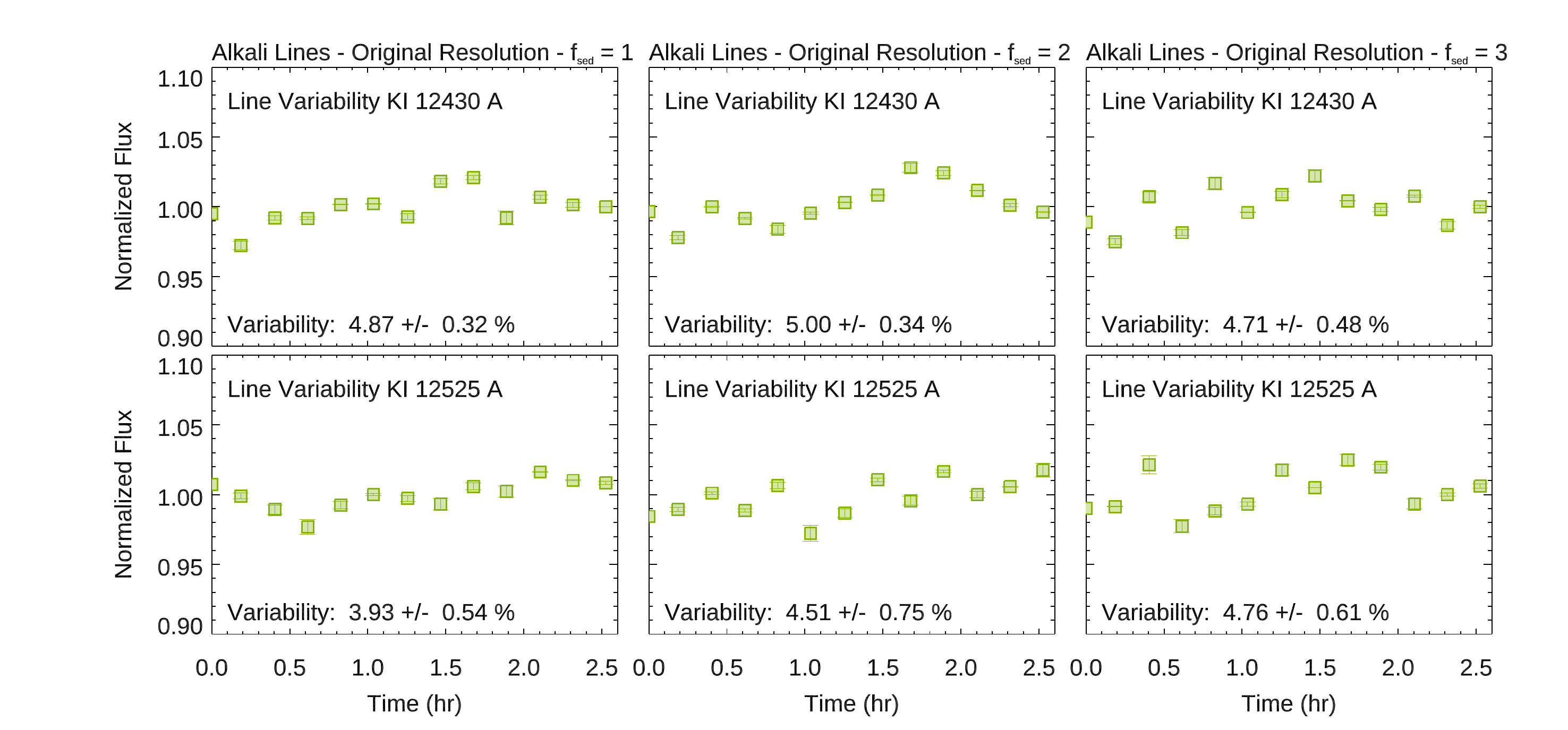}
    \caption{{Variability of the K\,I lines and their blue and red continuum as measured the modeled spectra for $f_{sed}$ = 1, 2 and 3.}}
    \label{var_model_line_continuum_Ks}
\end{figure*}

{Finally, as an illustration, we tested the effect of the cloud properties on the retrieved variability for the K\,I lines. 
Fig.~\ref{var_model_line_continuum_Ks} shows the retrieved amplitude of the K\,I line as a function of $f_\mathrm{sed}$ for a combination of 1800~K and 1650~K clouds as in our best-fit slope model. Changes in $f_\mathrm{sed}$ correspond to a change in the cloud properties,  and thus should correspond to changes in the retrieved variability. Indeed, Fig.~\ref{var_model_line_continuum_Ks} shows that the average retrieved variability of the model K\,I line changes slightly with the reduction of the optical thickness across our model atmospheres, even though, the variability amplitude for the three $f_{sed}$ values is similar within the error-bars.}

{Note that \citet{Zhou2020} found a subdued variability in the alkali lines of VHS 1256b, but their target was a cooler, L7 atmosphere with different cloud structure than our target. Changes in the temperature of the atmosphere affect the cloud structure and expected variability both in the $J$-band and Spitzer channels \citep{Vos2017,Vos2020} as well as in the alkali lines \citep[see also][for T and Y atmospheres]{2014ApJ...789L..14M}. Our result thus does not contradict that of \citet{Zhou2020}, but complements it with another spectral type. Future JWST observations that constrain the changes of alkali variability versus continuum variability as a function of atmospheric temperature would be important to map the changes in cloud structures as these atmospheres cool down.}

{Our observations  highlight the importance of high resolution spectroscopy to understand the atmospheric variability and 3D structures of brown dwarfs and giant exoplanets ground-based, with multi-object spectrographs like Keck\,I/MOSFIRE, or EMIR at the Gran Telescopio de Canarias (GTC) telescope, but also from space-based telescopes like HST/WFC3. In the near future, the James Webb Space Telescope (JWST) will be launched, and it is expected to produce ground-braking discoveries in the field of brown dwarfs and exoplanets. NIRSpec (Near Infrarred Spectrograph) and NIRISS (Near Infrared Imager and Slitless Spectrograph) on-board JWST will provide high signal-to-noise and resolution, and broad-wavelength spectroscopic observations, that will enable the detection of variability in multiple pressure layers, allowing us to probe the vertical structure of brown dwarf and imaged exoplanet atmospheres with an unprecedented accuracy. }

\section{Conclusions}\label{conclusions}

\begin{enumerate}

    \item We have used MOSFIRE at the Keck\,I telescope to monitor over $\sim$2.5~hr 2M2208+2921, an L3 young brown dwarf, member of the $\beta$-Pictoris young moving group, and an analog to the $\beta$~Pictoris~b directly-imaged giant exoplanet.
    
    \item We found significant spectro-photometric variability amplitude in the $J$-band using MOSFIRE spectroscopy with a minimum variability amplitude of 3.22$\pm$0.42\%.
    
    \item The ratio between the maximum and the minimum spectra of 2M2208+2921 show a slightly wavelength dependence, with the variability amplitude descending toward redder wavelengths. It also shows potentially enhanced variability amplitude in the K\,I doublet and Na\,I alkali lines.
    
    \item More detailed analysis of the variability amplitude of the continuum and the flux inside the K\,I, and Na\,I lines further suggests the enhanced variability amplitude inside those lines. The enhanced variability partially dissapaears if we degrade the resolution of the spectra to R$\sim$100, especially for the Na\,I line, coinciding with the spectral resolution of HST/WFC3 + G141 grism, explaining why enhanced variability amplitude has not been found in previous works using low-resolution data {for brown dwarfs of similar spectral type}.
    
    \item We use radiative-transfer models to predict the different heterogeneous layers of clouds that might be introducing the spectro-photometric variability detected and their composition.    
    
    \item Using radiative-transfer models, we produced simulated $J$-band spectra for an object with the same $\mathrm{T_{eff}}$ and log~$g$ than 2M2208+2921, and with the same $J$-band variability amplitude, and rotational period. We measured the variability amplitude of the K\,I doublet and Na\,I alkali lines and their respective continuums, finding  an enhanced variability for the alkali lines, {in agreement with our observations}.
    
    \item {Using the \textit{Aeolus} code to produce brown dwarf maps, we are able to reproduce that the $J$-band light curve has smaller variability amplitude than the K\,I or the Na\,I lines light curves, in agreement with our observations. }

    \item We produce an artistic representation reproducing the vertical structure of 2M2208+2921, the different layers of clouds and their composition as proposed by the relative-transfer models, and the different pressure levels that each spectral characteristic (the $J$-band, the K\,I lines and the Na\,I line) traces in the atmosphere of 2M2208+2921,  analog to the {$\beta$-Pictoris~b exoplanet}.

\end{enumerate}

\acknowledgments
{We thank our anonymous referee for the constructive comments provided for our manuscript, that helped to improve it.}

The authors wish to recognize and acknowledge the very significant cultural role and reverence that the summit of Mauna Kea has always had within the indigenous Hawaiian community.  We are most fortunate to have the opportunity to conduct observations from this mountain.

We would like to acknowledge the PypeIt Development team for developing a pipeline that was able to reduce our challenging MOSFIRE data with extremely wide slits, in particular to Dr. Joe Hennawi for his efficient support.

We acknowledge the MOSFIRE/Keck I Instrument Scientist, Dr. Josh Walawender, for his advises and recommendations on the preparation of the observations, and the reduction of the data. Thanks to his idea of taking "skylines spectra" of our mask with narrower slits we could calibrate in wavelength the spectra presented in this paper.

We acknowledge W. M. Keck Observatory Chief Scientist, Dr. John O'Meara, for investing some of his granted time on taking the "skylines spectra" of our masks that made wavelength calibration possible.

We acknowledge Dr. Daniel Apai and his group for their comments and suggestions on the analysis and interpretation of these data.

\facilities{MOSFIRE (W. M. Keck Observatory)}


\software{astropy \citep{2013A&A...558A..33A}}
\software{Pypeit \citep{Prochaska2019,Prochaska2020}}



\newpage

\appendix

\section{Correlation between parameters}

\begin{figure}
    \centering
    \includegraphics[width=0.48\textwidth]{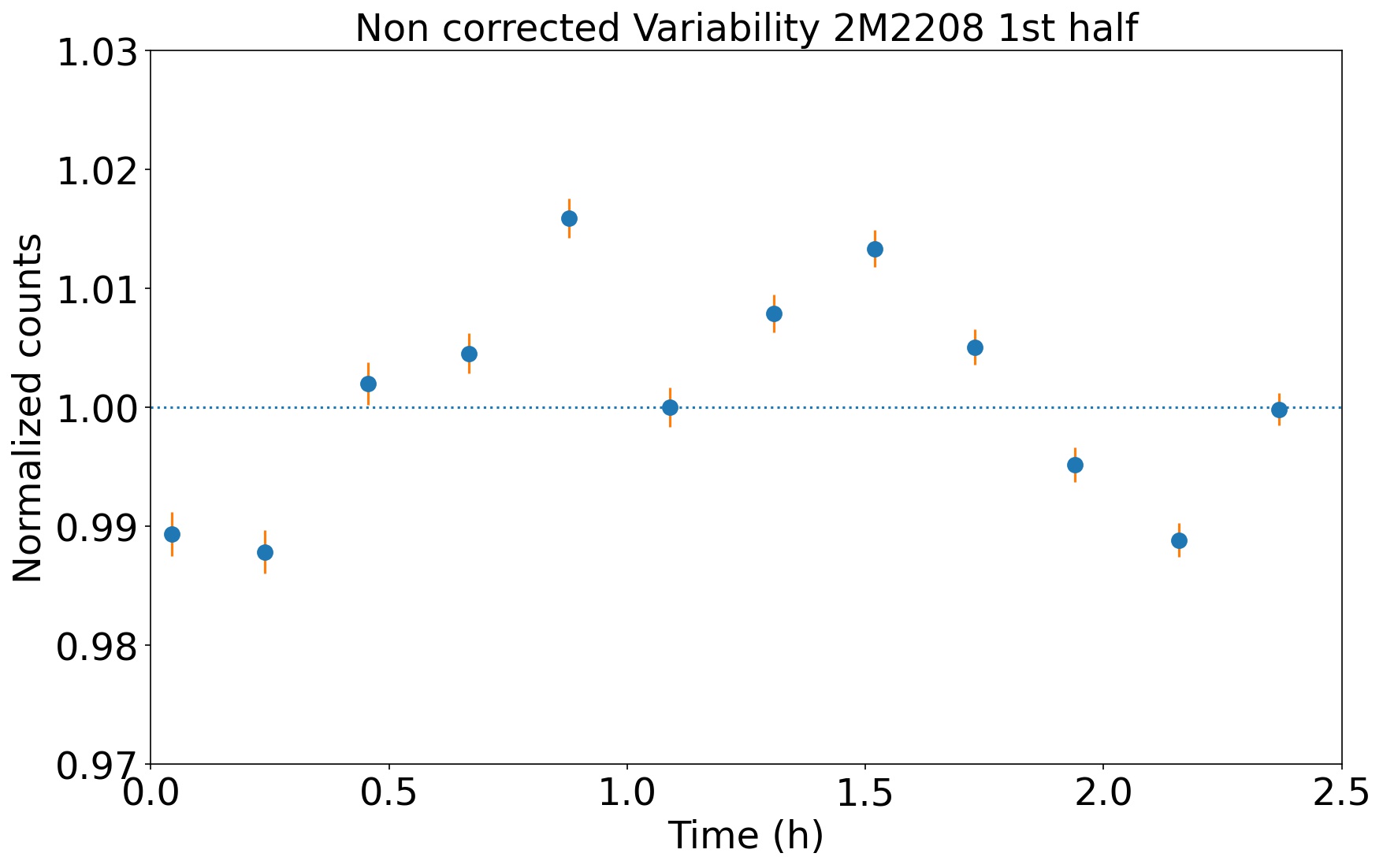}
    \includegraphics[width=0.48\textwidth]{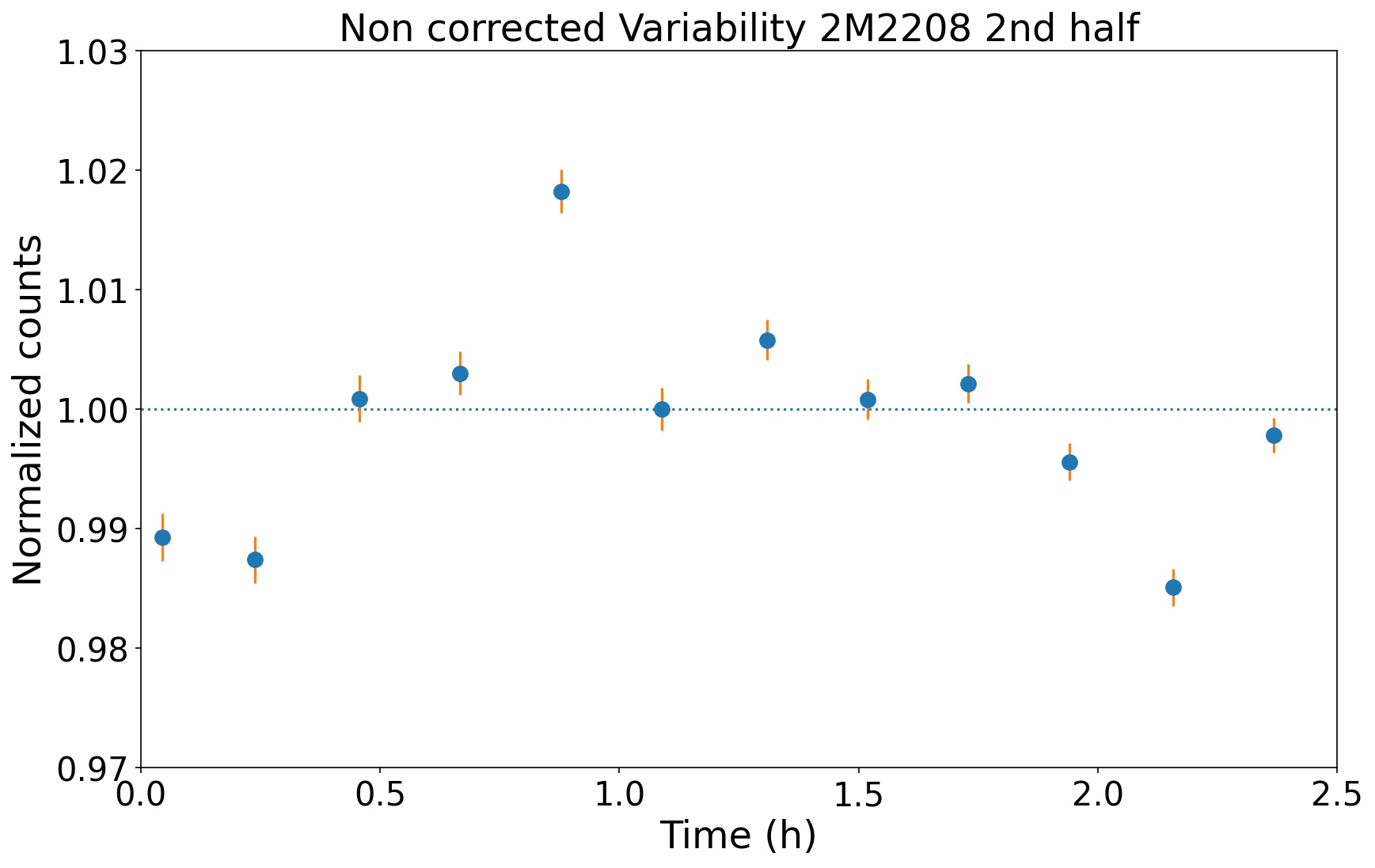}
    \caption{Light curves obtained using only the first half wavelength range spectrum and the second half wavelength range.}
    \label{Correlation_LC_halves}
\end{figure}

\begin{figure}
    \centering
    \includegraphics[width=0.48\textwidth]{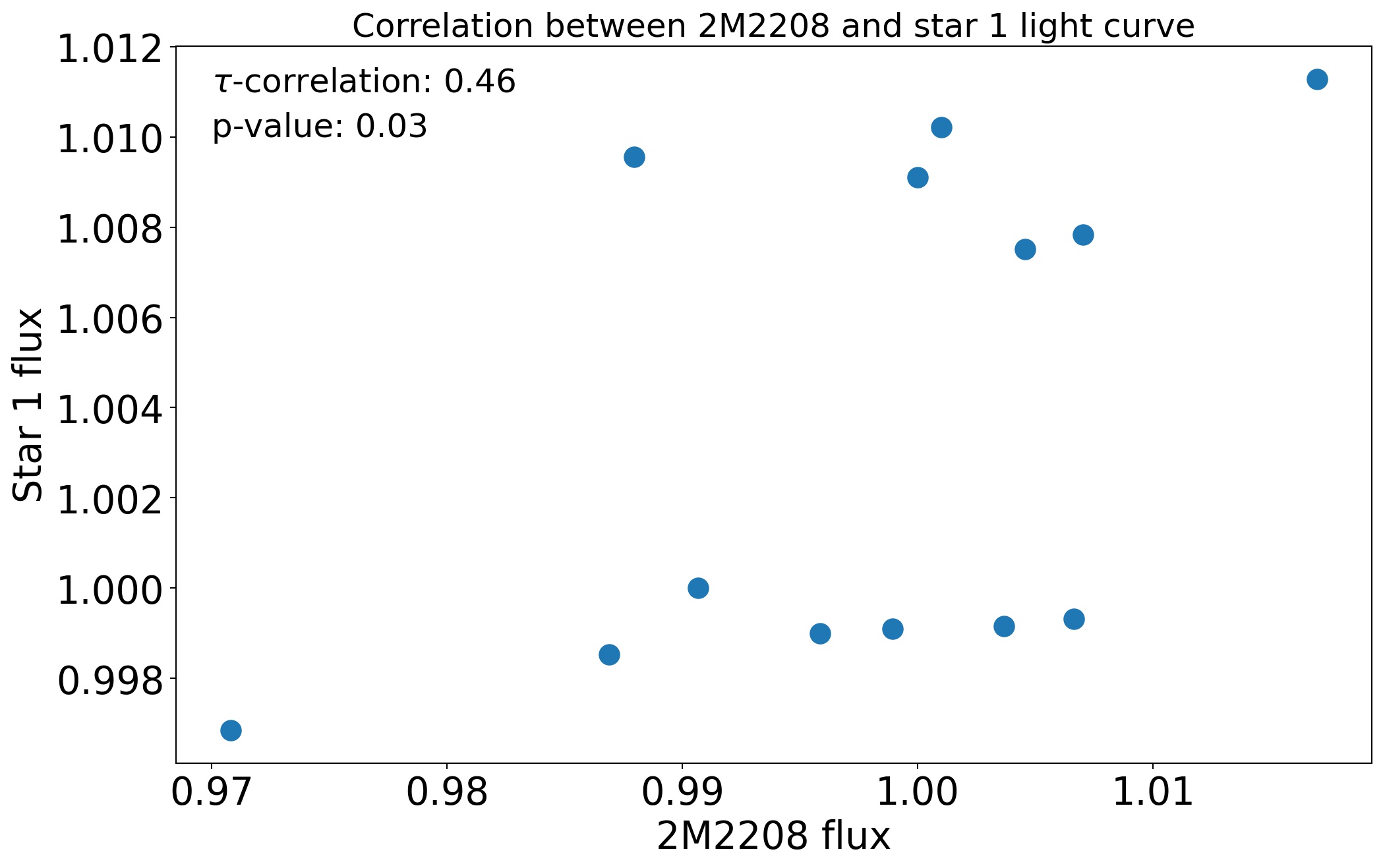}
    \includegraphics[width=0.48\textwidth]{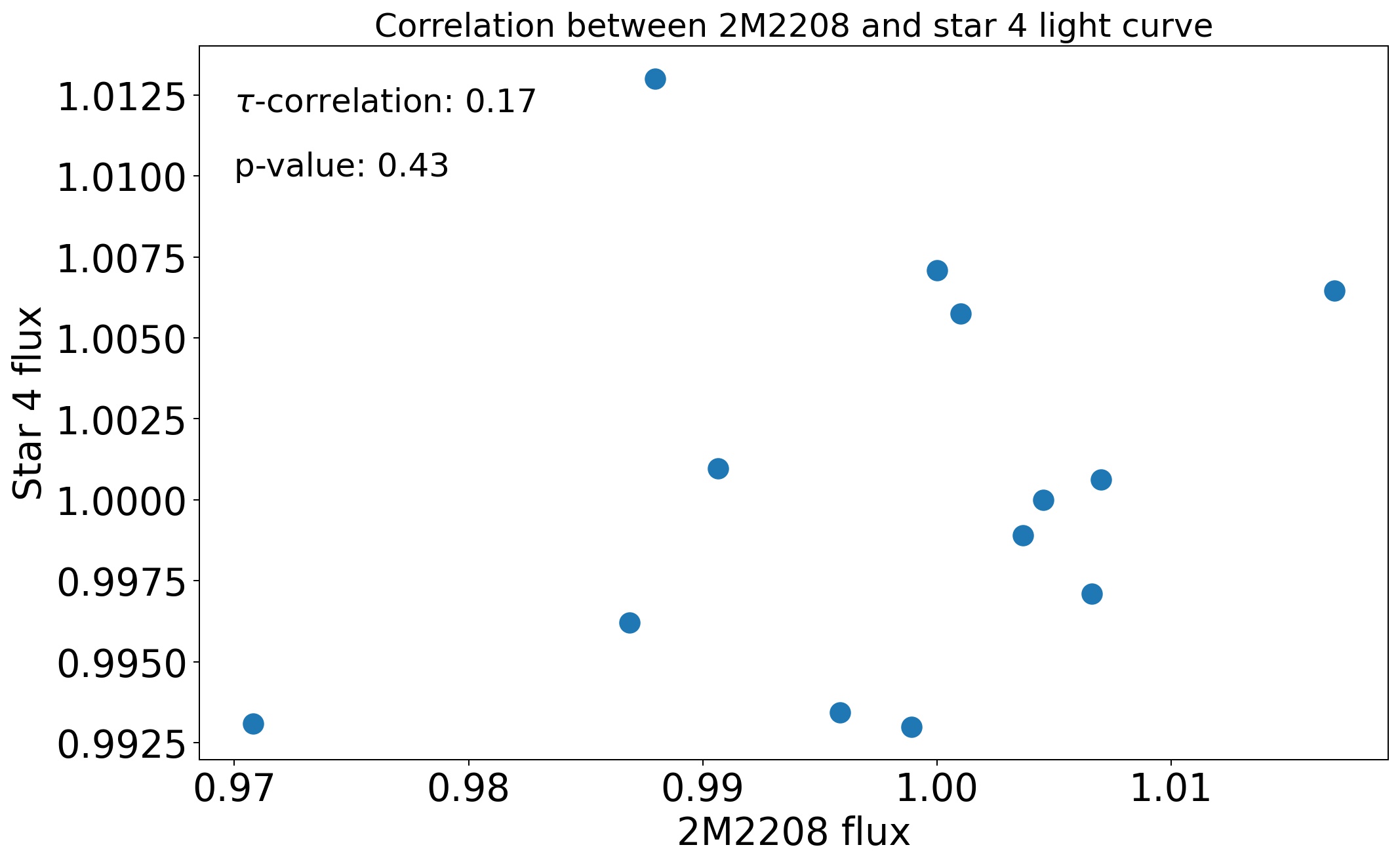}
    \includegraphics[width=0.48\textwidth]{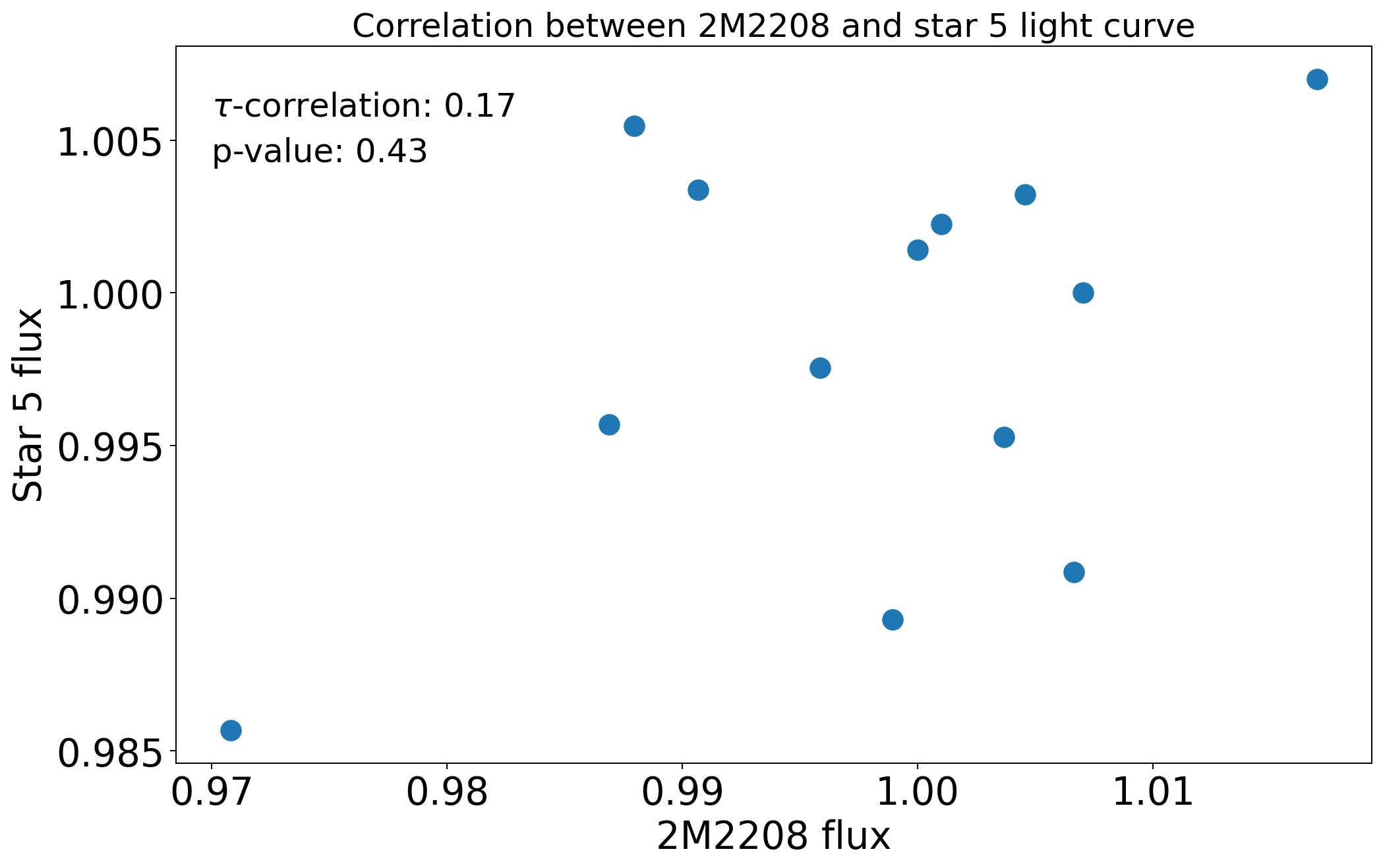}
    \includegraphics[width=0.48\textwidth]{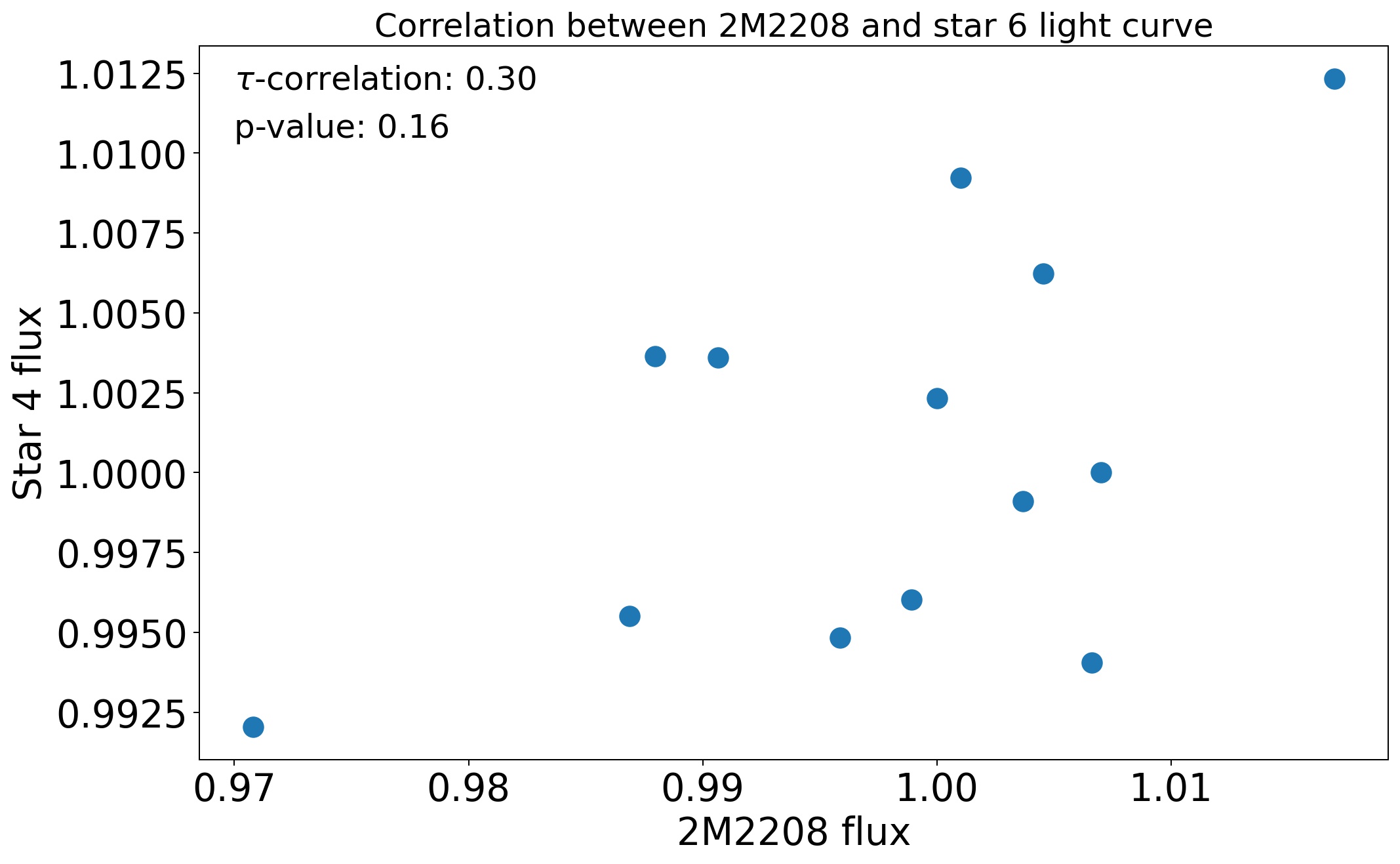}
    \includegraphics[width=0.48\textwidth]{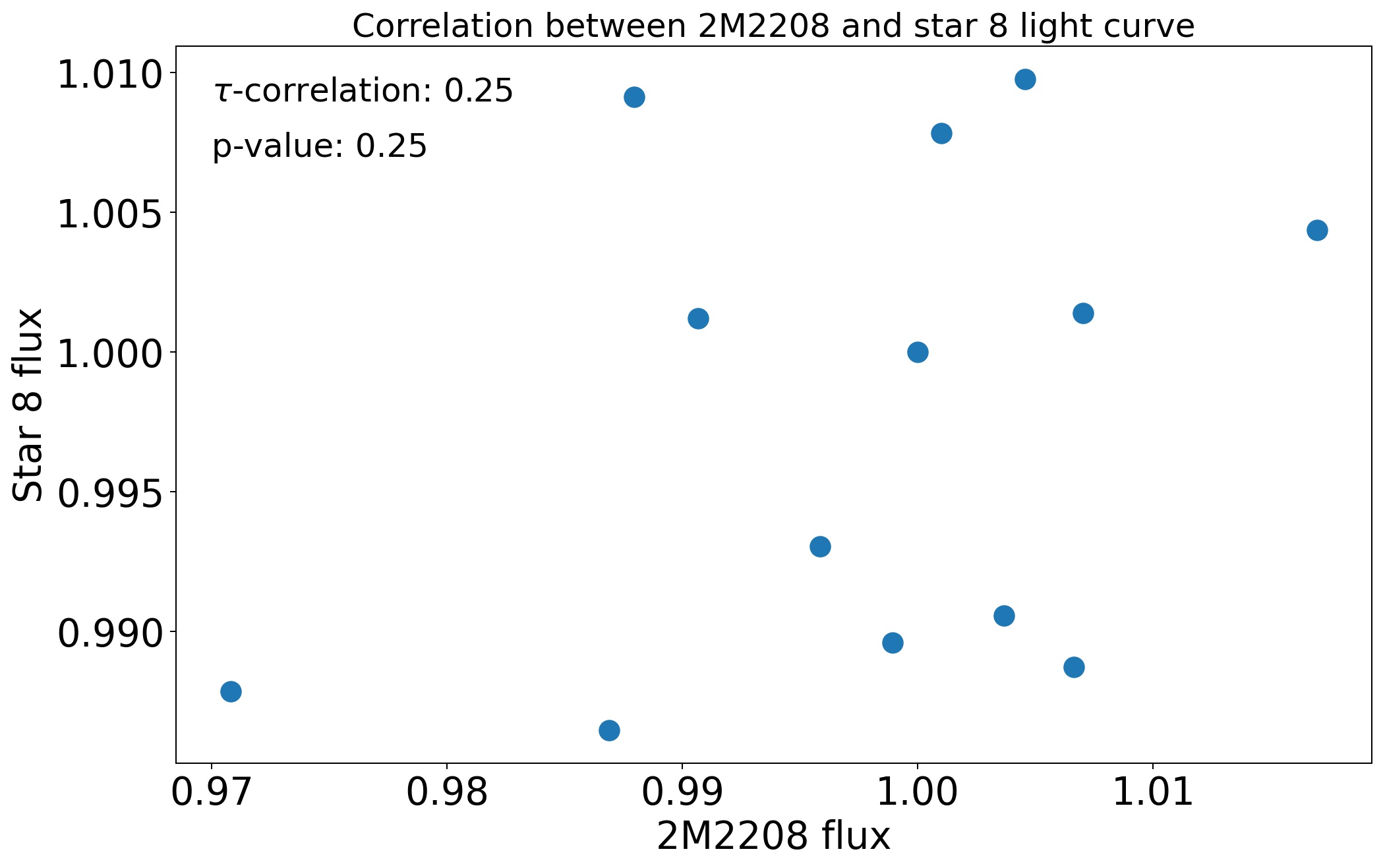}
    \caption{Correlation between the target's non-corrected light curve, and the non-corrected calibration stars light curves.}
    \label{corr_LC_stars}
\end{figure}

\begin{figure}
    \centering
    \includegraphics[width=0.48\textwidth]{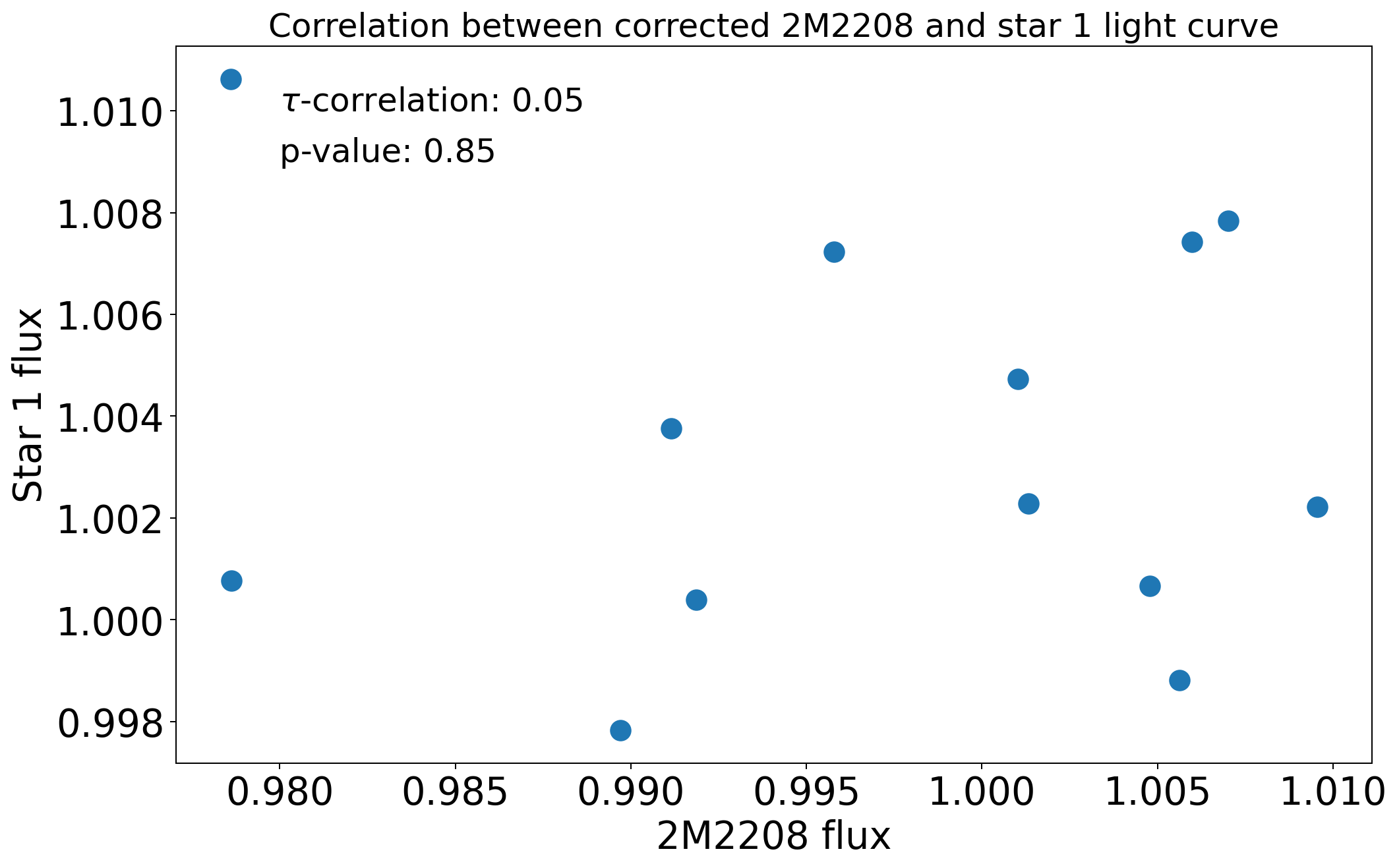}
    \includegraphics[width=0.48\textwidth]{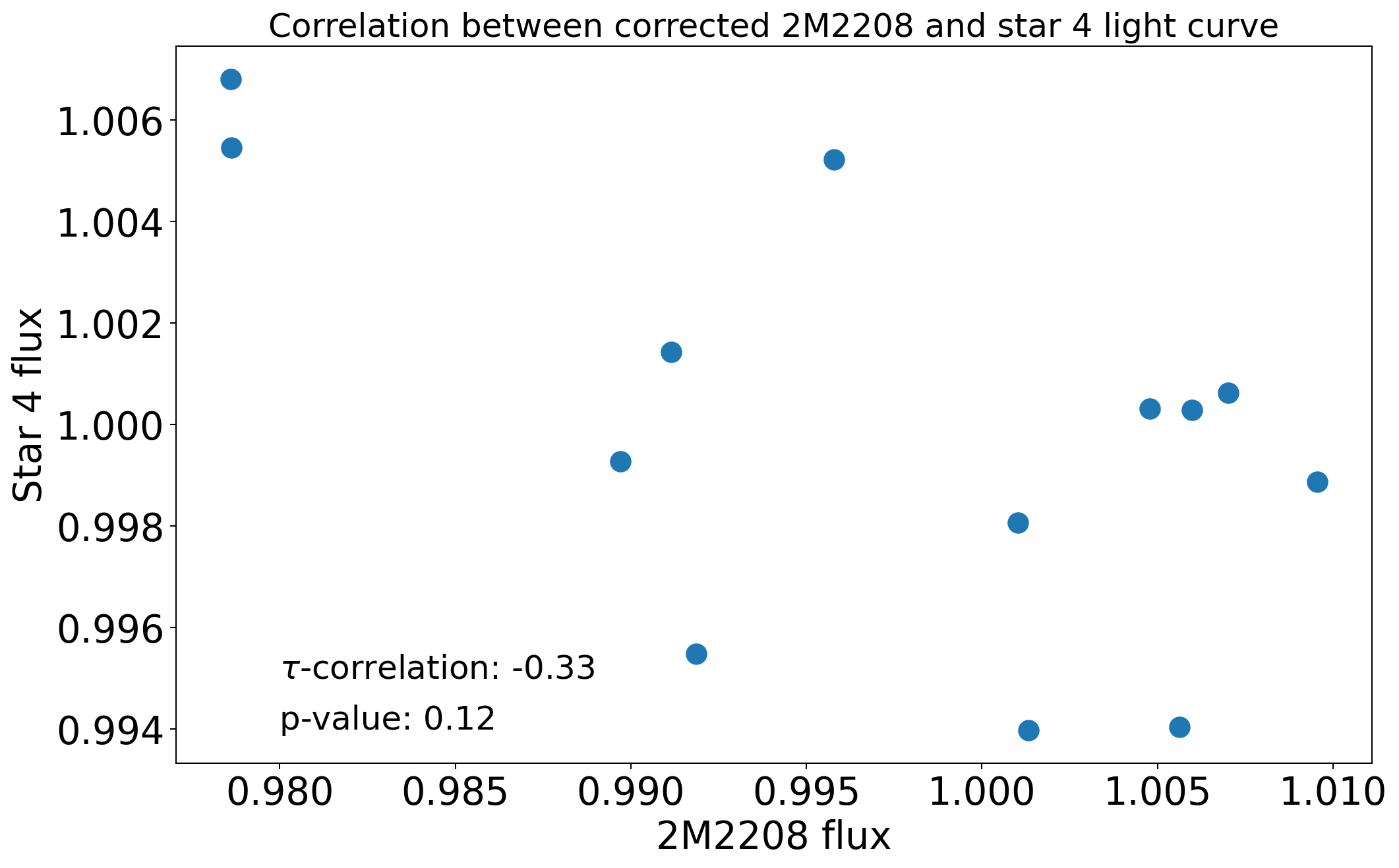}
    \includegraphics[width=0.48\textwidth]{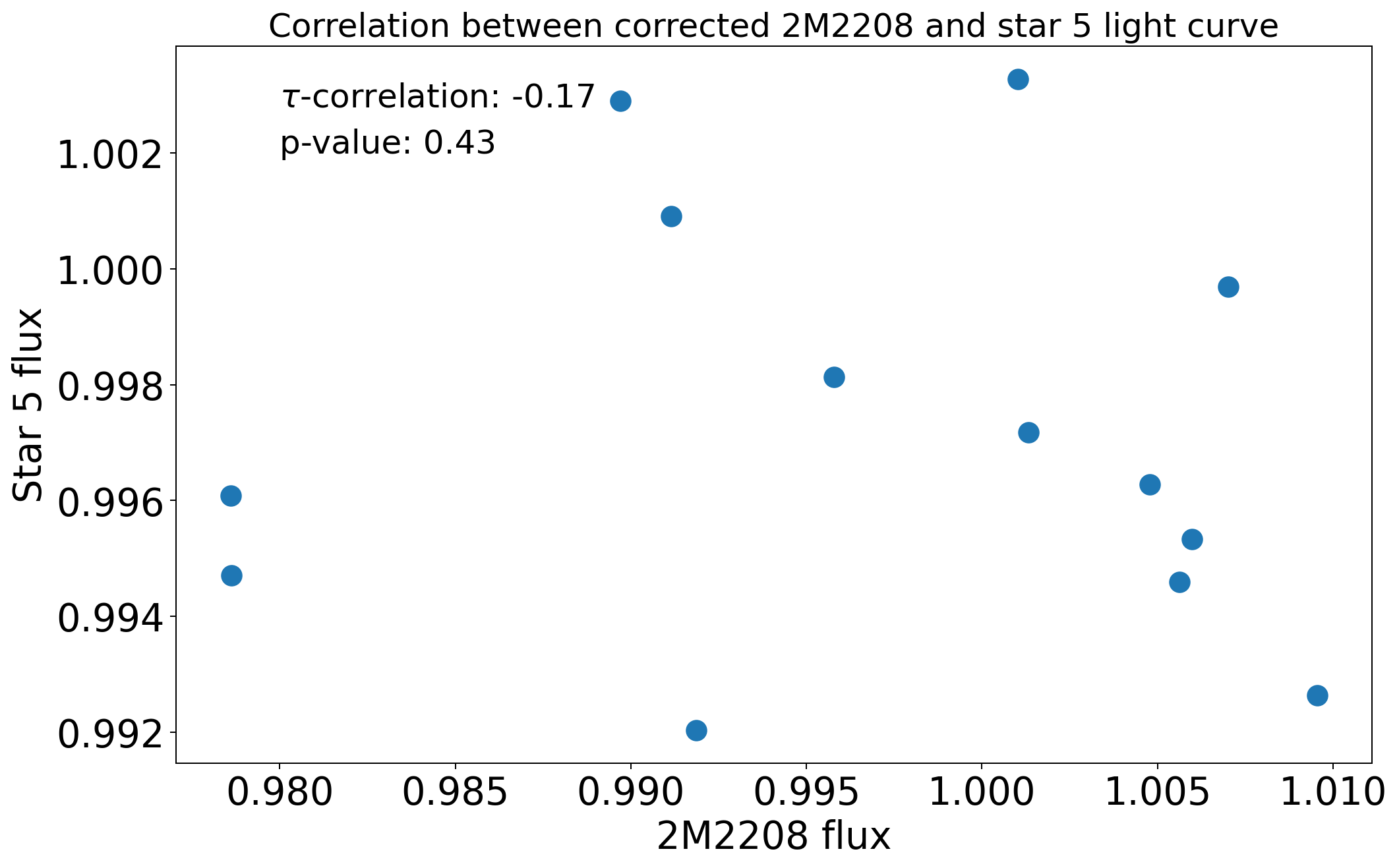}
    \includegraphics[width=0.48\textwidth]{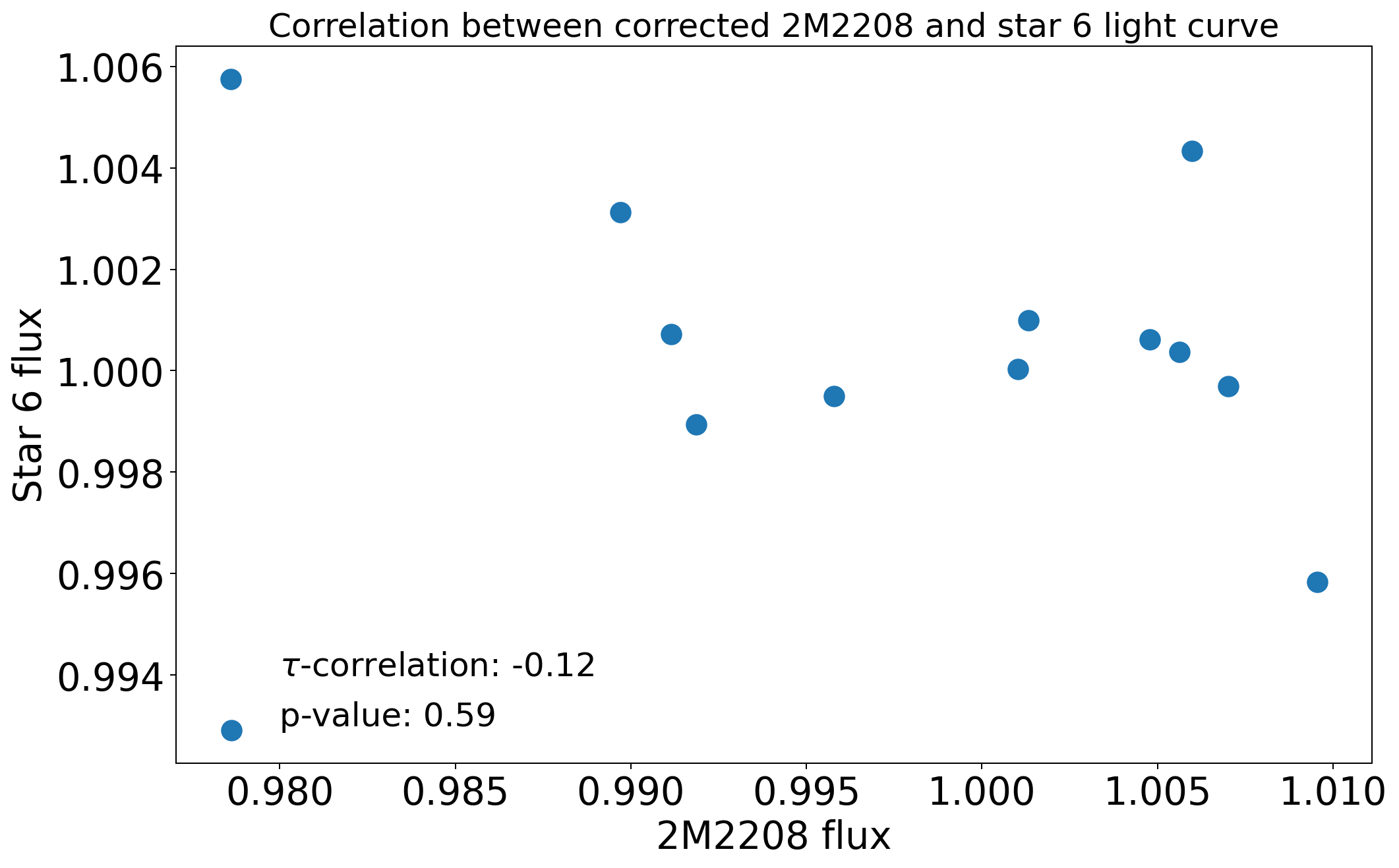}
    \includegraphics[width=0.48\textwidth]{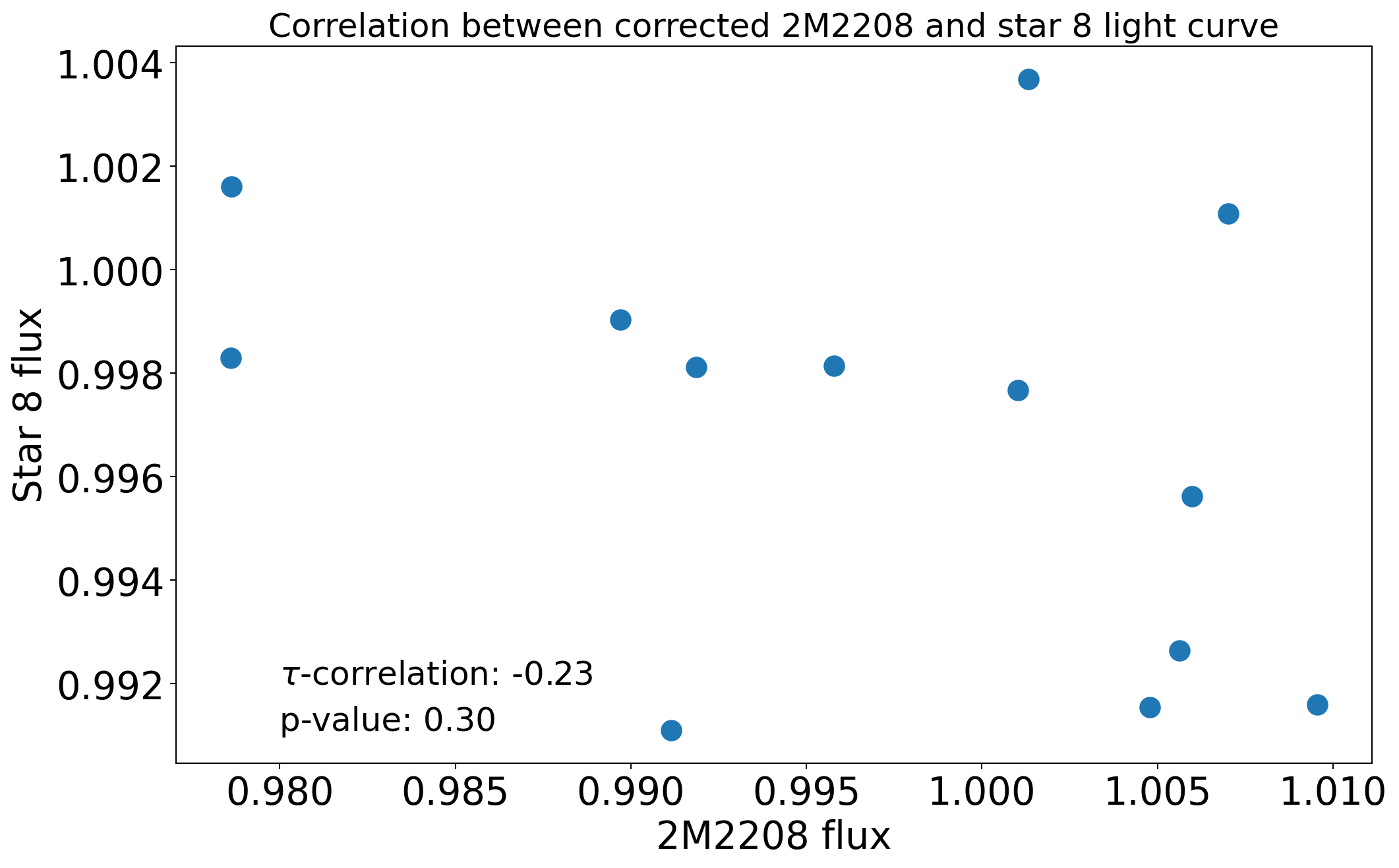}
    \caption{Correlation between the target's corrected light curve, and the corrected calibration stars light curves.}
    \label{corr_LC_stars_corrected}
\end{figure}

\begin{figure}
    \centering
    \includegraphics[width=0.48\textwidth]{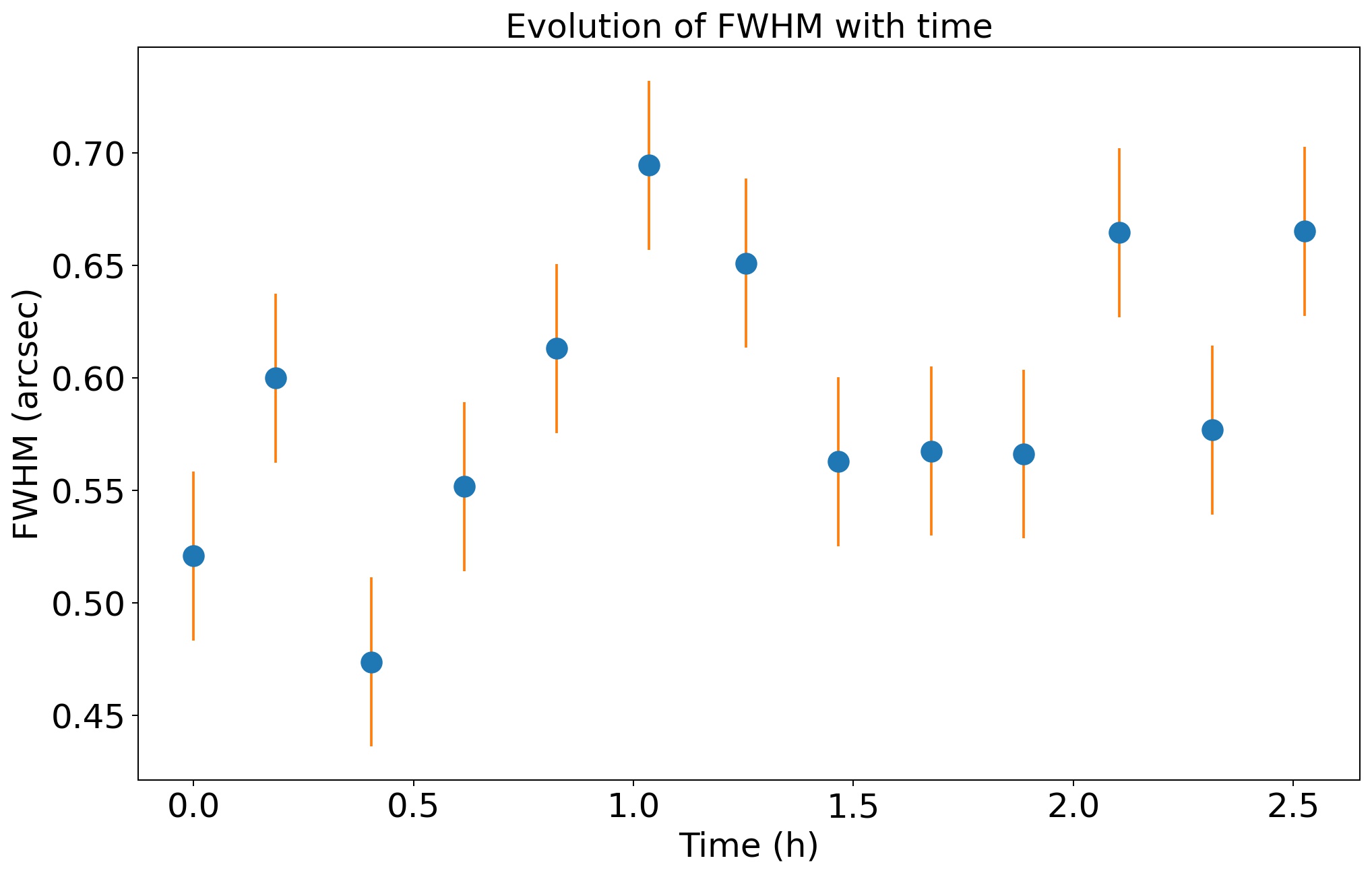}
    \includegraphics[width=0.48\textwidth]{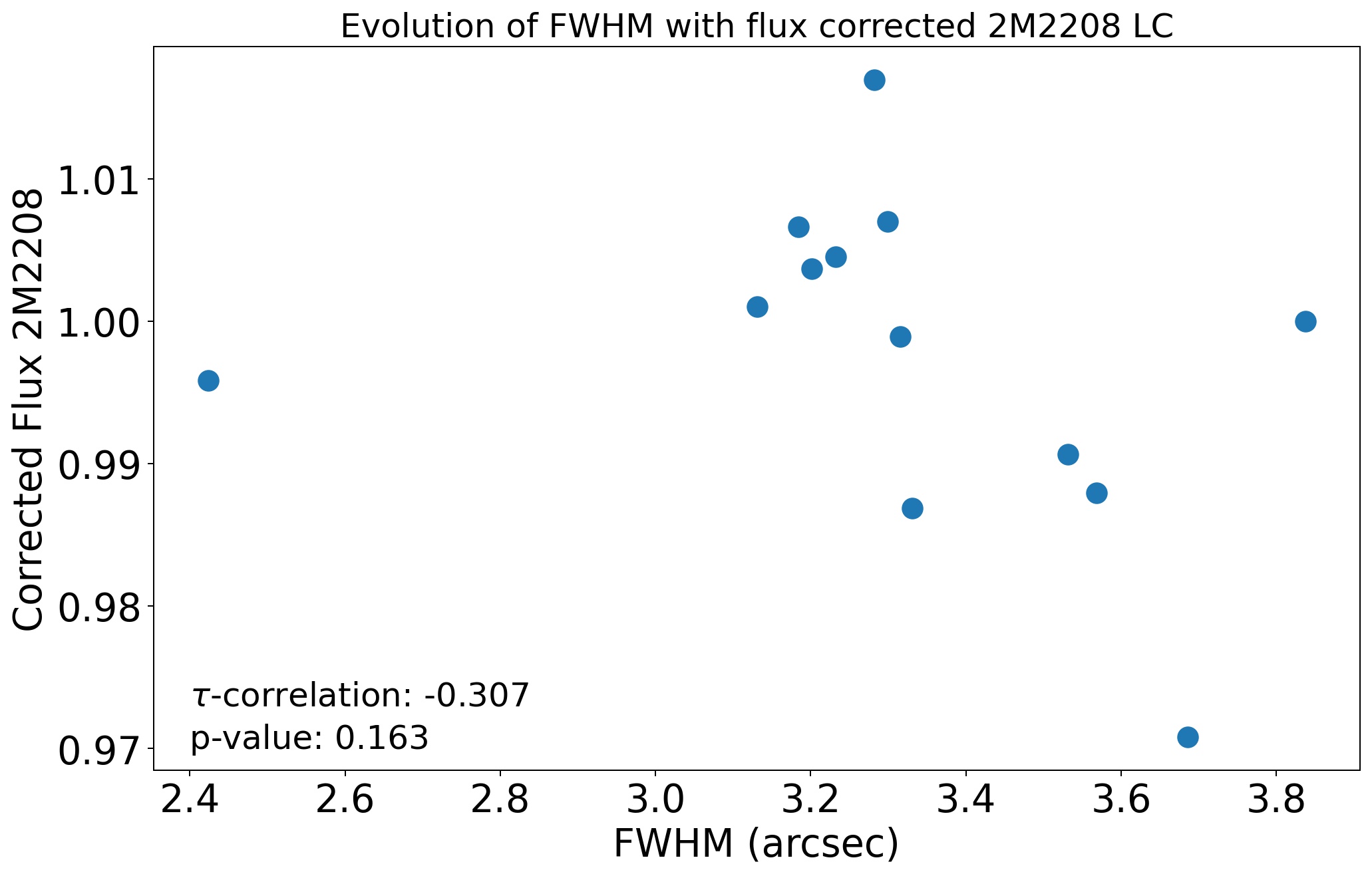}
    \caption{Left: Evolution of the FWHM with time. Right: Correlation between FWHM and 2M2208 light curve.}
    \label{evol_FWHM}
\end{figure}

\begin{figure}
    \centering
    \includegraphics[width=0.48\textwidth]{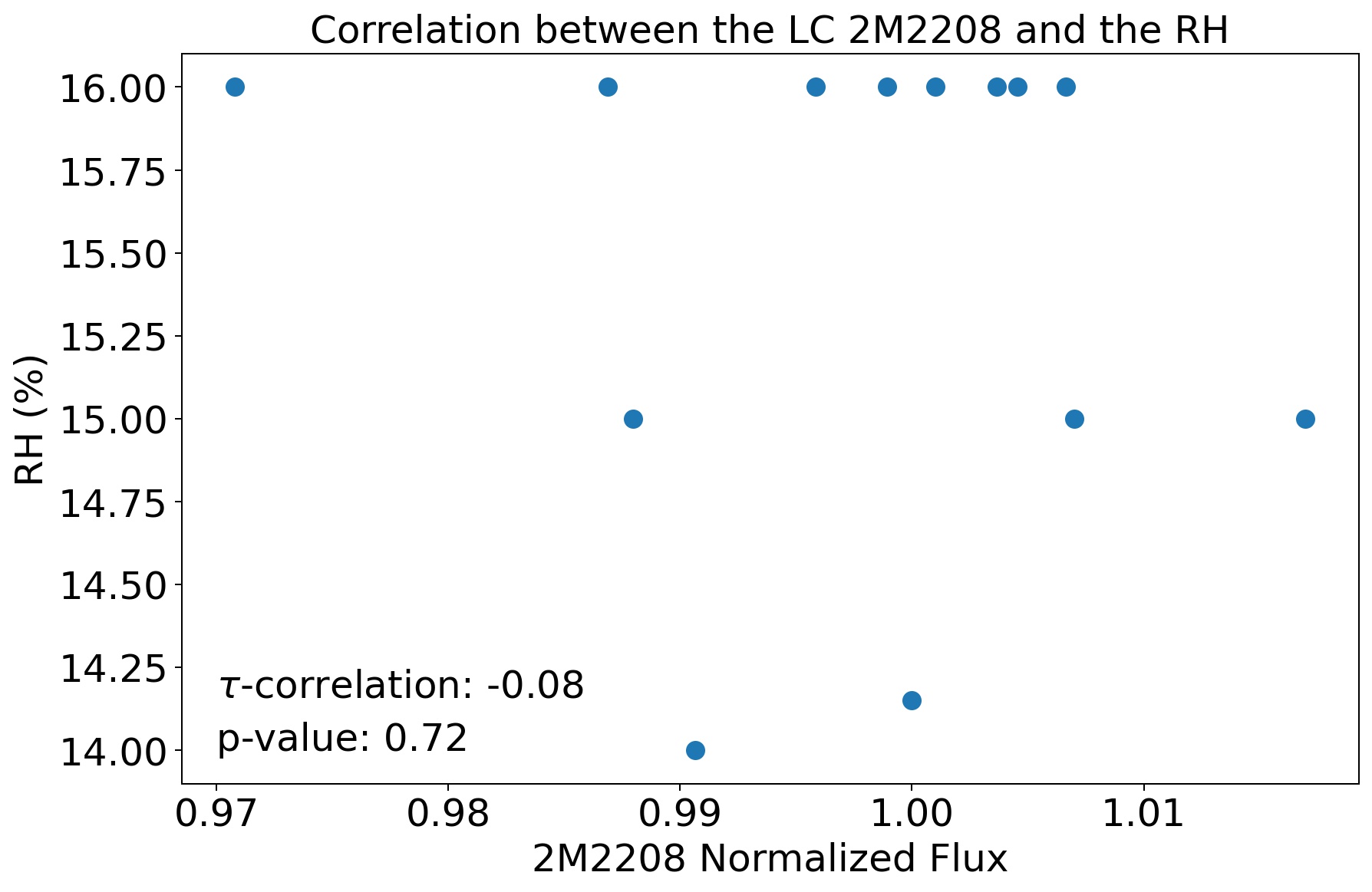}
    \includegraphics[width=0.465\textwidth]{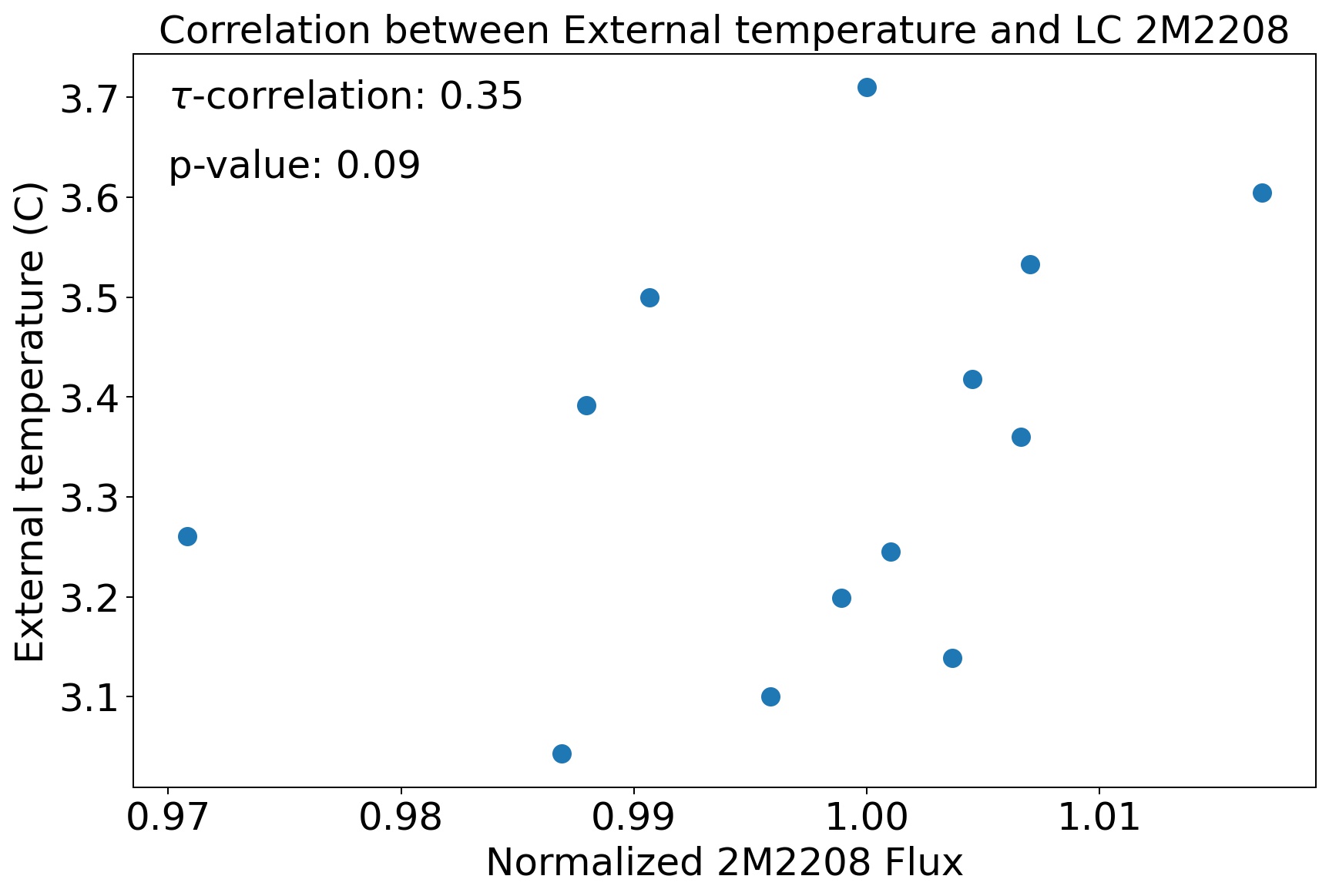}
    \includegraphics[width=0.48\textwidth]{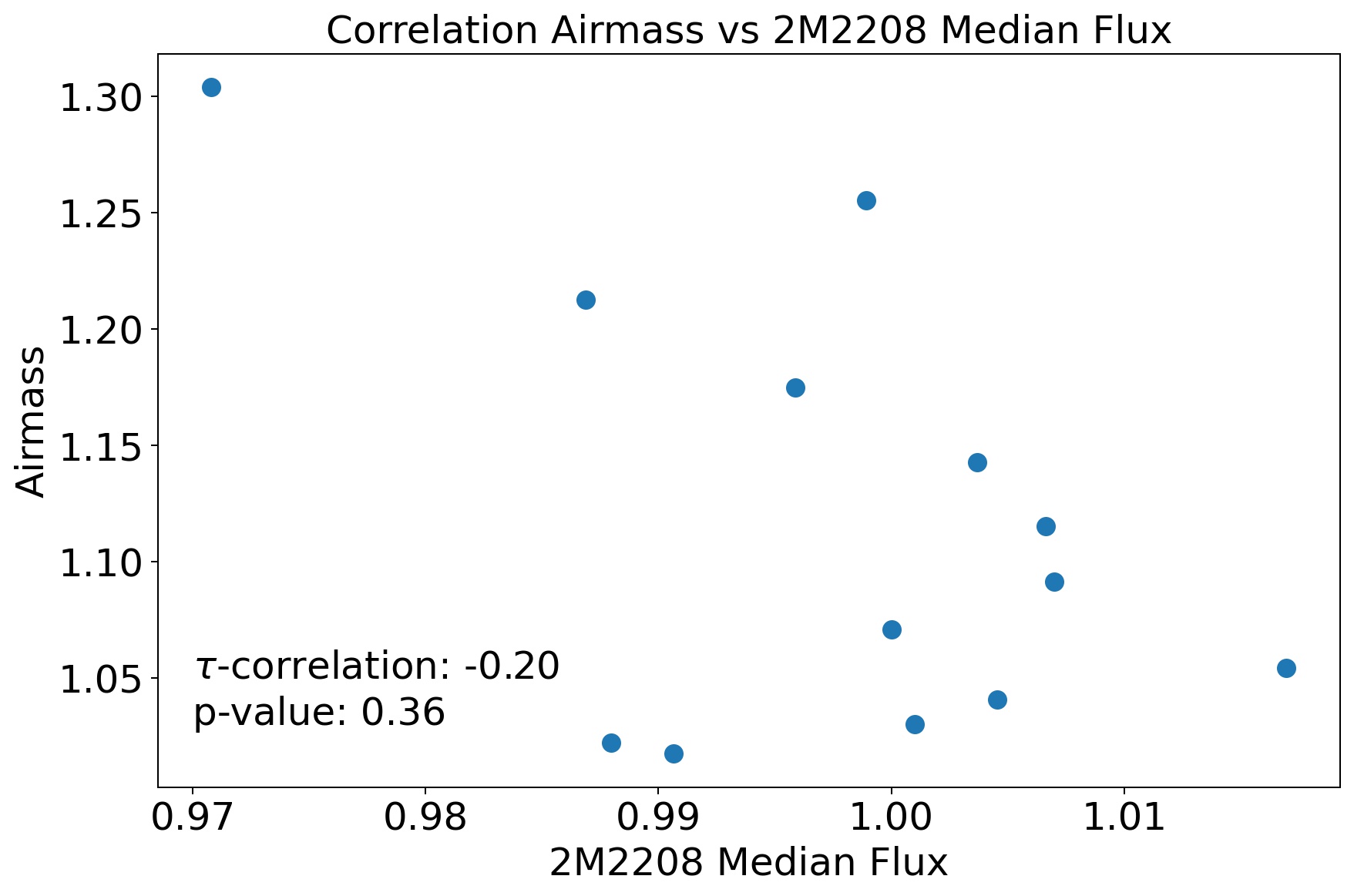}
    \caption{Left: Correlation between the target's light curve and relative external humidity (RH). Right: Correlation between the target's light curve and the external temperature. Bottom: Correlation between the target's light curve and the airmass.}
    \label{corr_RH_temp_airm}
\end{figure}

\section{$J$-band Light curves of the calibration stars before and after correction}

\begin{figure*}
    \centering
    \includegraphics[width=0.46\textwidth]{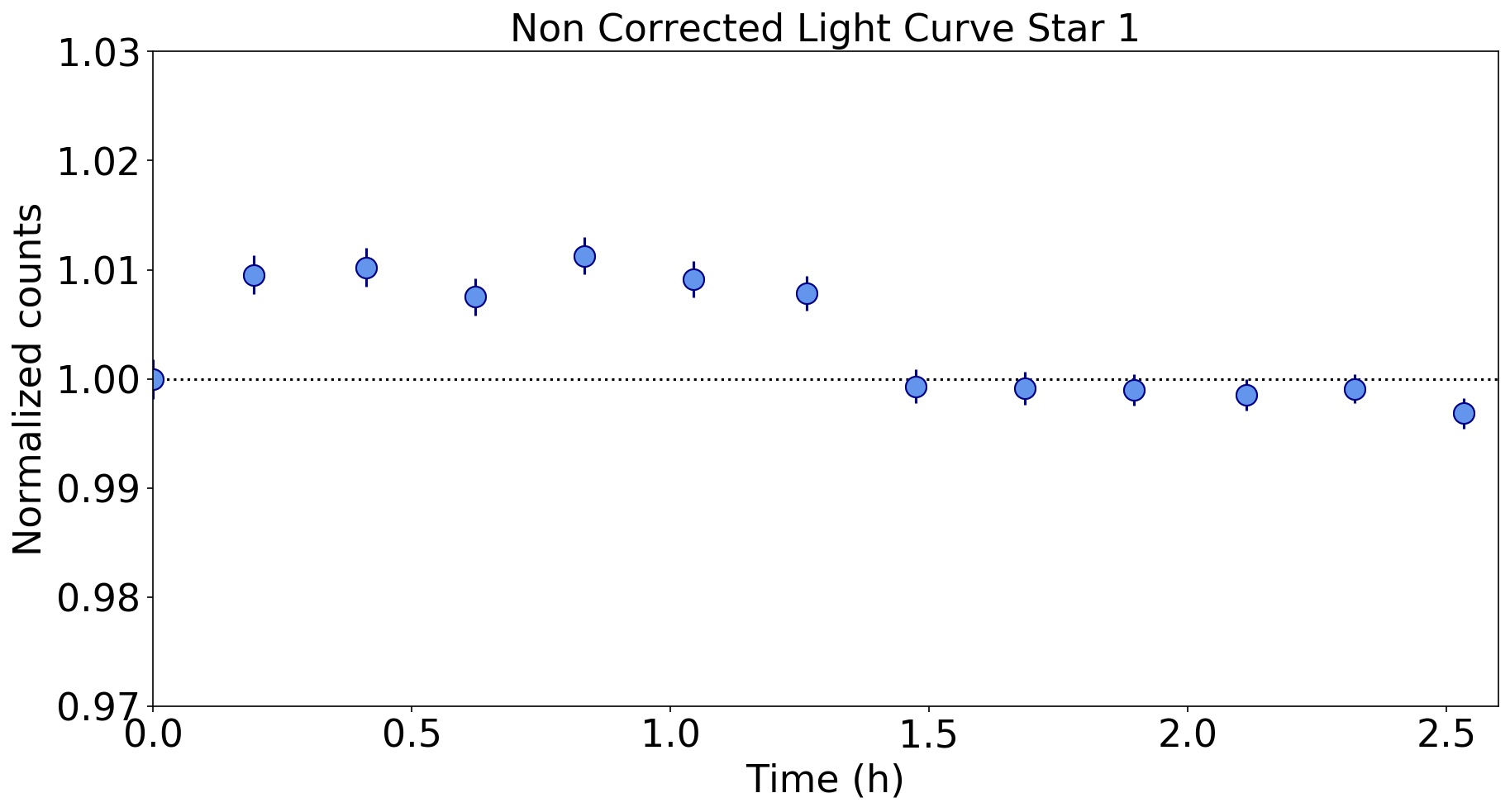}
    \includegraphics[width=0.46\textwidth]{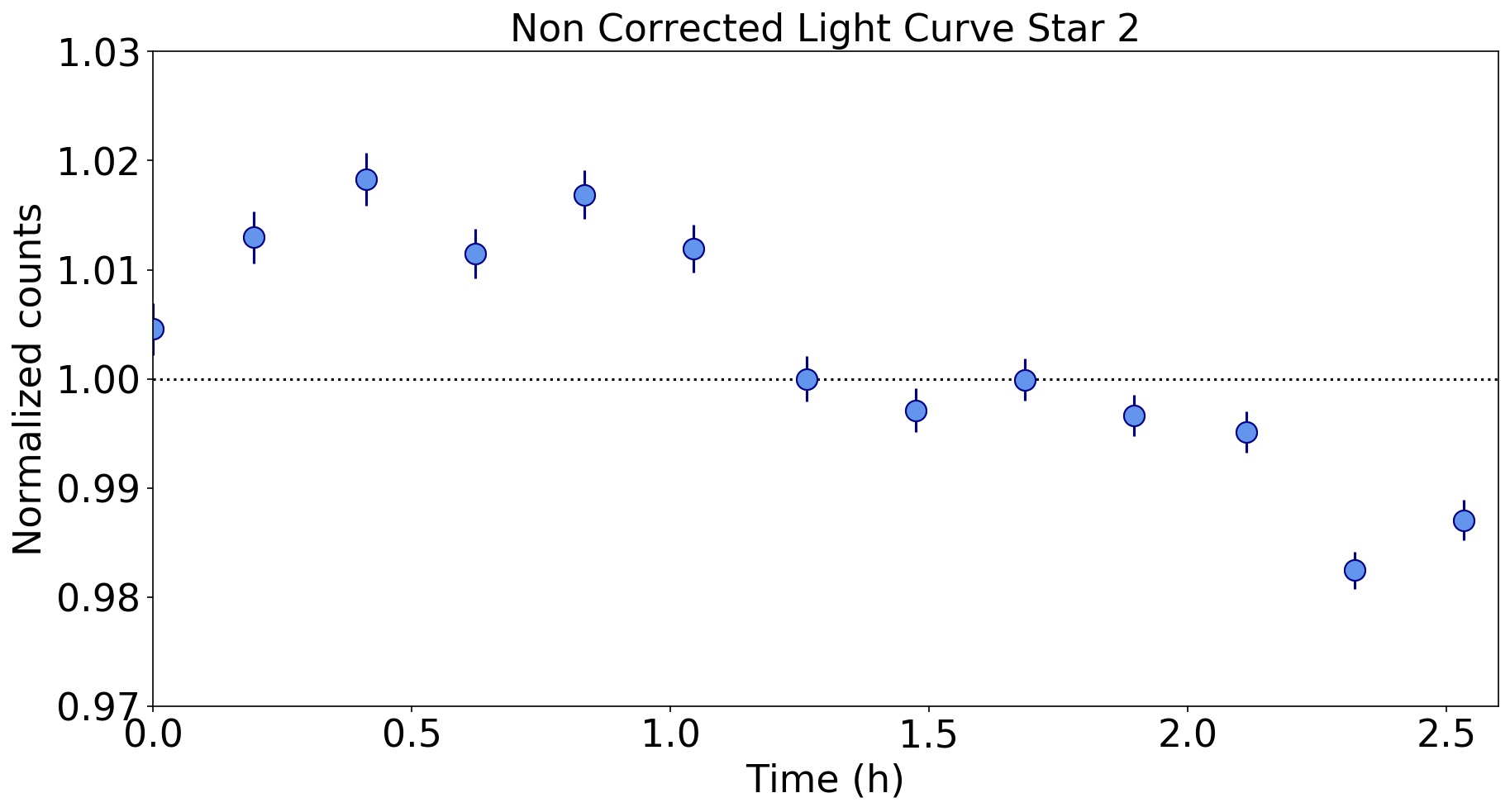}
    \includegraphics[width=0.46\textwidth]{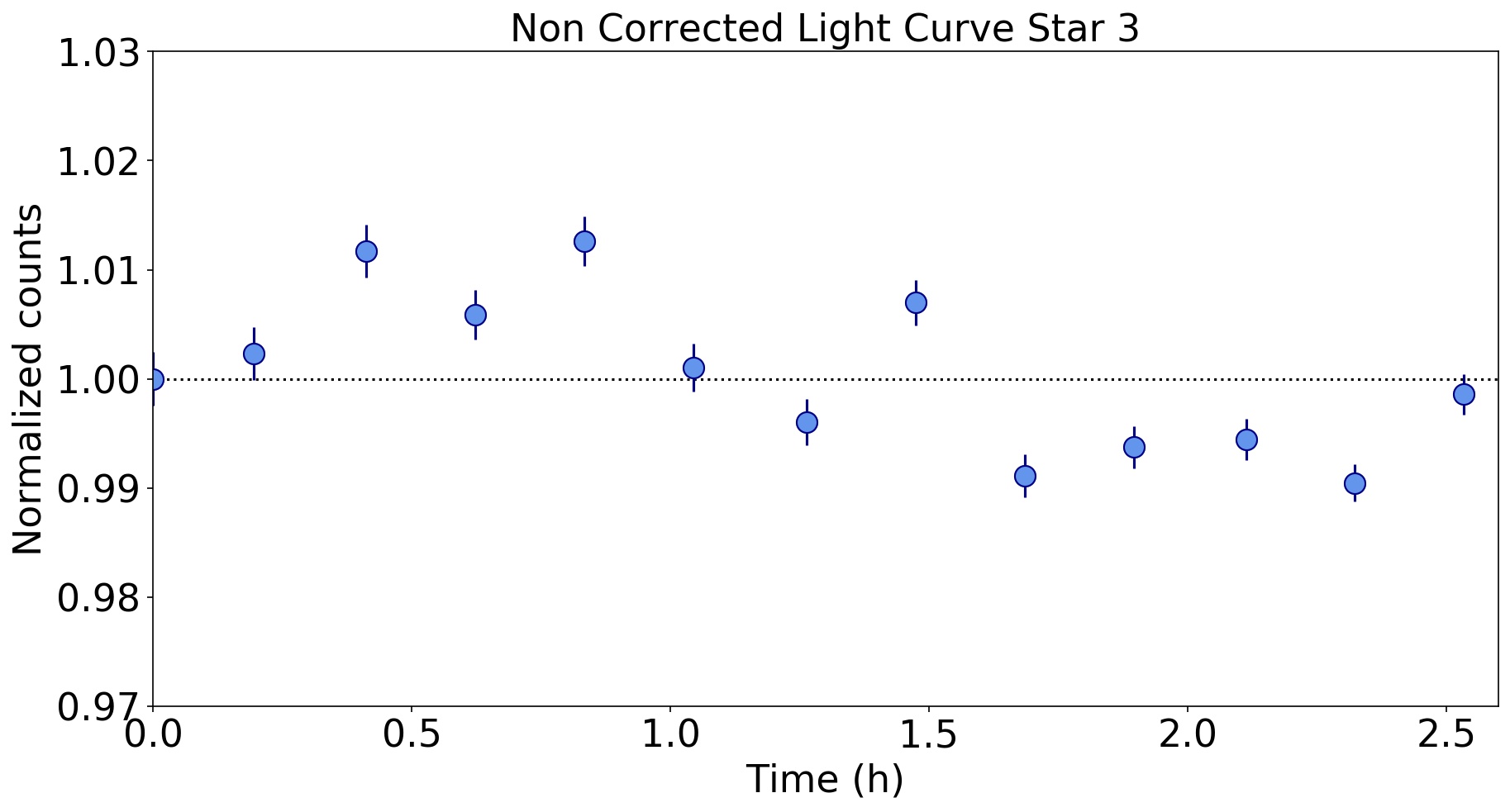}
    \includegraphics[width=0.46\textwidth]{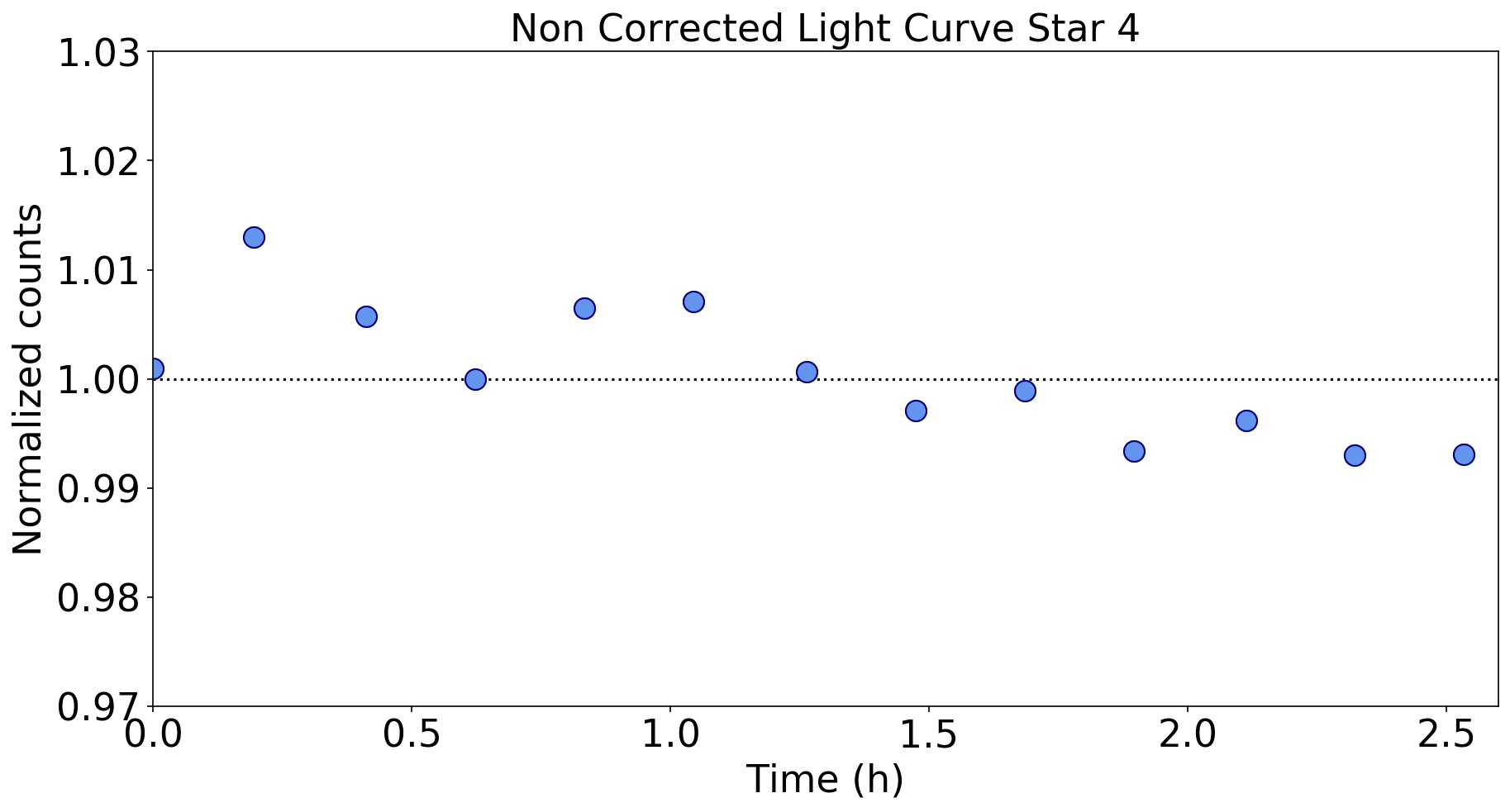}
    \includegraphics[width=0.46\textwidth]{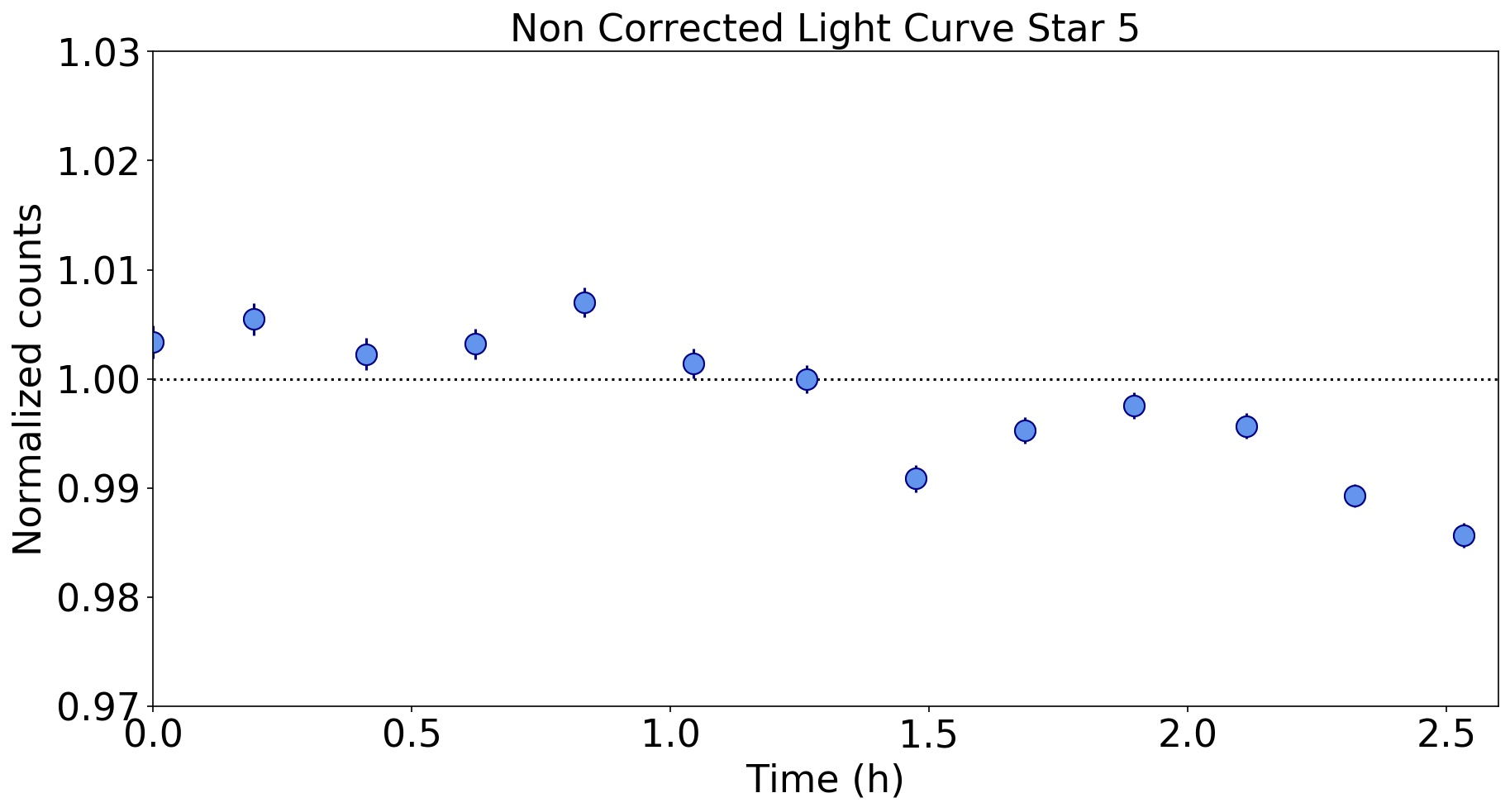}
    \includegraphics[width=0.46\textwidth]{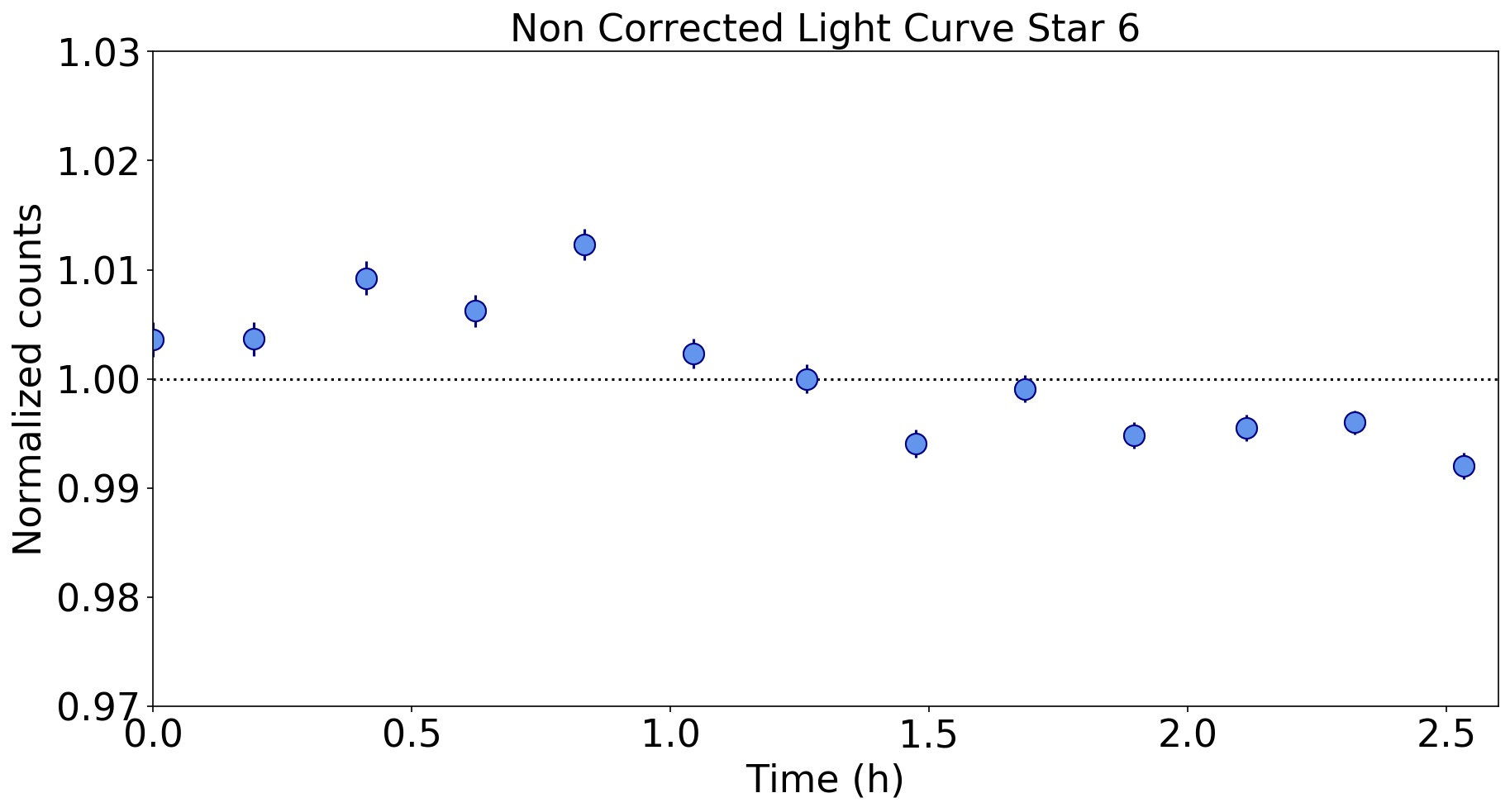}
    \includegraphics[width=0.46\textwidth]{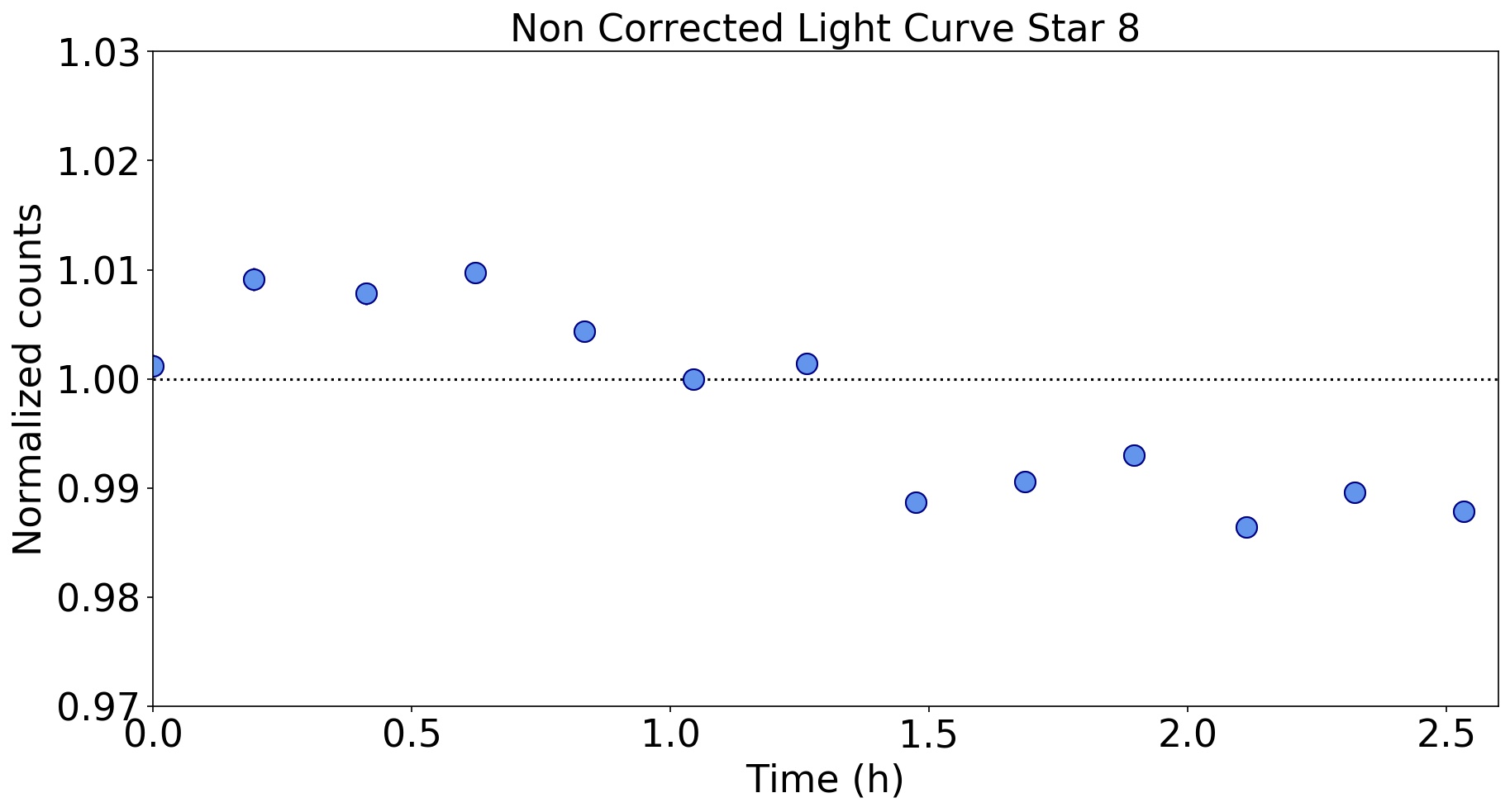}
    \includegraphics[width=0.46\textwidth]{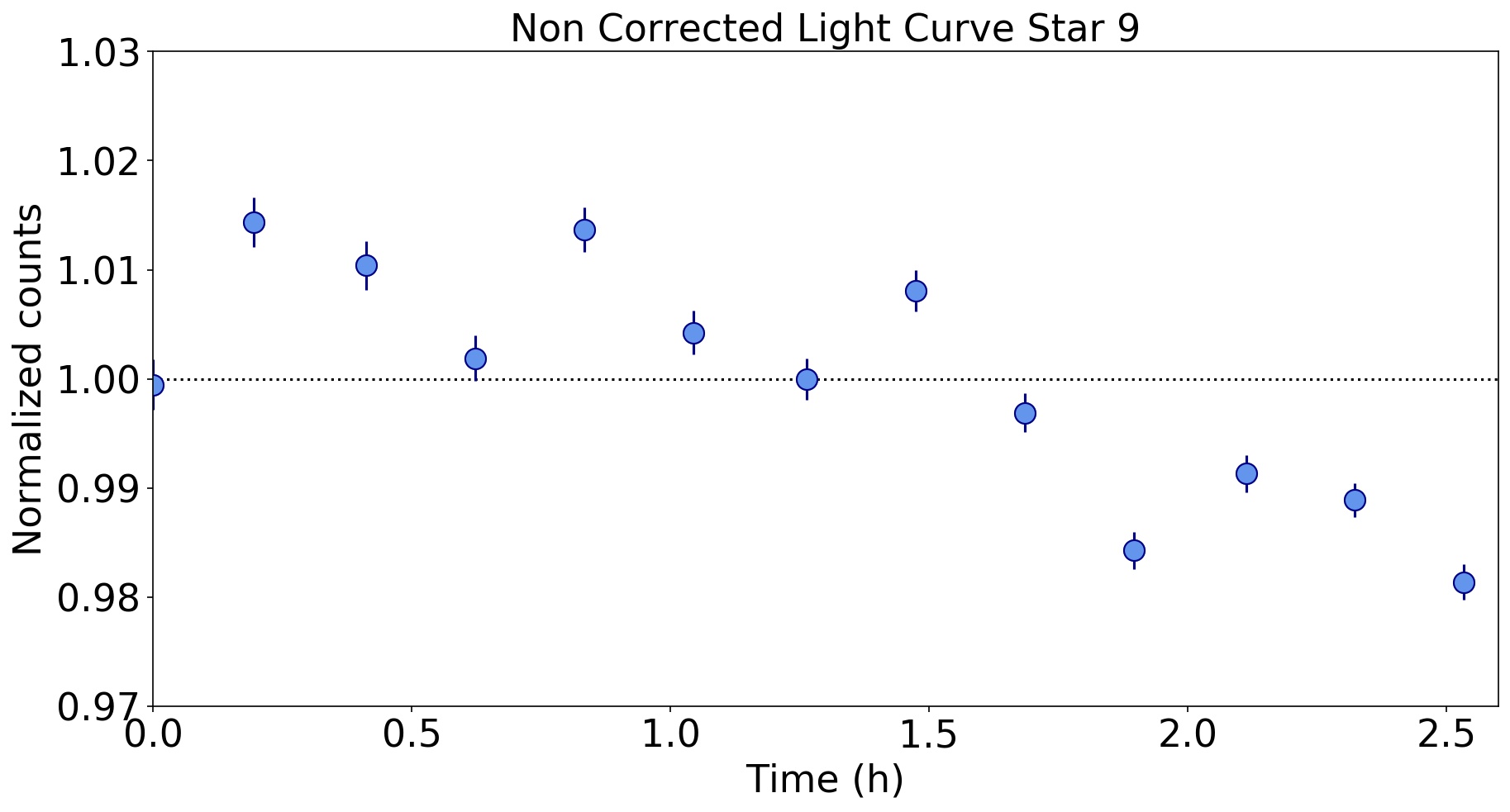}
    \includegraphics[width=0.46\textwidth]{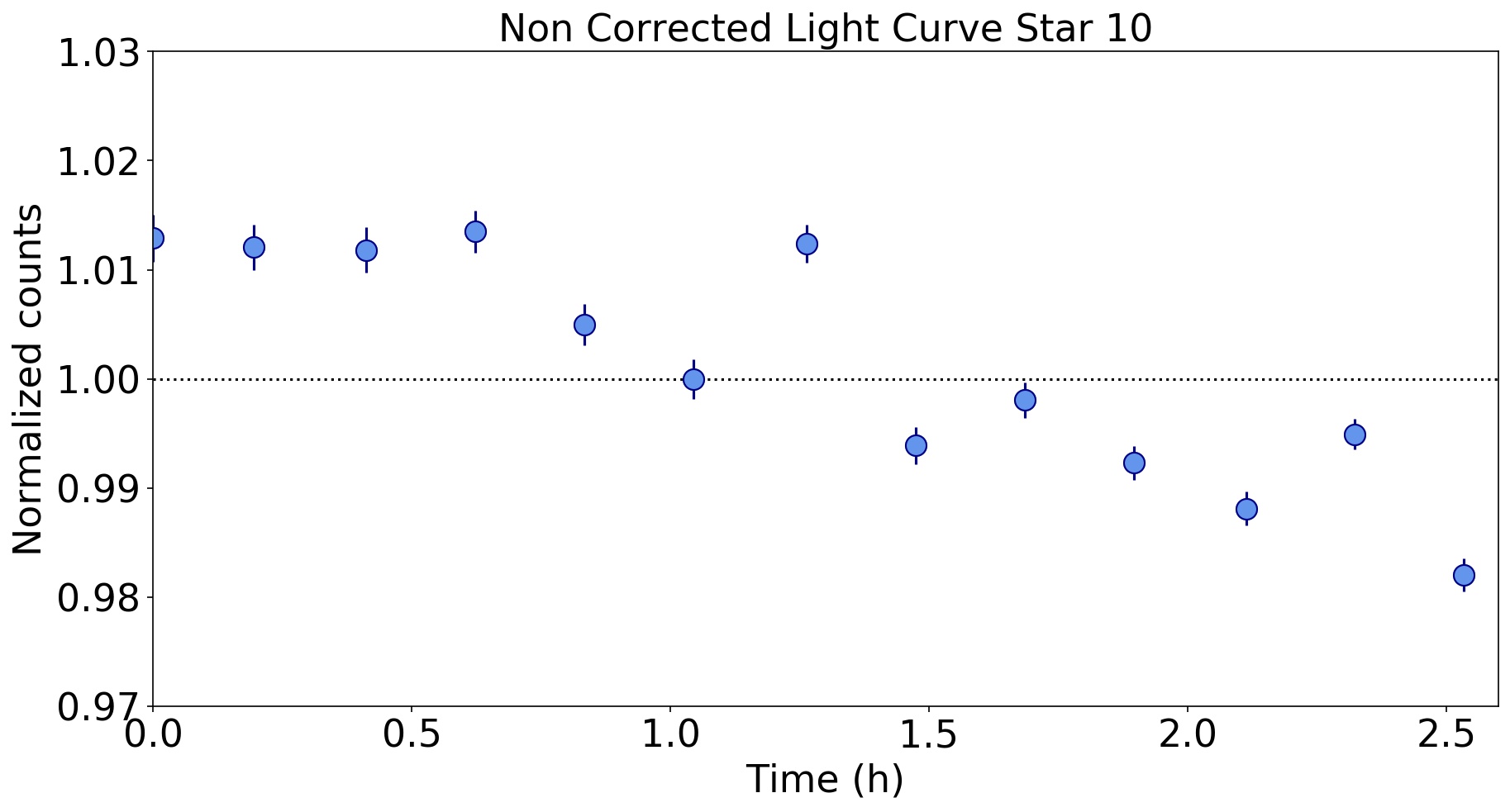}
    \caption{Normalized non-corrected light curves of  the calibration stars on the field of 2M2208+2921.}
    \label{non_corr_LCs}
\end{figure*}

\begin{figure*}
    \centering
    \includegraphics[width=0.46\textwidth]{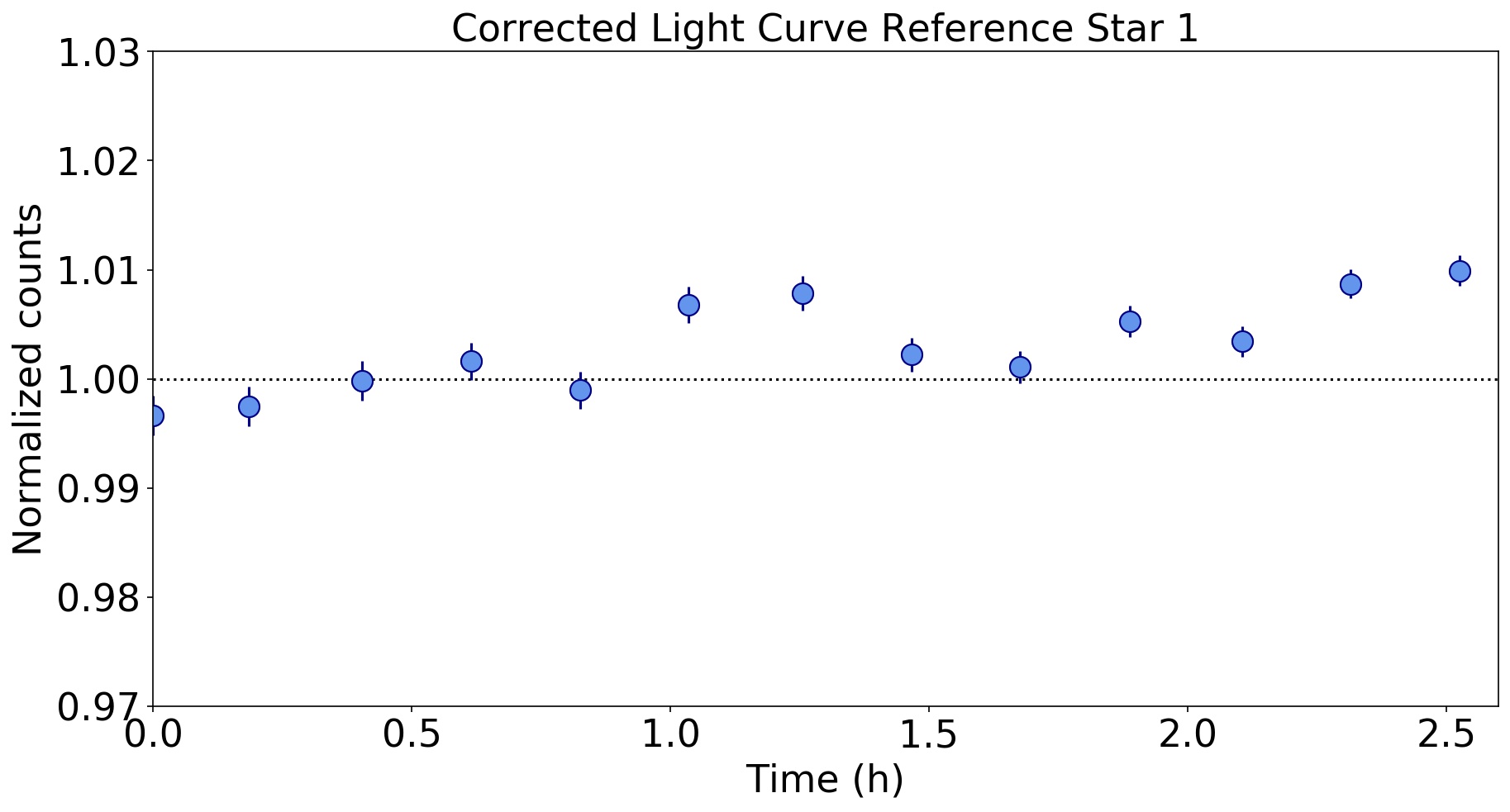}
    \includegraphics[width=0.46\textwidth]{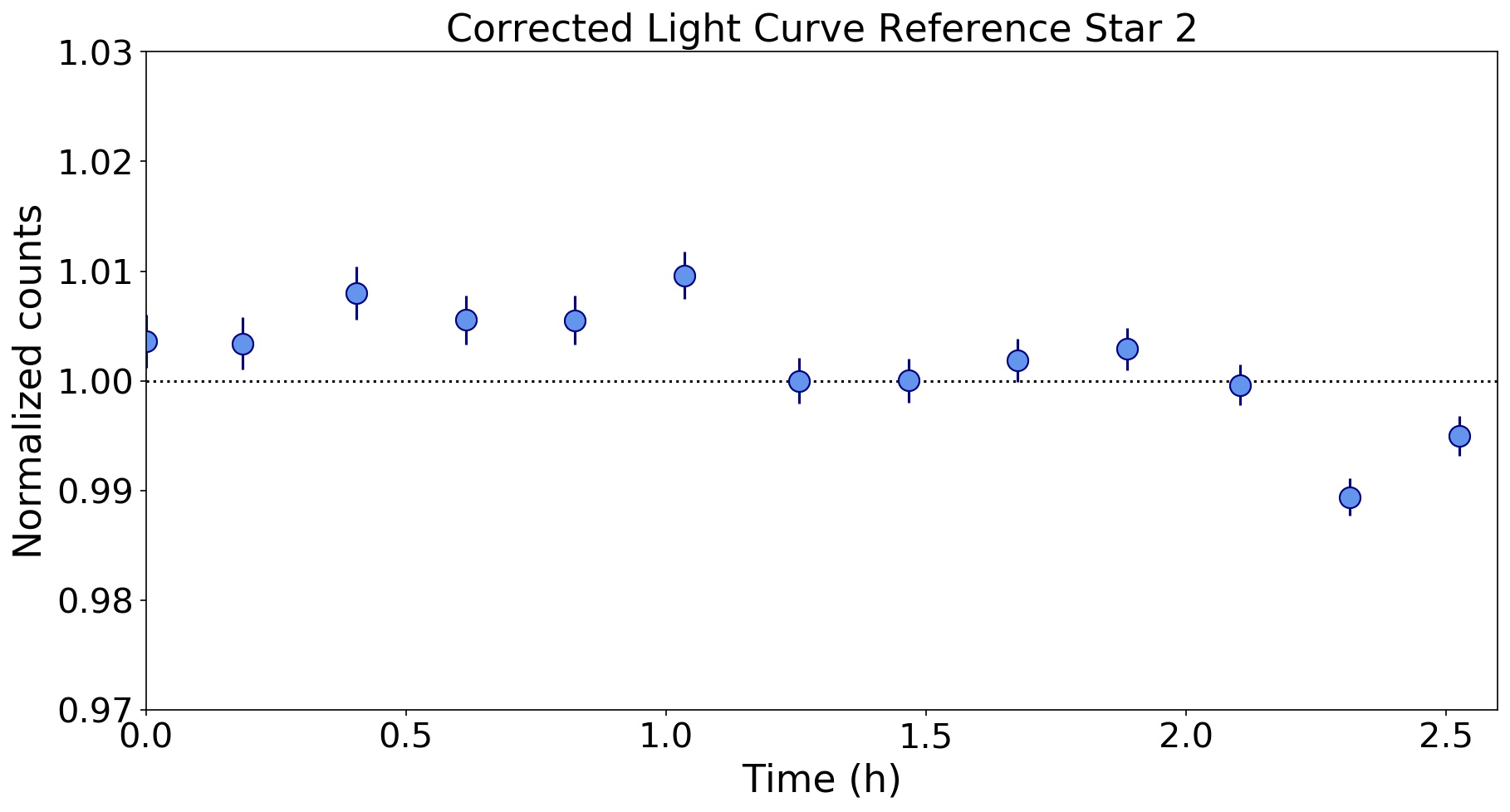}
    \includegraphics[width=0.46\textwidth]{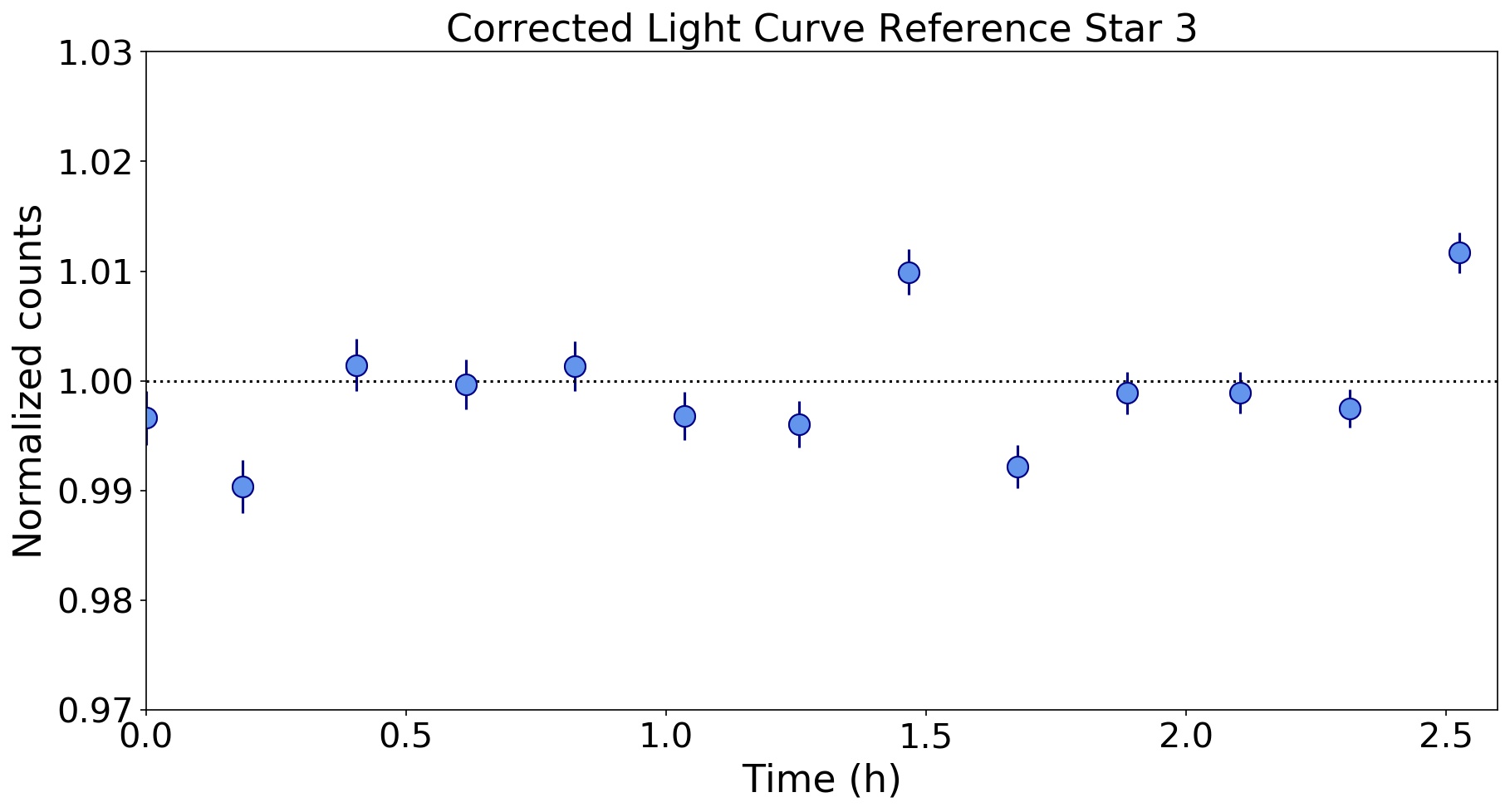}
    \includegraphics[width=0.46\textwidth]{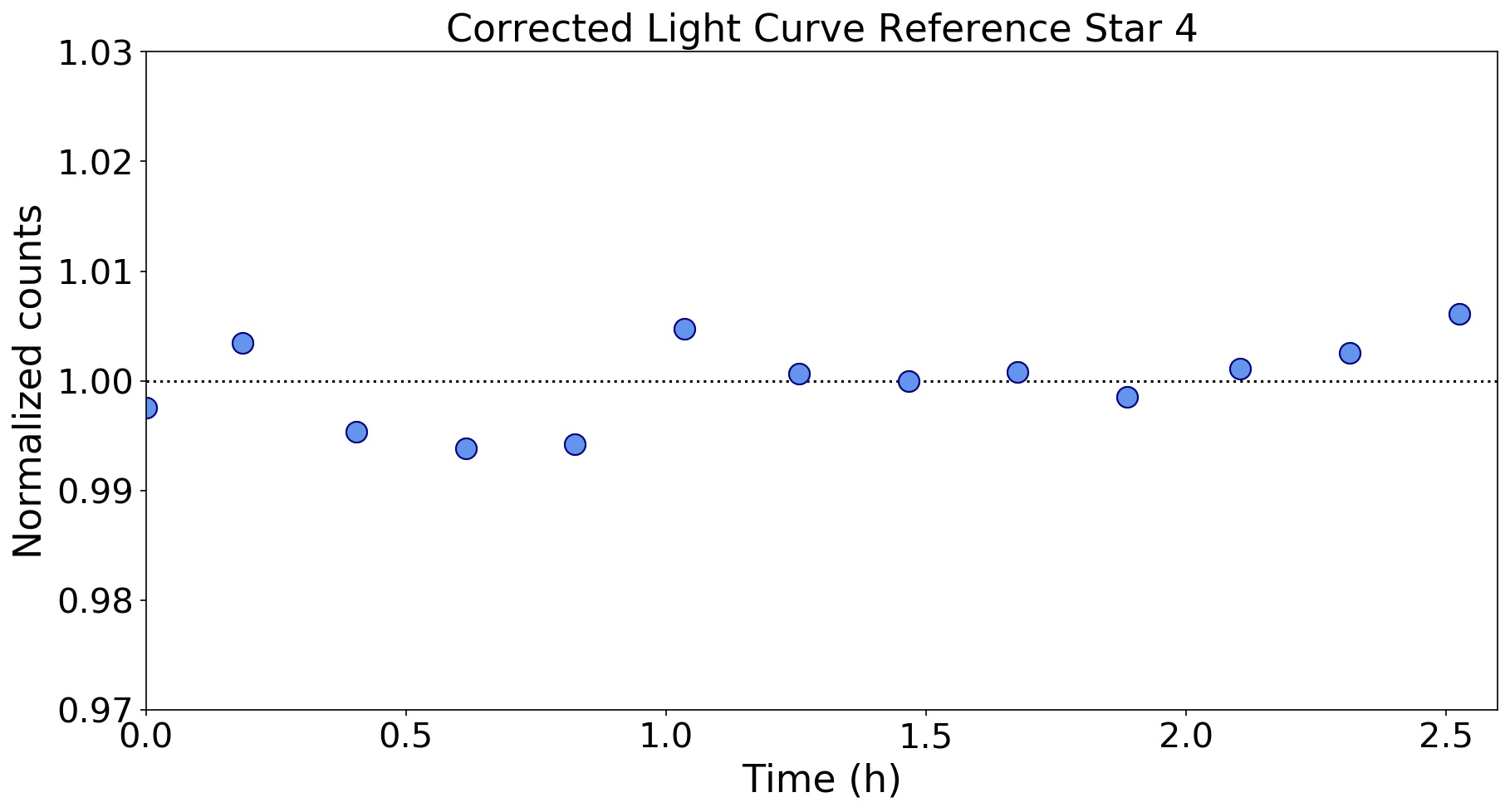}
    \includegraphics[width=0.46\textwidth]{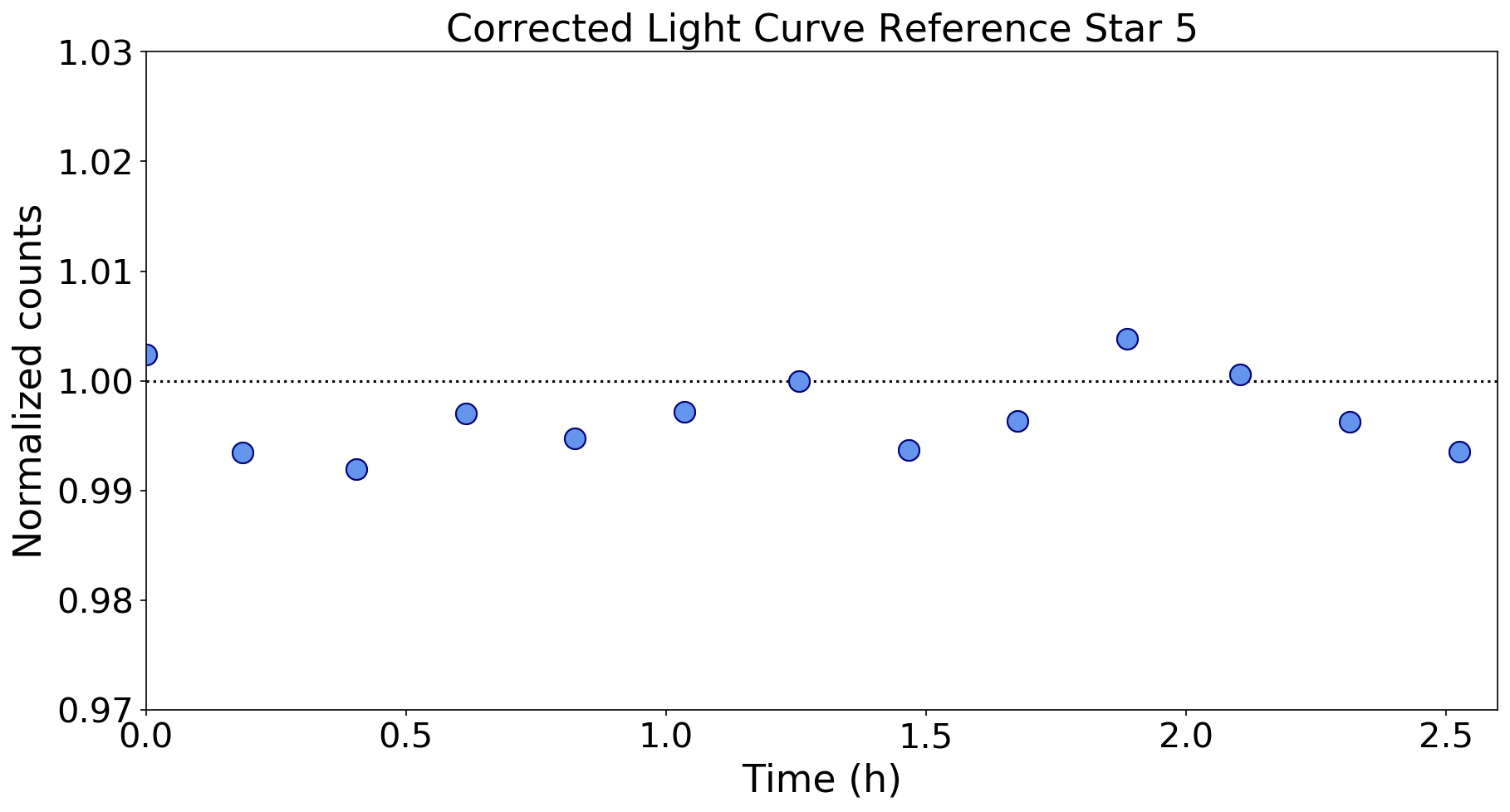}
    \includegraphics[width=0.46\textwidth]{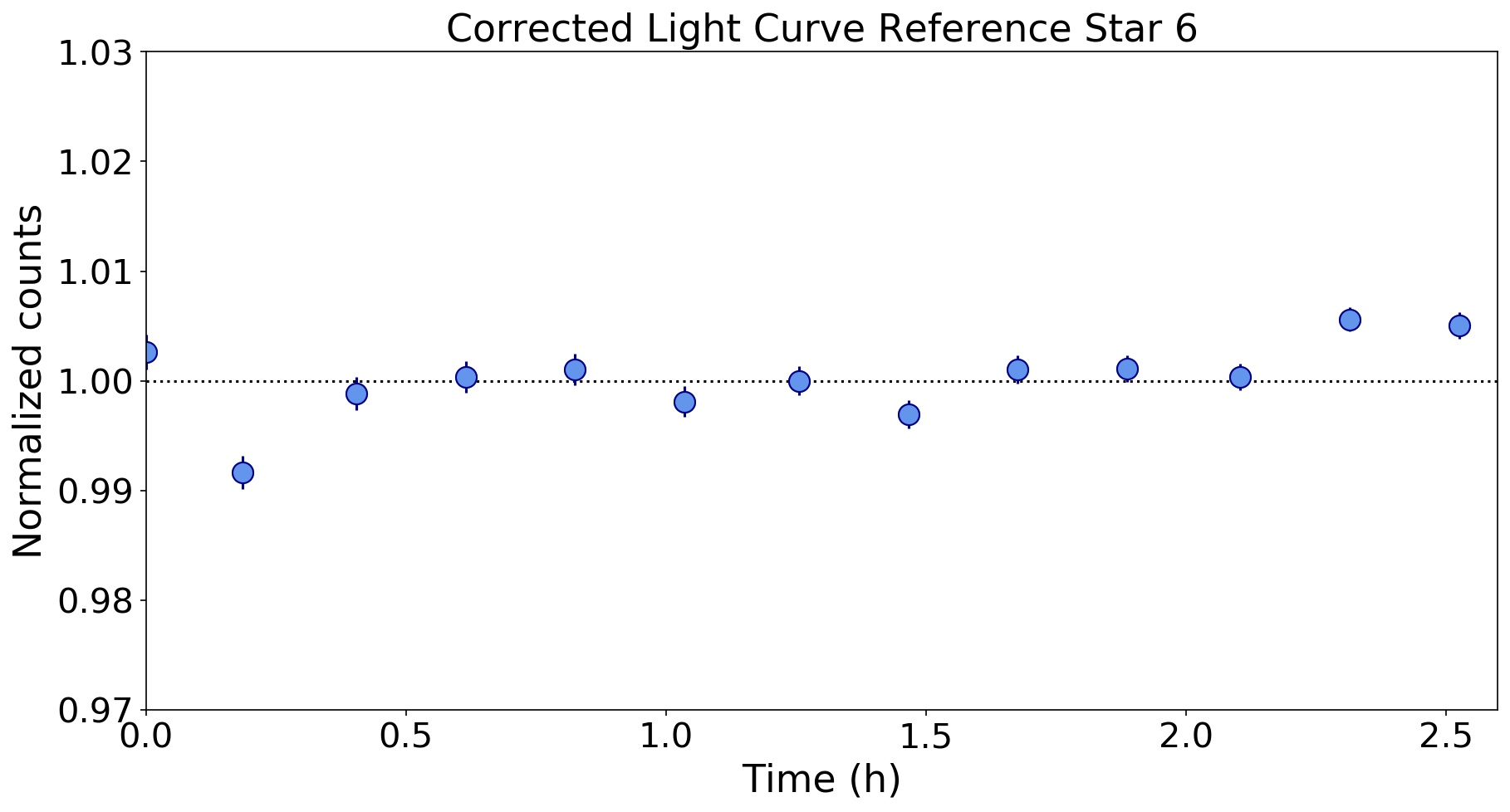}
    \includegraphics[width=0.46\textwidth]{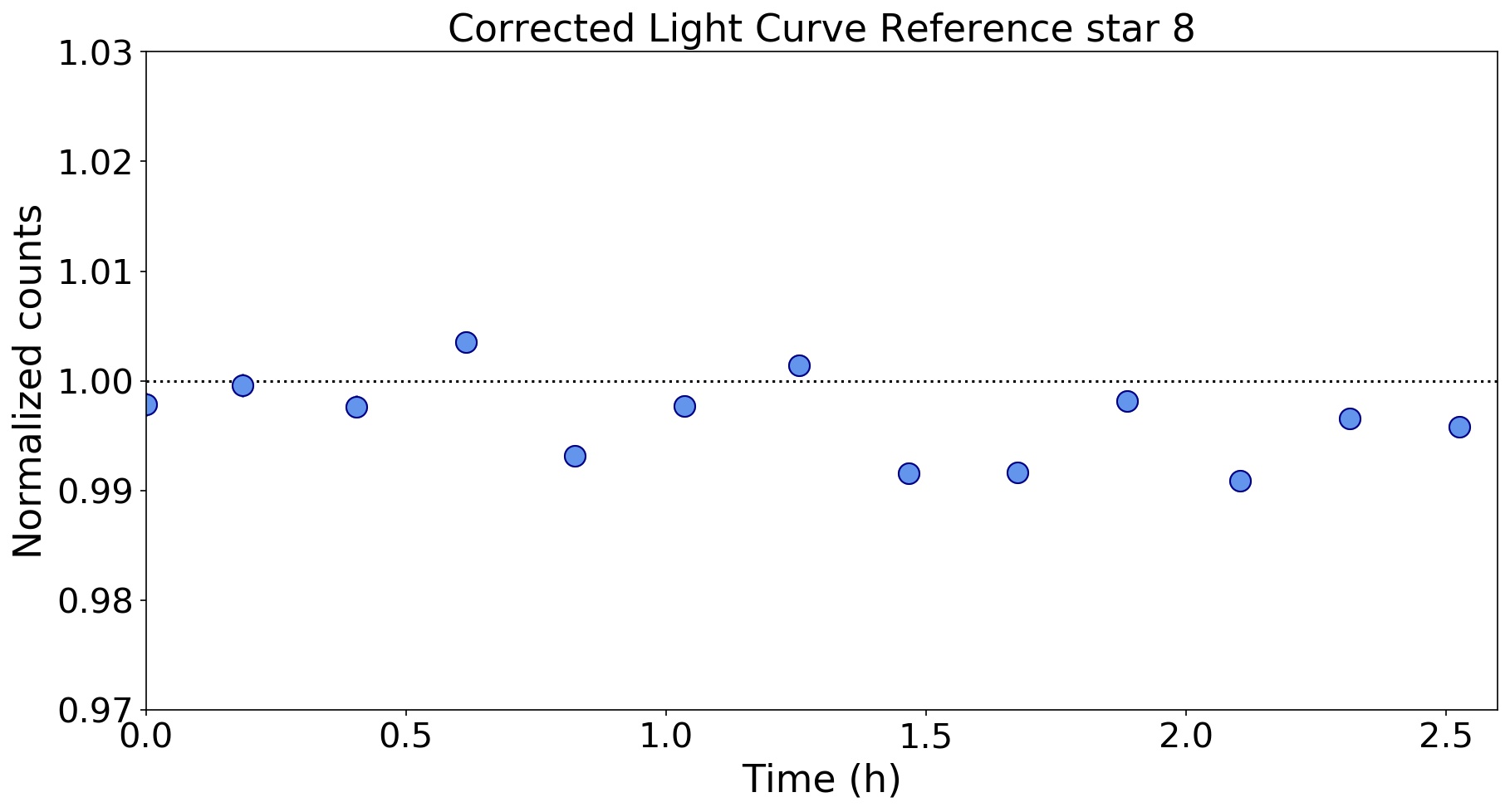}
    \includegraphics[width=0.46\textwidth]{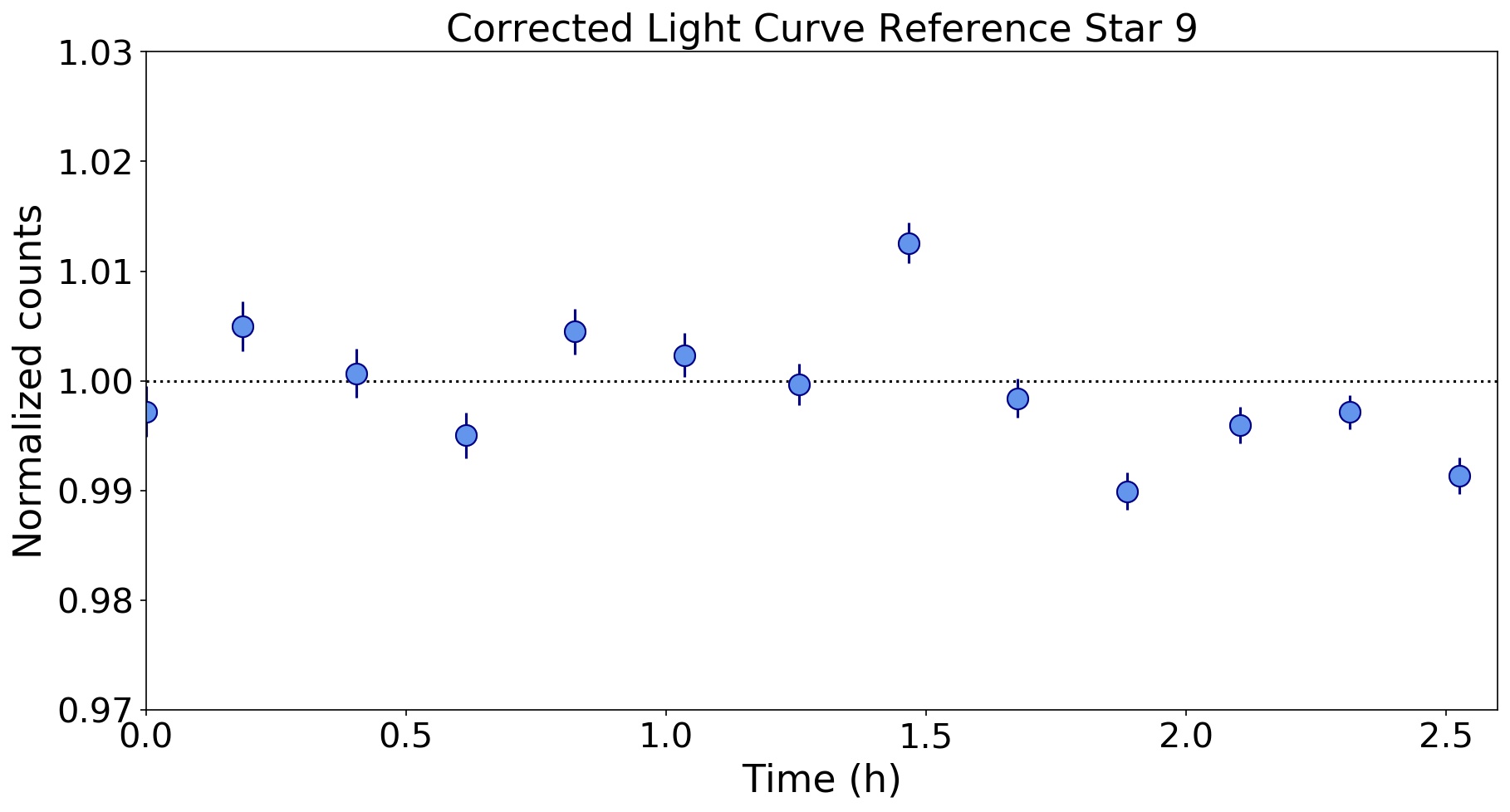}
    \includegraphics[width=0.46\textwidth]{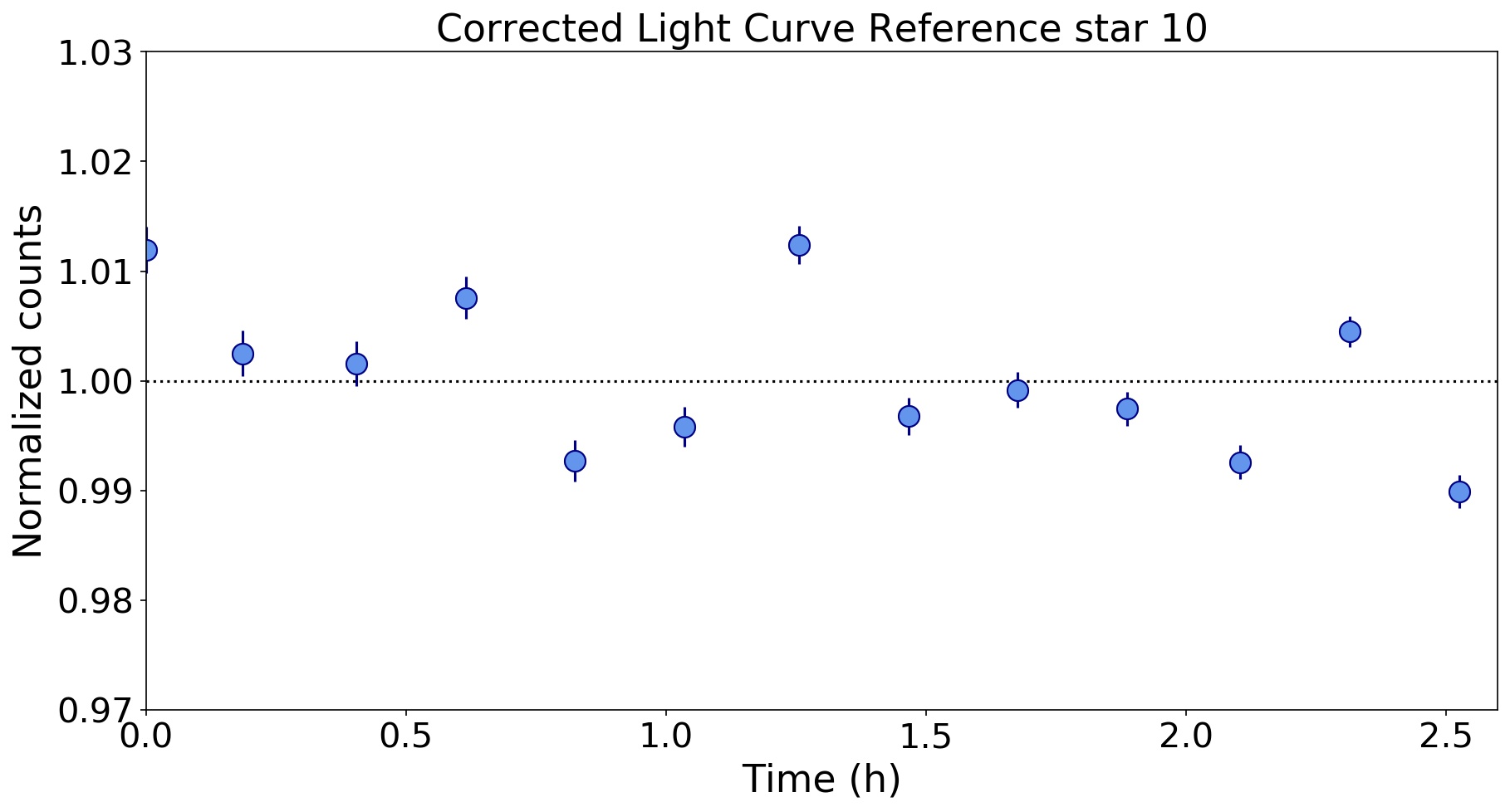}
    \caption{Normalized corrected light curves of the  calibration stars on the field of 2M2208+2921.}
    \label{corr_LCs}
\end{figure*}

\section{Light curves of the calibration stars at the wavelength of the K\,I doublet and the Na\,I alkali lines}

\begin{figure*}
    \centering
    \includegraphics[width=0.56\textwidth]{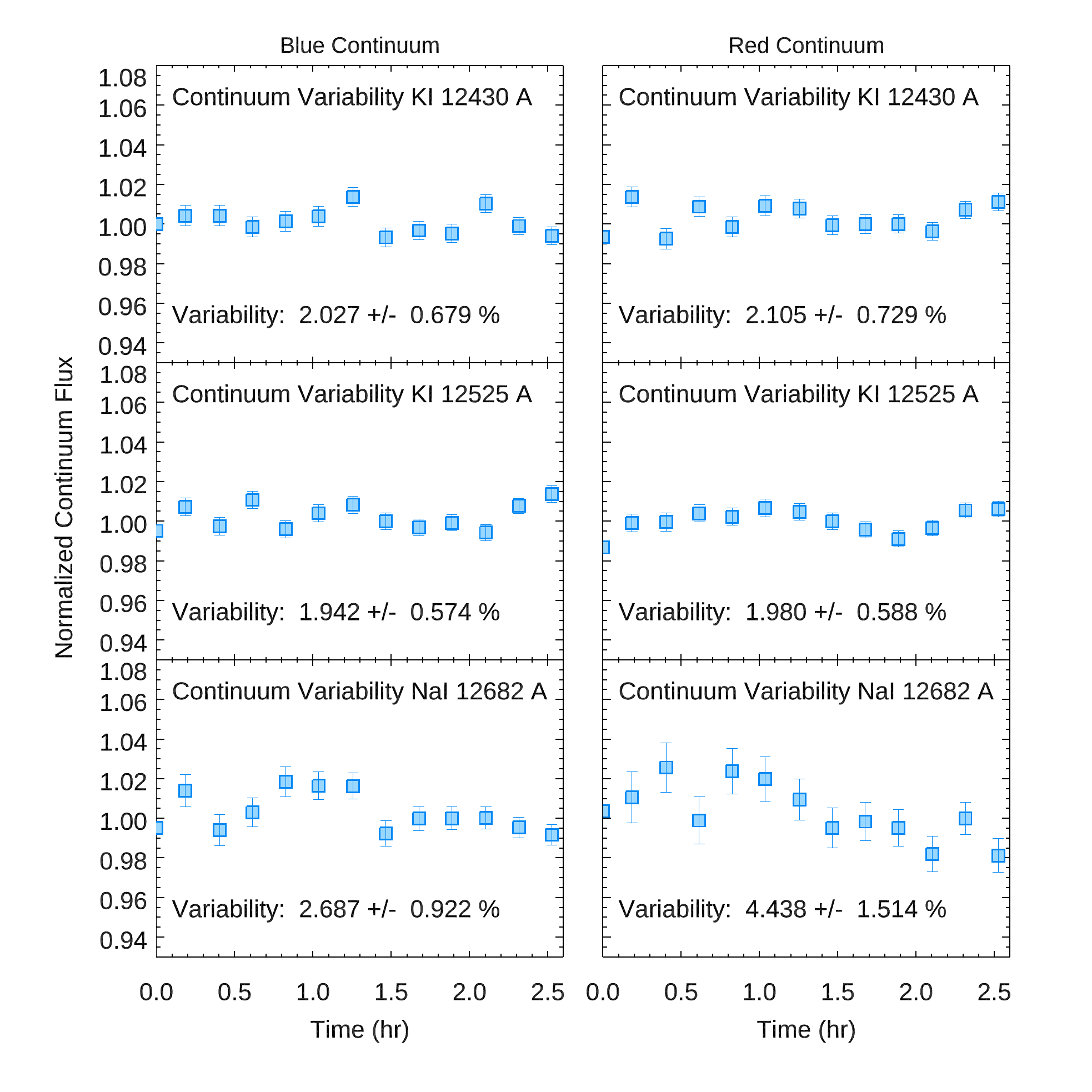}
    \includegraphics[width=0.41\textwidth]{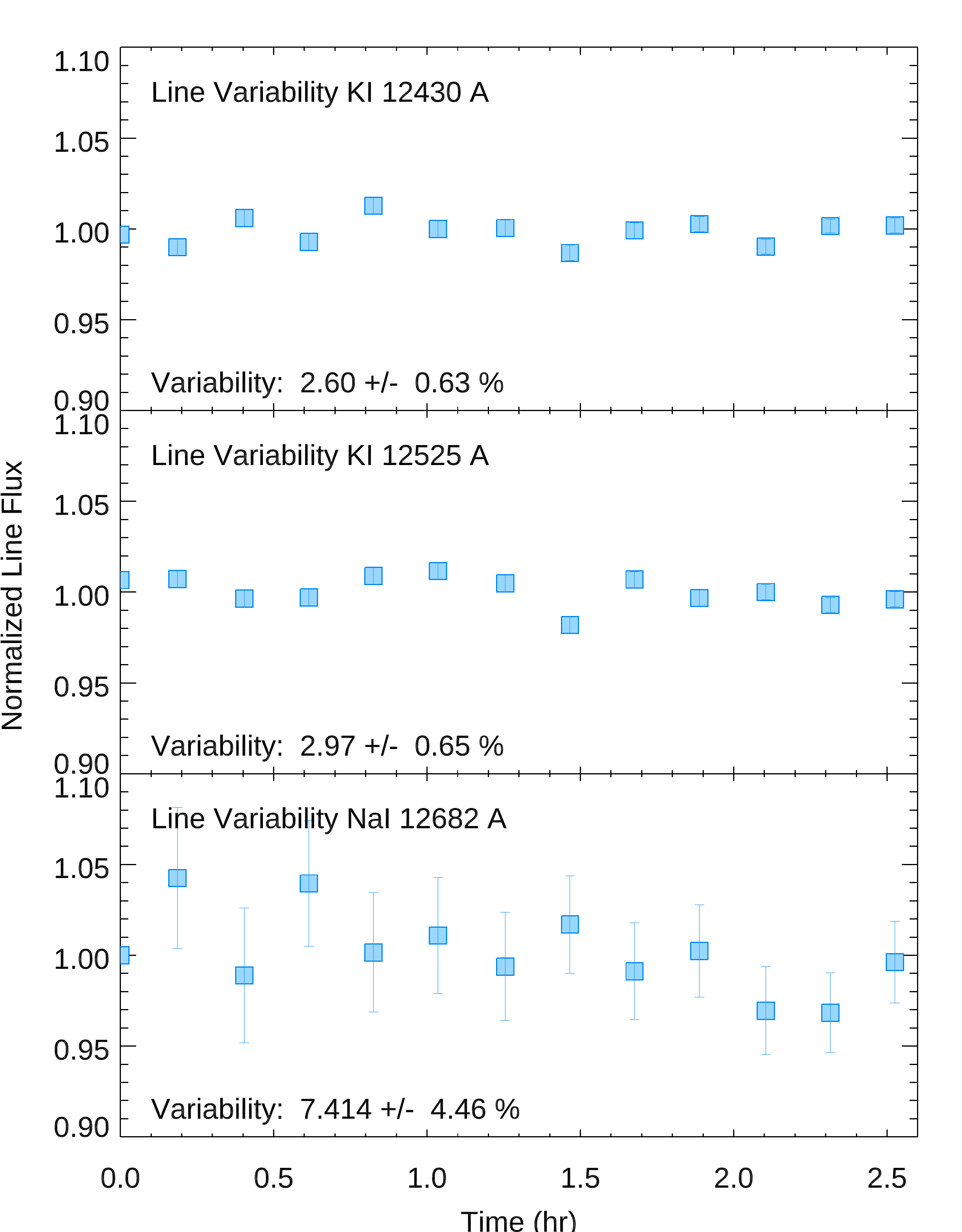}
    \caption{Variability inside the wavelength range of the blue and red continuum, and inside the alkali lines wavelength range for the calibration star 1 for spectra with the original $J$-band MOSFIRE resolution.}
    \label{corr_LCs_obj1}
\end{figure*}

\begin{figure*}
    \centering
    \includegraphics[width=0.56\textwidth]{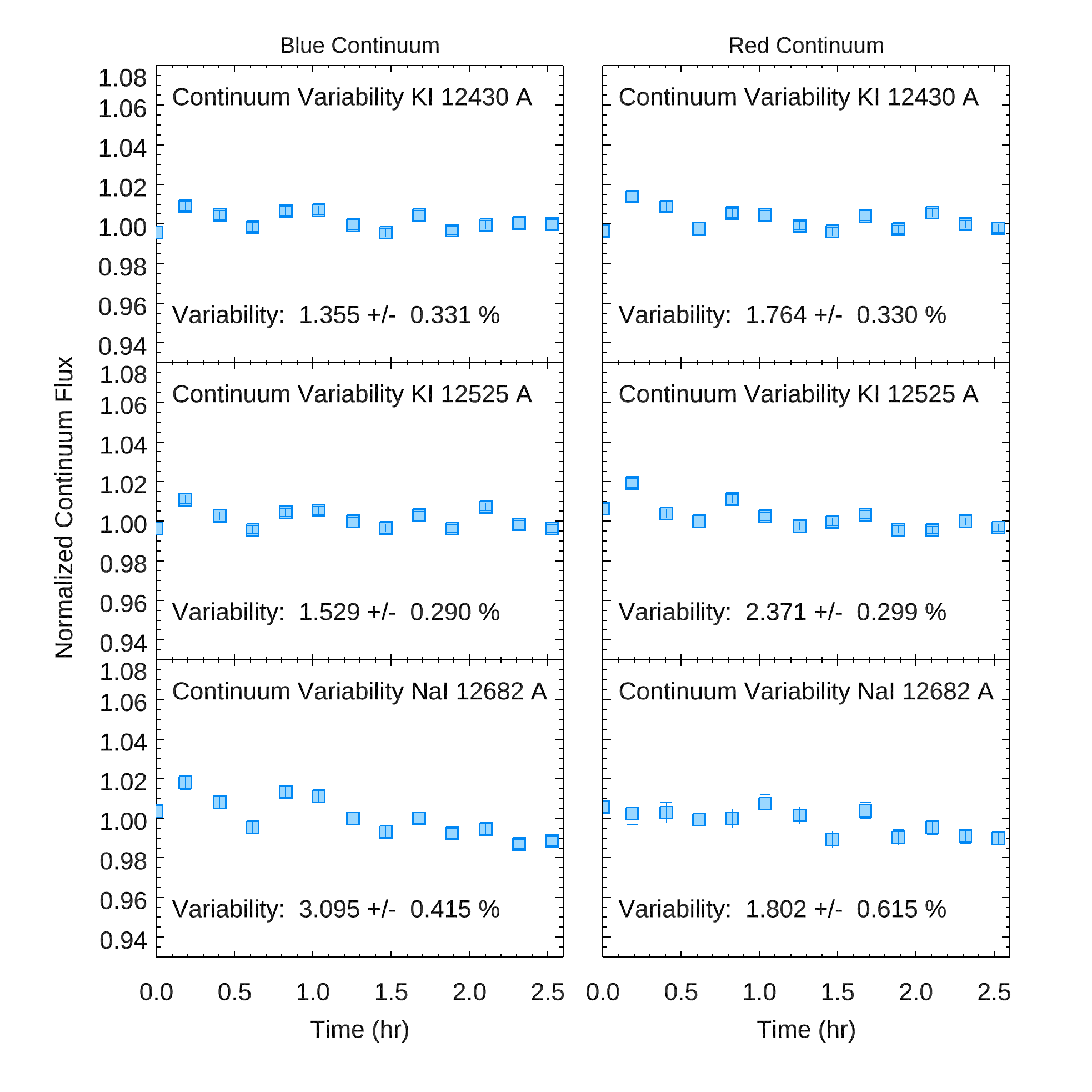}
    \includegraphics[width=0.41\textwidth]{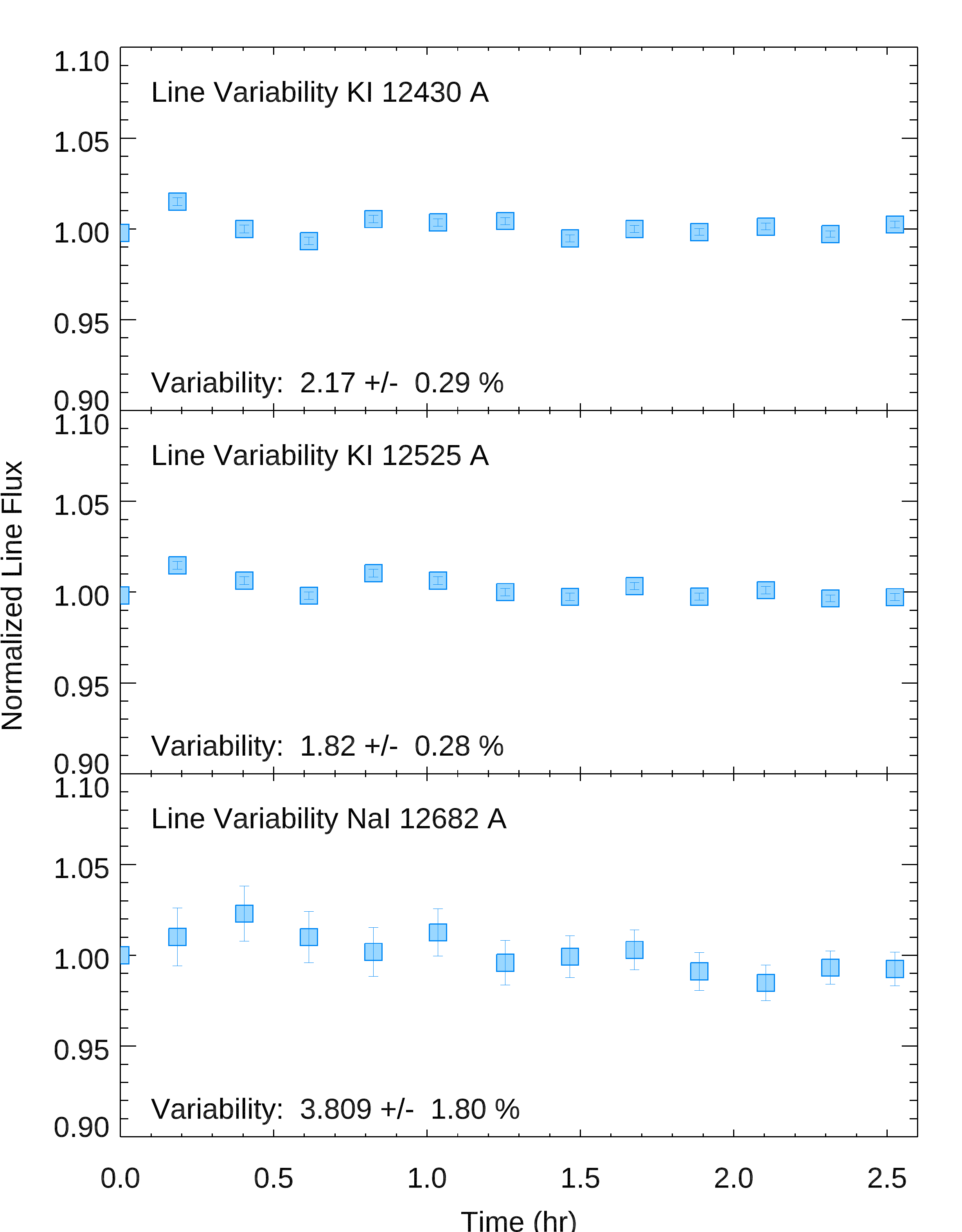}
    \caption{Variability inside the wavelength range of the blue and red continuum, and inside the alkali lines wavelength range for the calibration star 4 for spectra with the original $J$-band MOSFIRE resolution.}
    \label{corr_LCs_obj4}
\end{figure*}

\begin{figure*}
    \centering
    \includegraphics[width=0.56\textwidth]{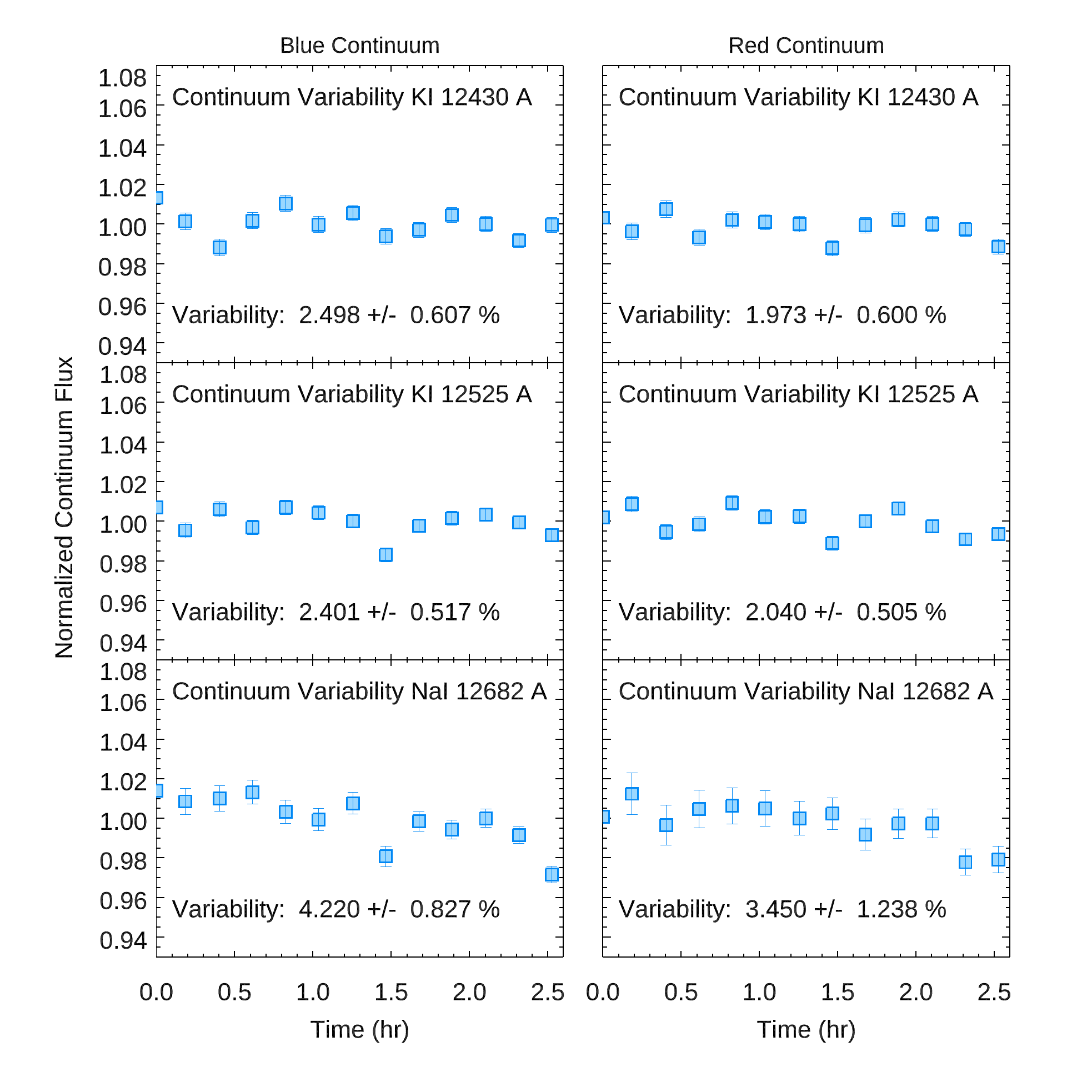}
    \includegraphics[width=0.41\textwidth]{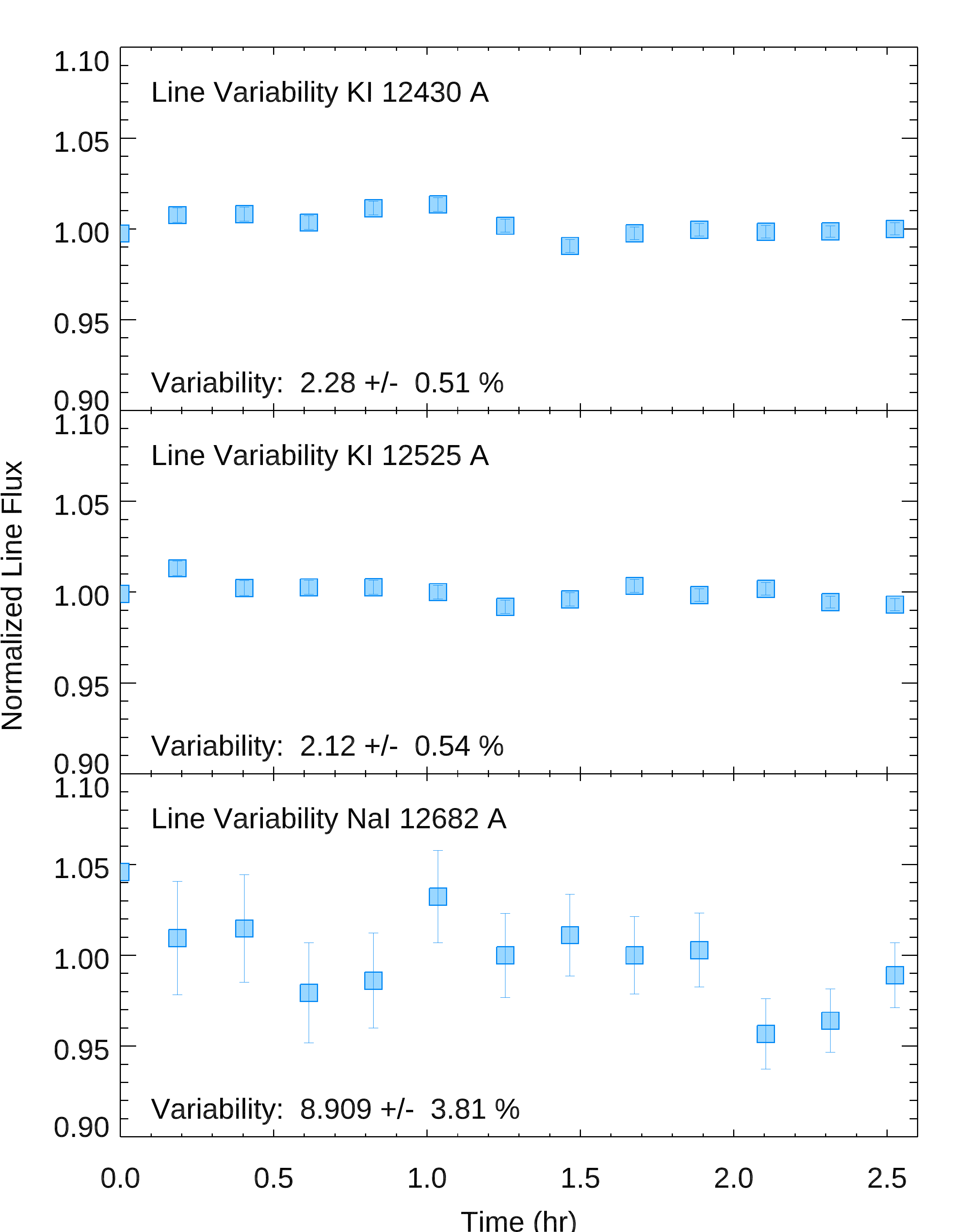}
    \caption{Variability inside the wavelength range of the blue and red continuum, and inside the alkali lines wavelength range for the calibration star 5 for spectra with the original $J$-band MOSFIRE resolution.}
    \label{corr_LCs_obj5}
\end{figure*}

\begin{figure*}
    \centering
    \includegraphics[width=0.56\textwidth]{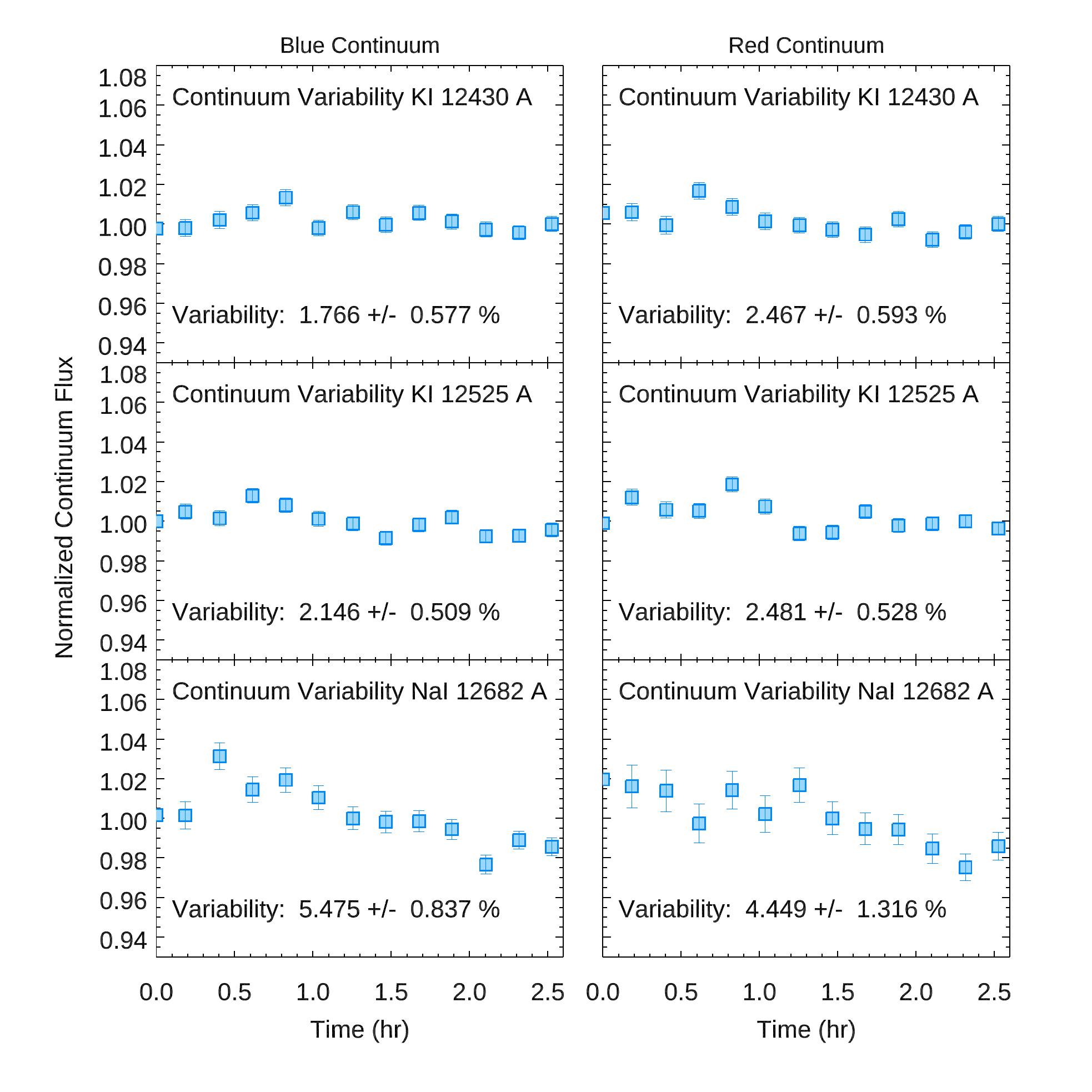}
    \includegraphics[width=0.41\textwidth]{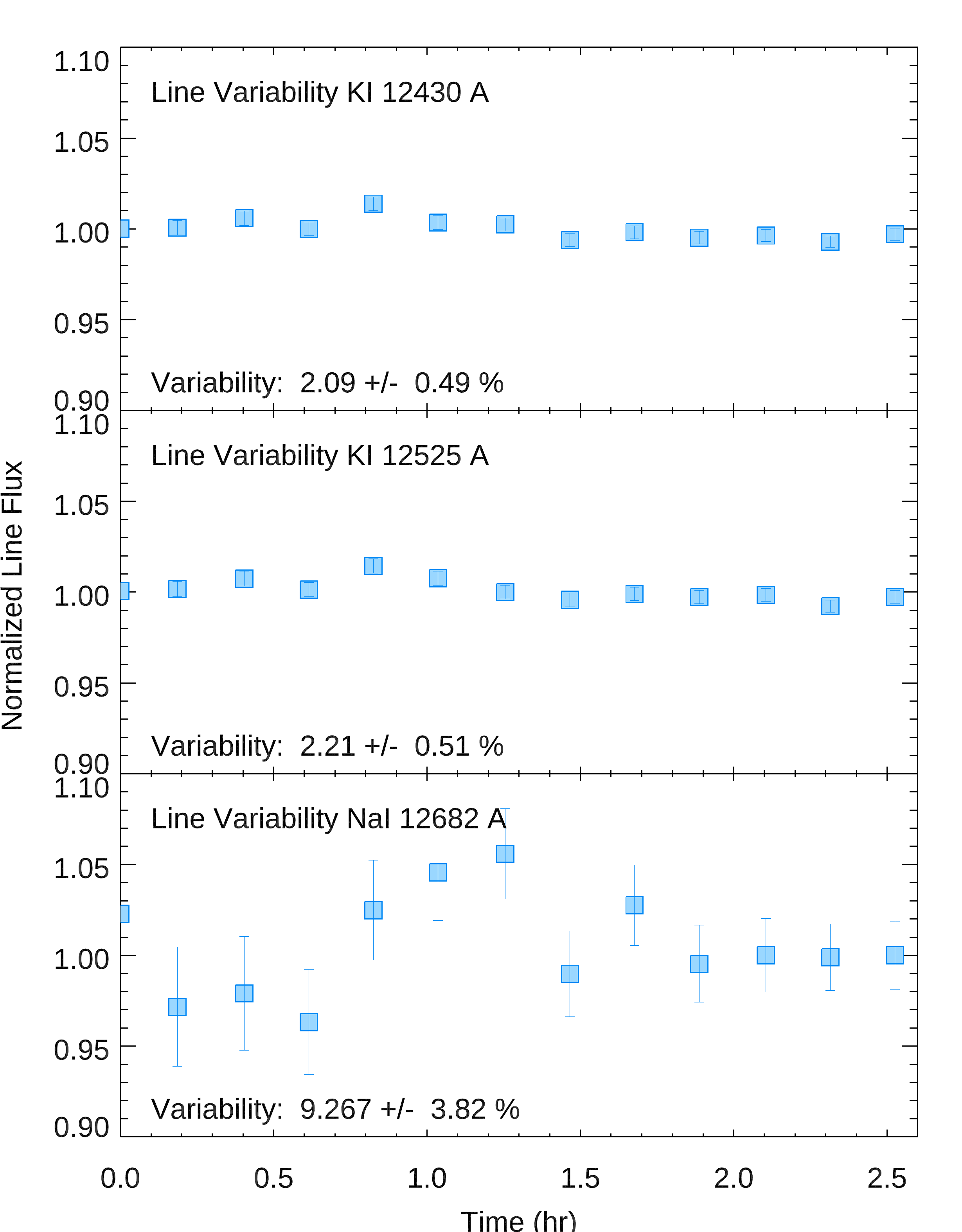}
    \caption{Variability inside the wavelength range of the blue and red continuum, and inside the alkali lines wavelength range for the calibration star 6 for spectra with the original $J$-band MOSFIRE resolution.}
    \label{corr_LCs_obj6}
\end{figure*}

\begin{figure*}
    \centering
    \includegraphics[width=0.56\textwidth]{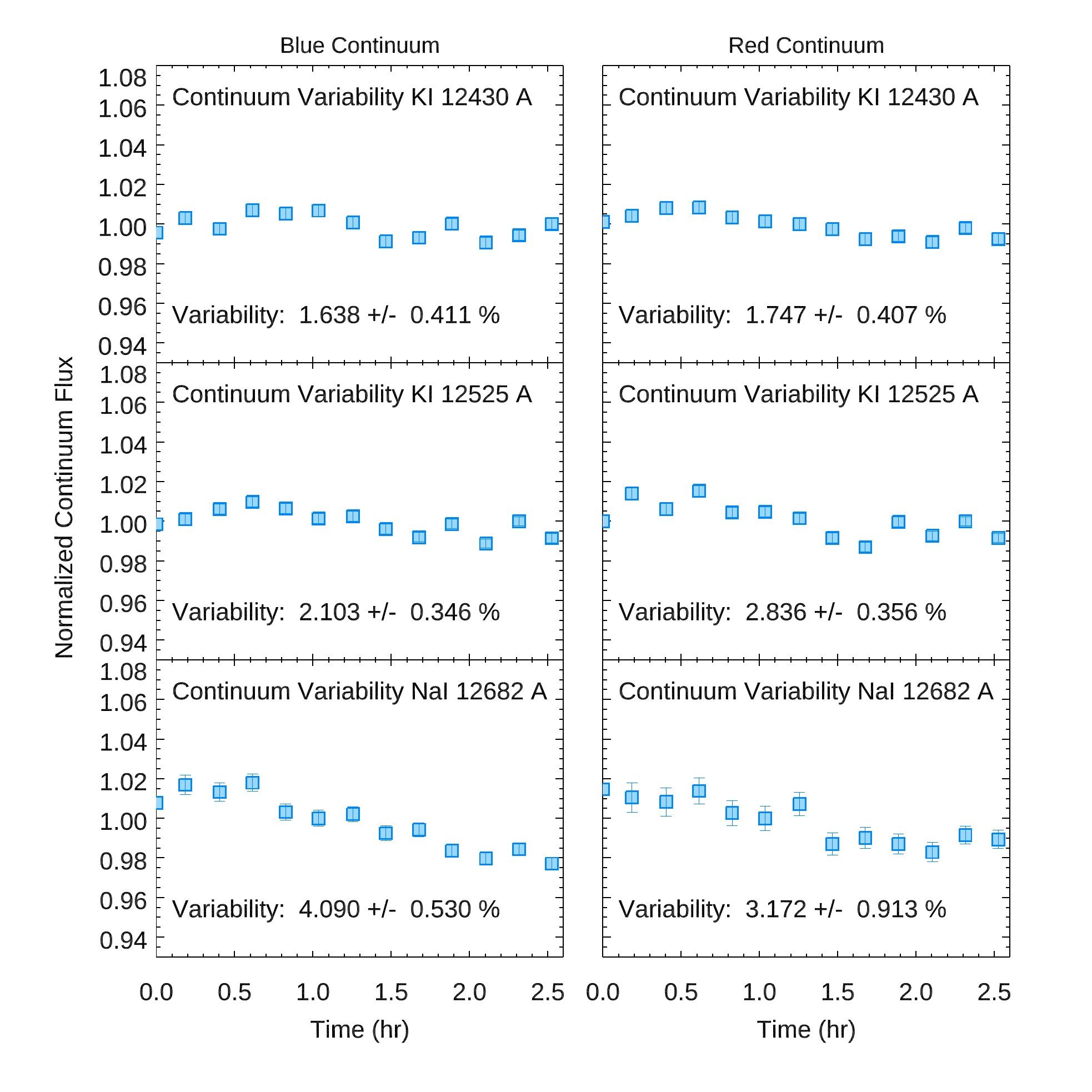}
    \includegraphics[width=0.41\textwidth]{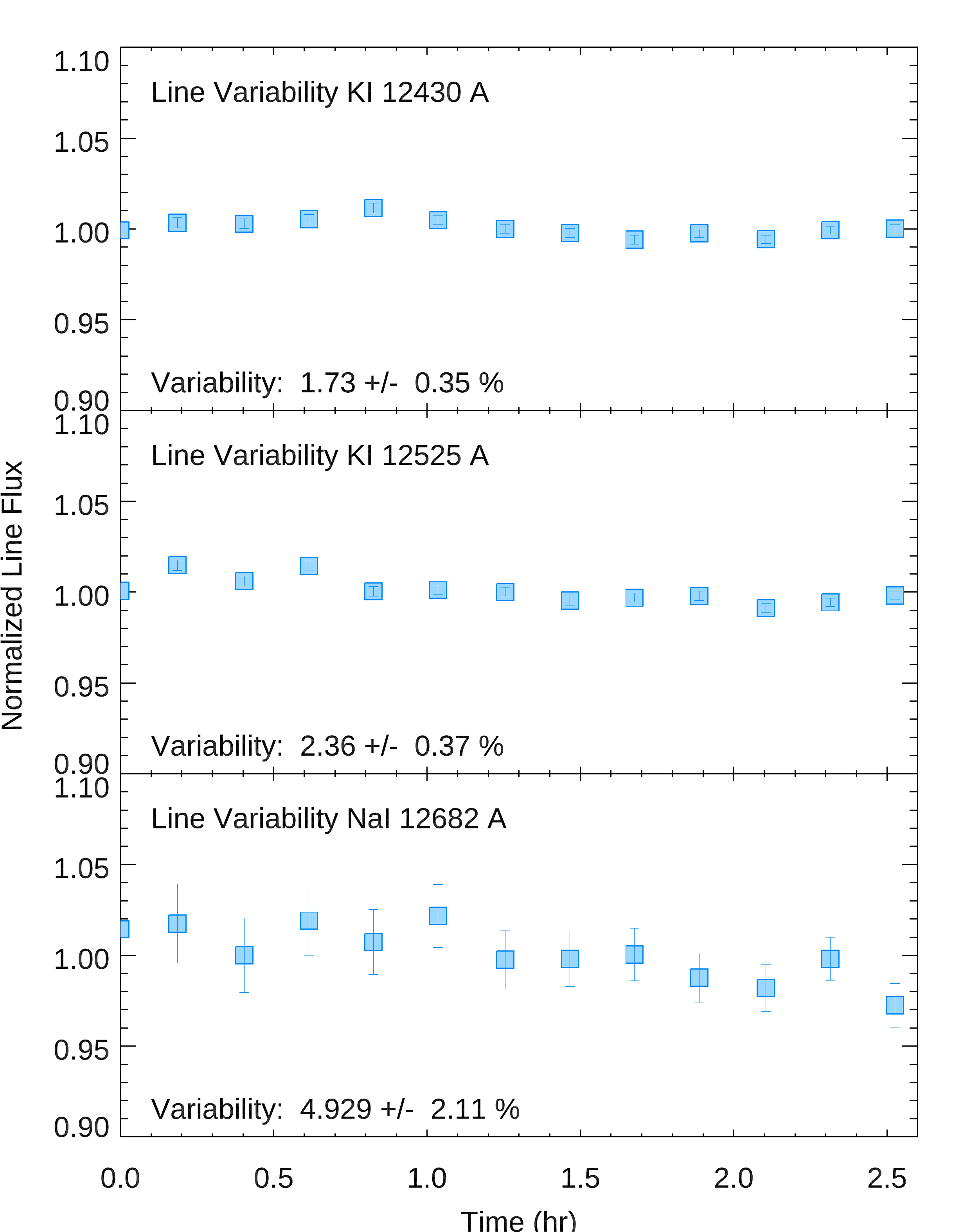}
    \caption{Variability inside the wavelength range of the blue and red continuum, and inside the alkali lines wavelength range for the calibration star 8 for spectra with the original $J$-band MOSFIRE resolution.}
    \label{corr_LCs_obj8}
\end{figure*}

\begin{figure*}
    \centering
    \includegraphics[width=0.56\textwidth]{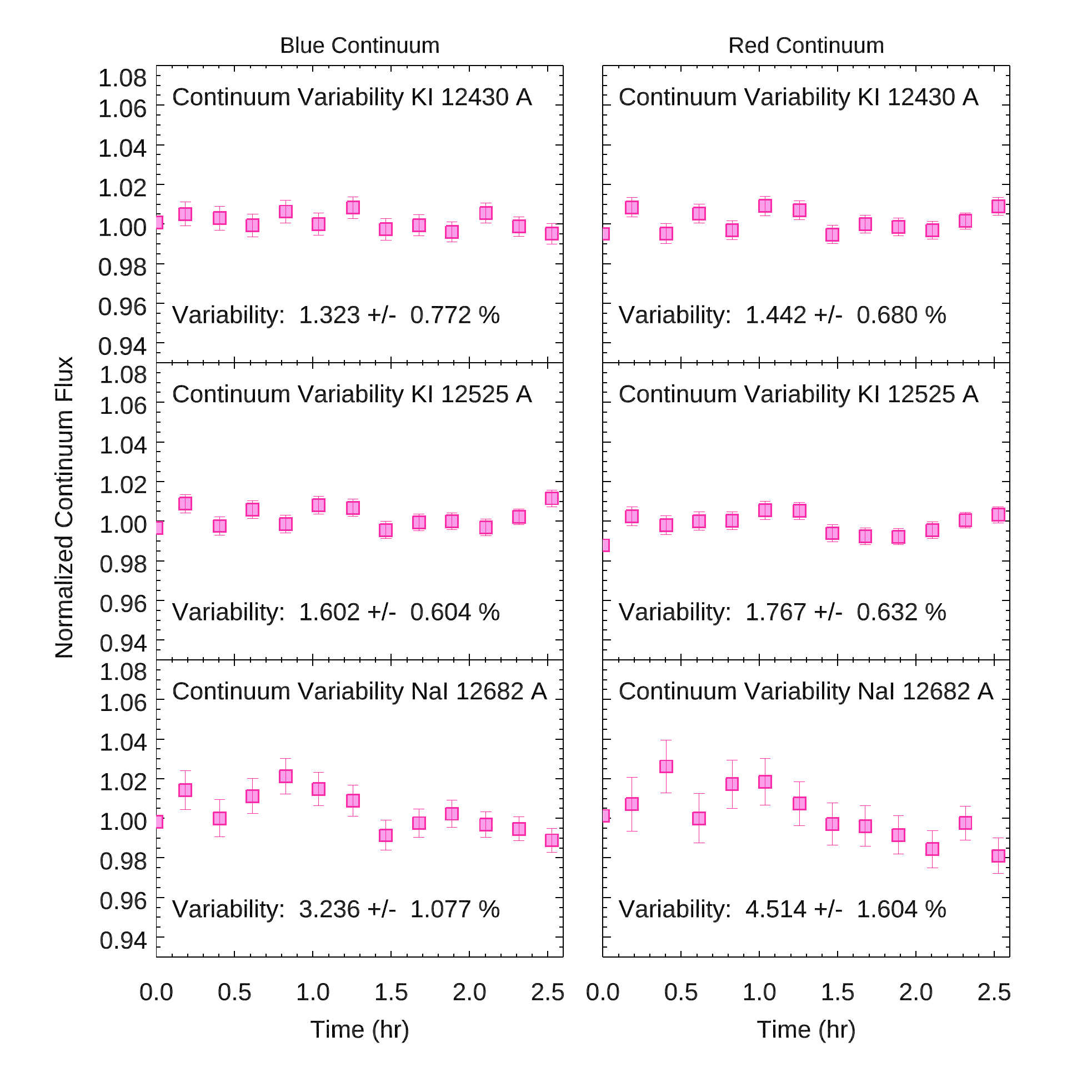}
    \includegraphics[width=0.41\textwidth]{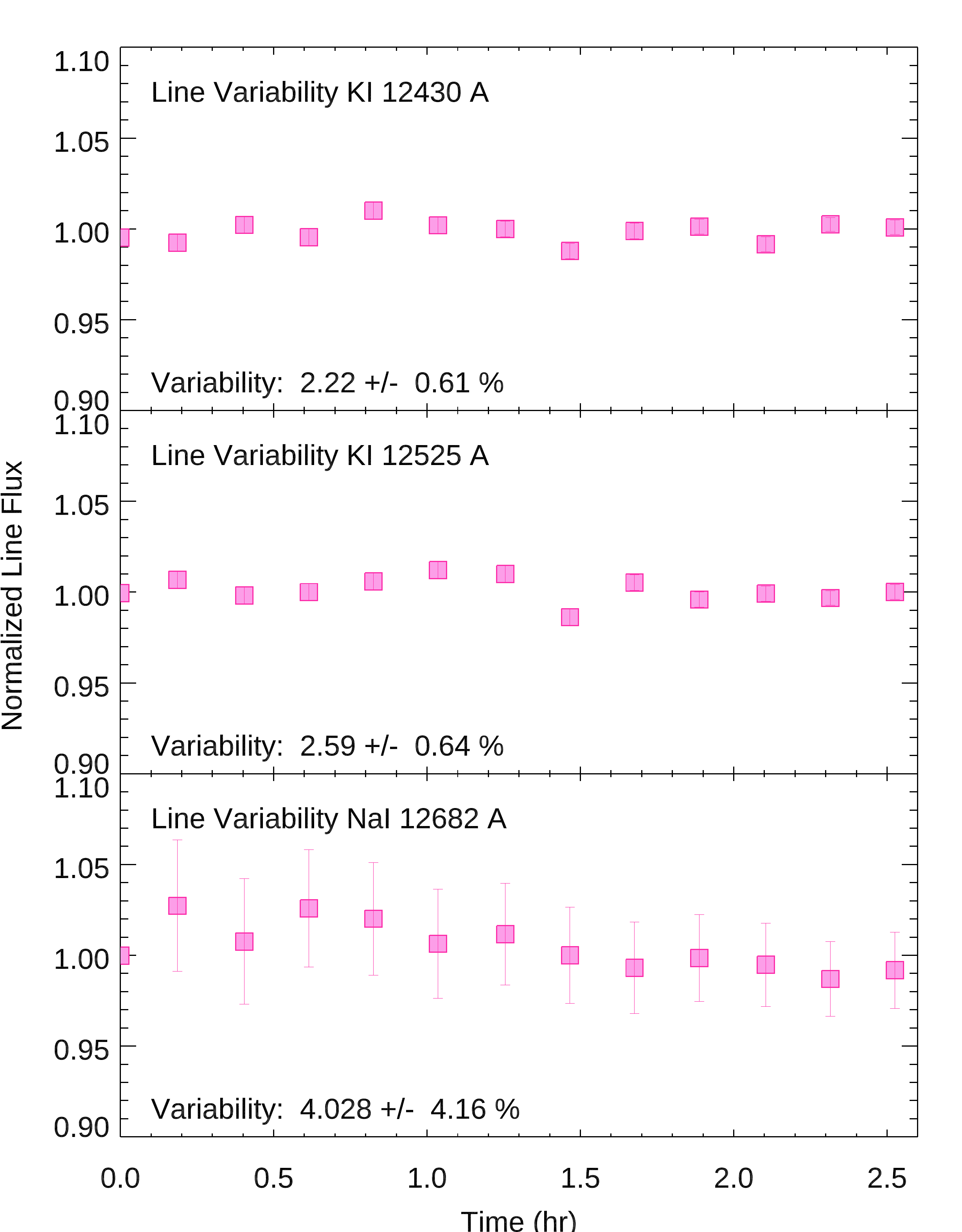}
    \caption{Variability inside the wavelength range of the blue and red continuum, and inside the alkali lines wavelength range for the calibration star 1 for spectra with R$\sim$100.}
    \label{corr_LCs_obj1_R100}
\end{figure*}

\begin{figure*}
    \centering
    \includegraphics[width=0.56\textwidth]{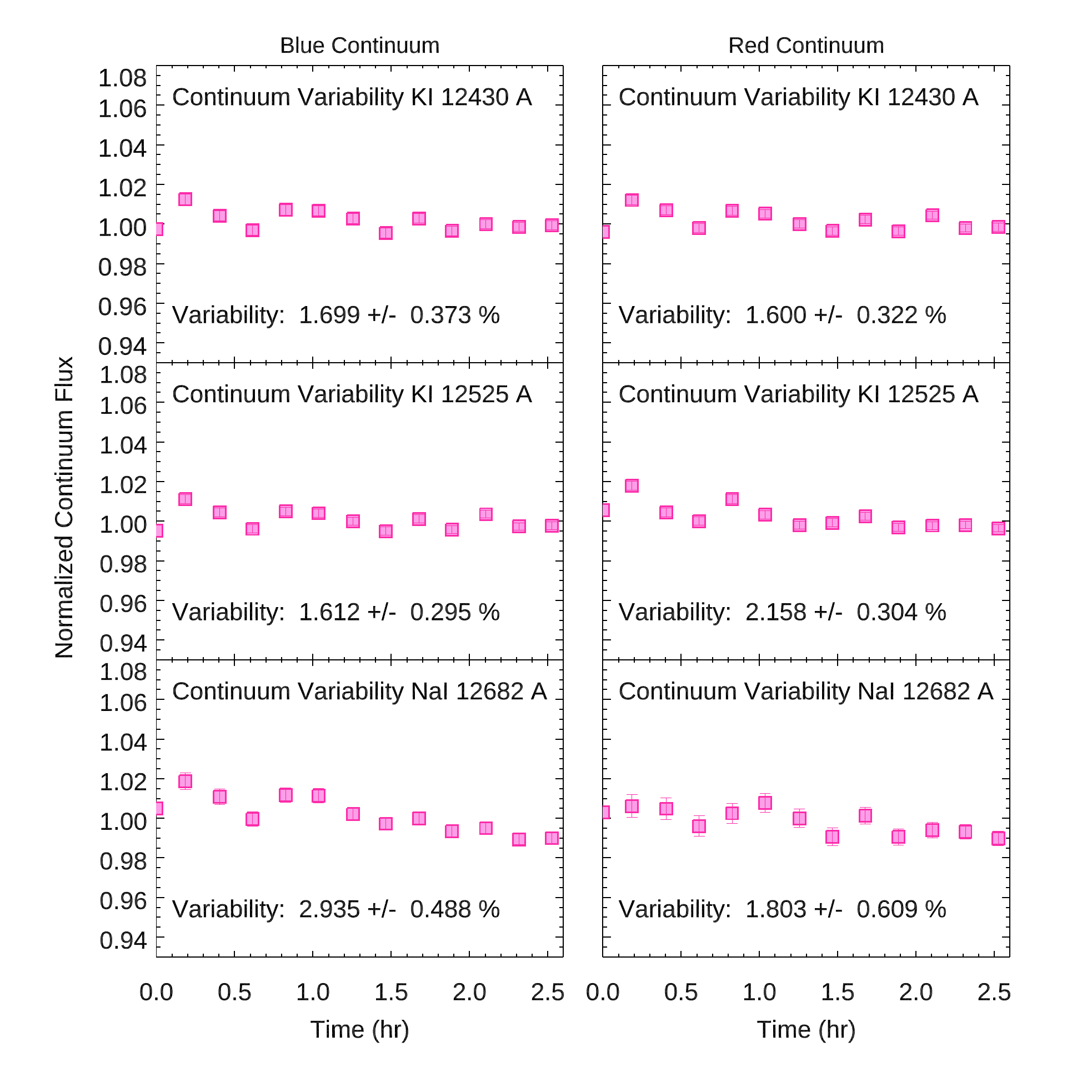}
    \includegraphics[width=0.41\textwidth]{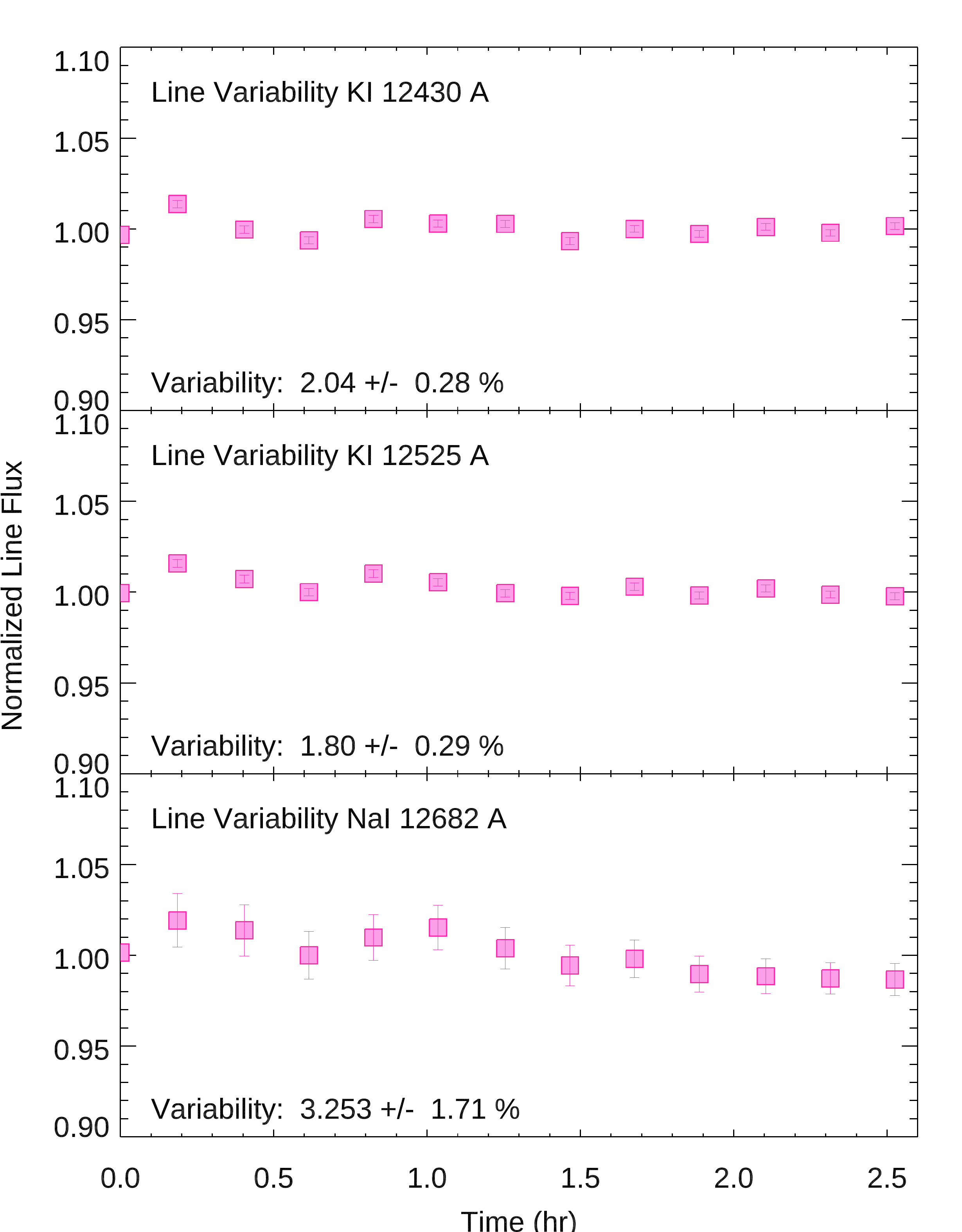}
    \caption{Variability inside the wavelength range of the blue and red continuum, and inside the alkali lines wavelength range for the calibration star 4 for spectra with R$\sim$100.}
    \label{corr_LCs_obj4_R100}
\end{figure*}

\begin{figure*}
    \centering
    \includegraphics[width=0.56\textwidth]{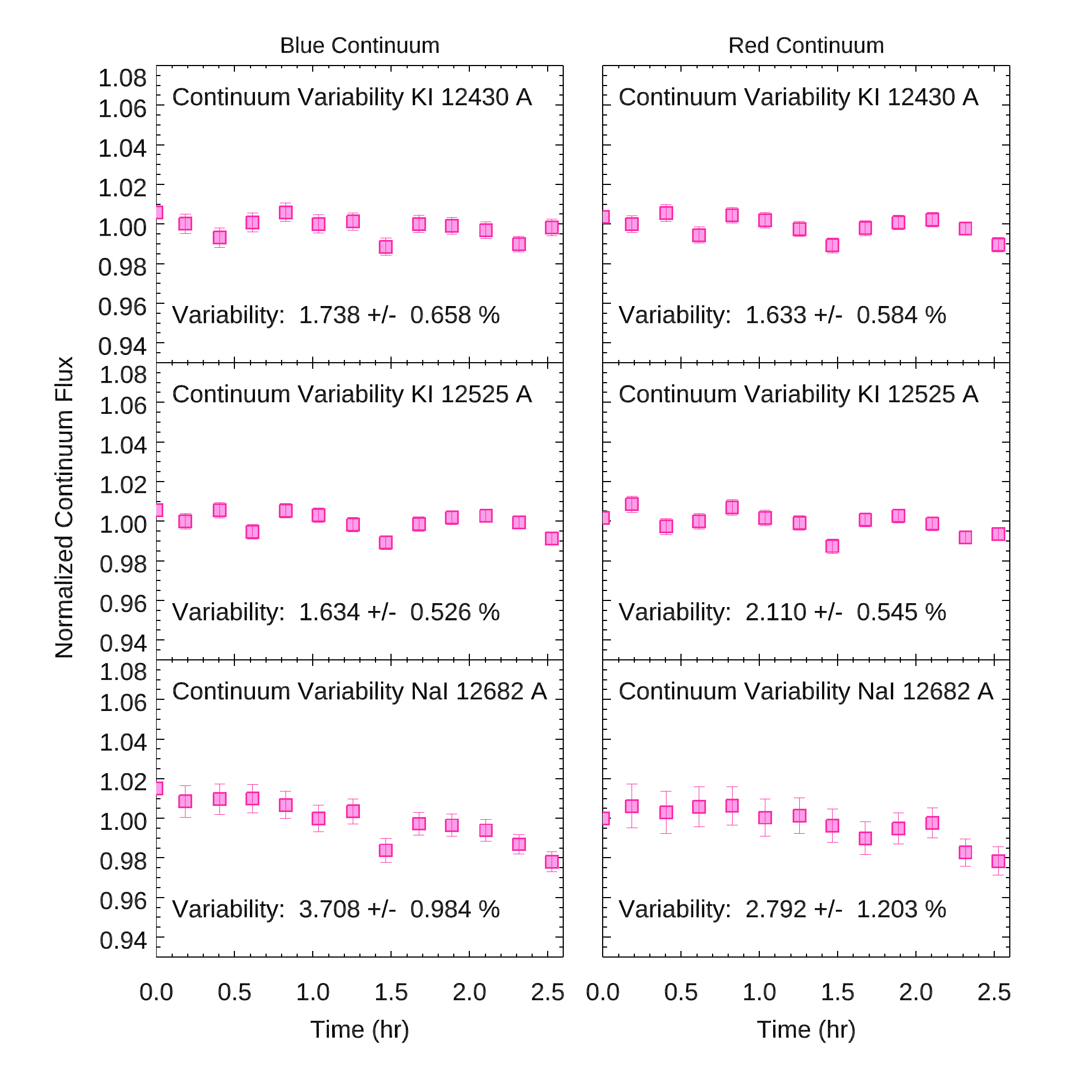}
    \includegraphics[width=0.41\textwidth]{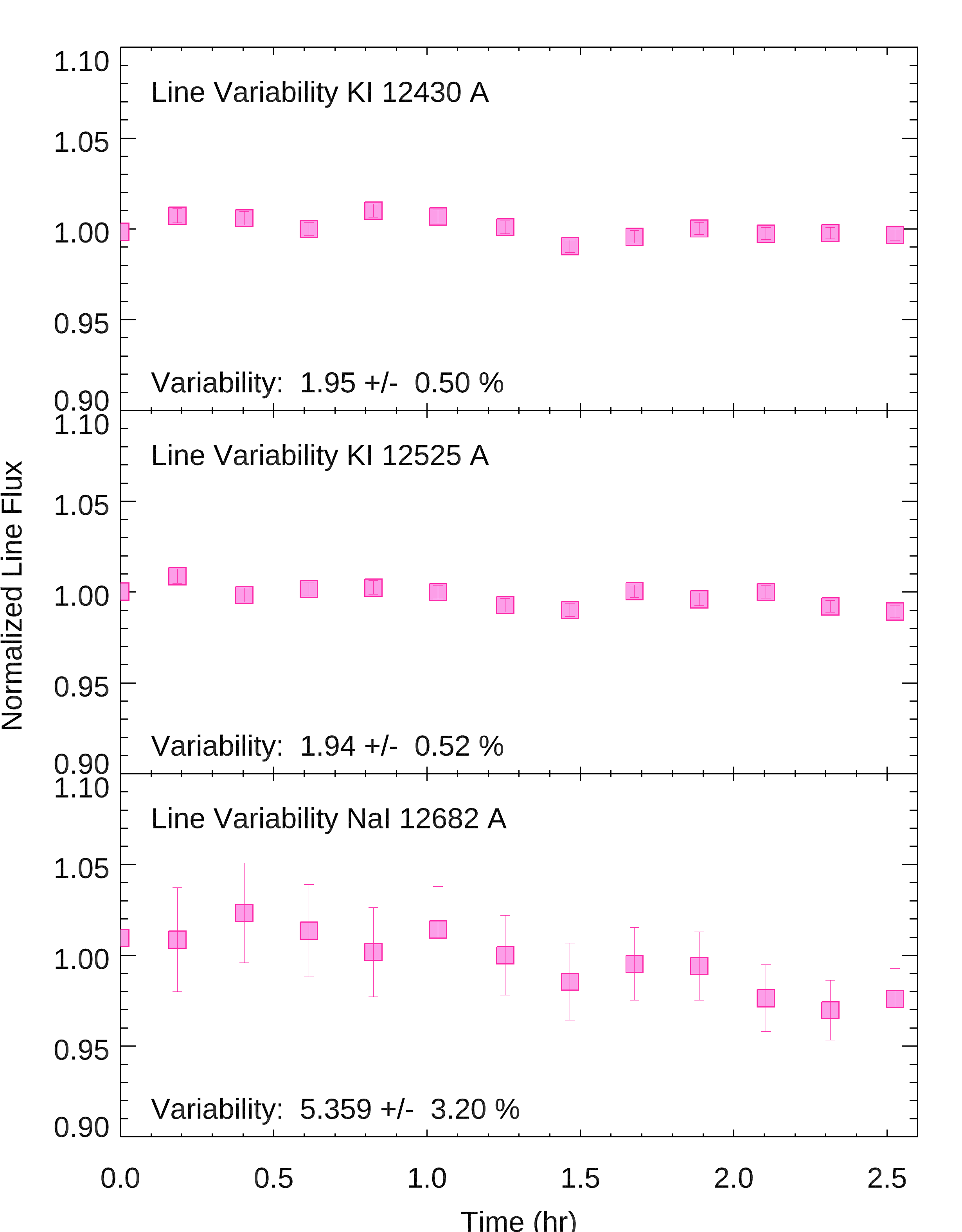}
    \caption{Variability inside the wavelength range of the blue and red continuum, and inside the alkali lines wavelength range for the calibration star 5 for spectra with R$\sim$100.}
    \label{corr_LCs_obj5_R100}
\end{figure*}

\begin{figure*}
    \centering
    \includegraphics[width=0.56\textwidth]{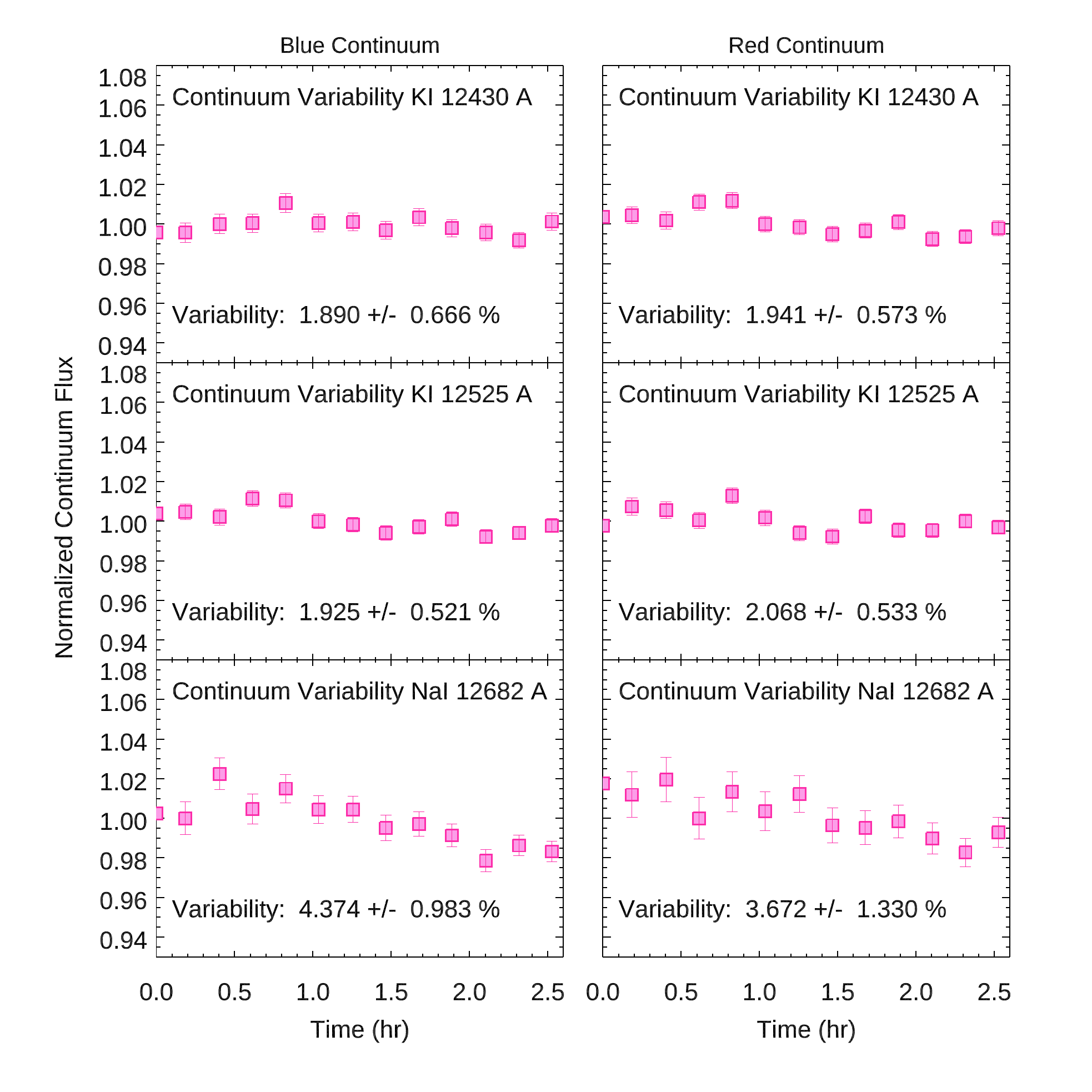}
    \includegraphics[width=0.41\textwidth]{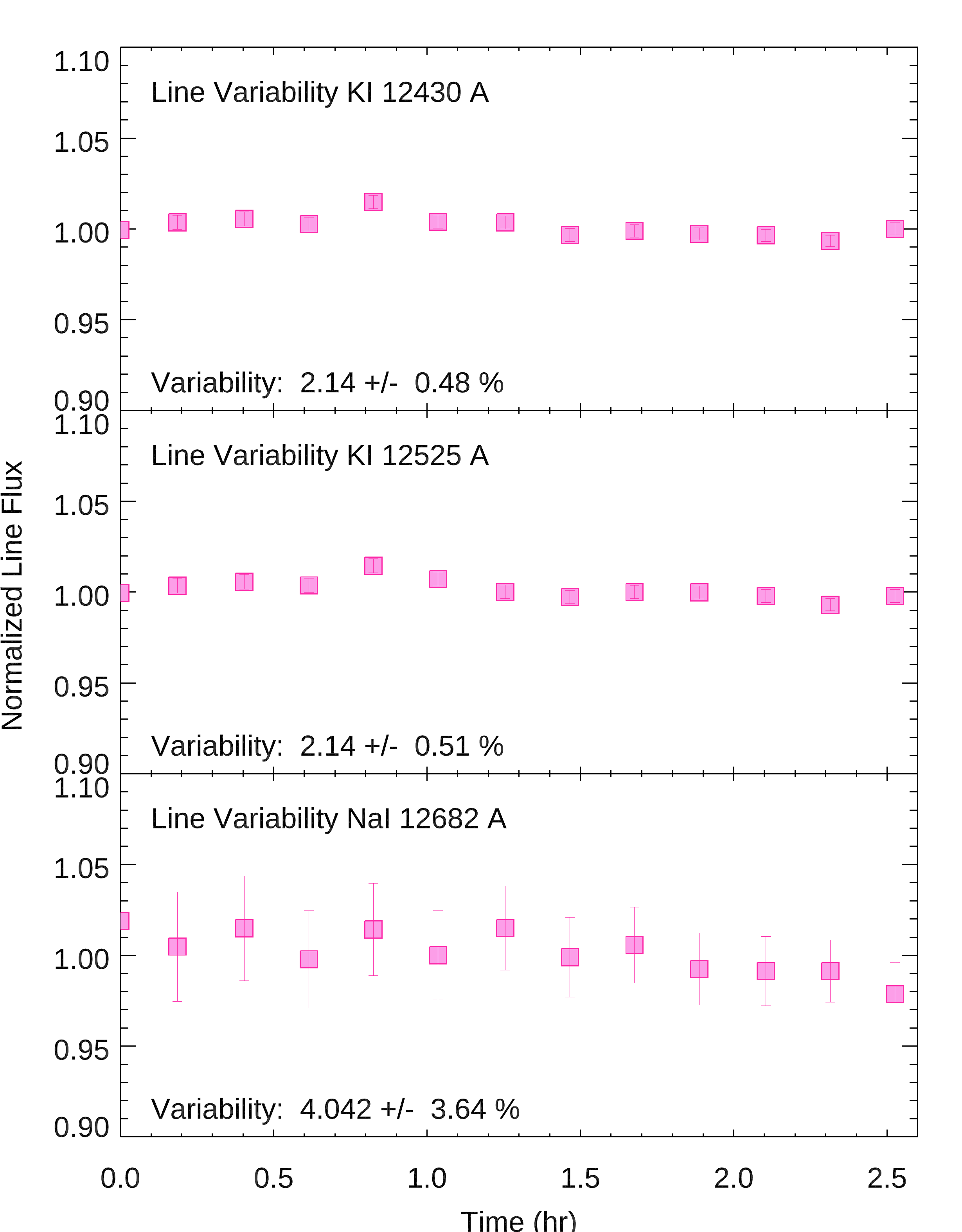}
    \caption{Variability inside the wavelength range of the blue and red continuum, and inside the alkali lines wavelength range for the calibration star 6 for spectra with R$\sim$100.}
    \label{corr_LCs_obj6_R100}
\end{figure*}

\begin{figure*}
    \centering
    \includegraphics[width=0.56\textwidth]{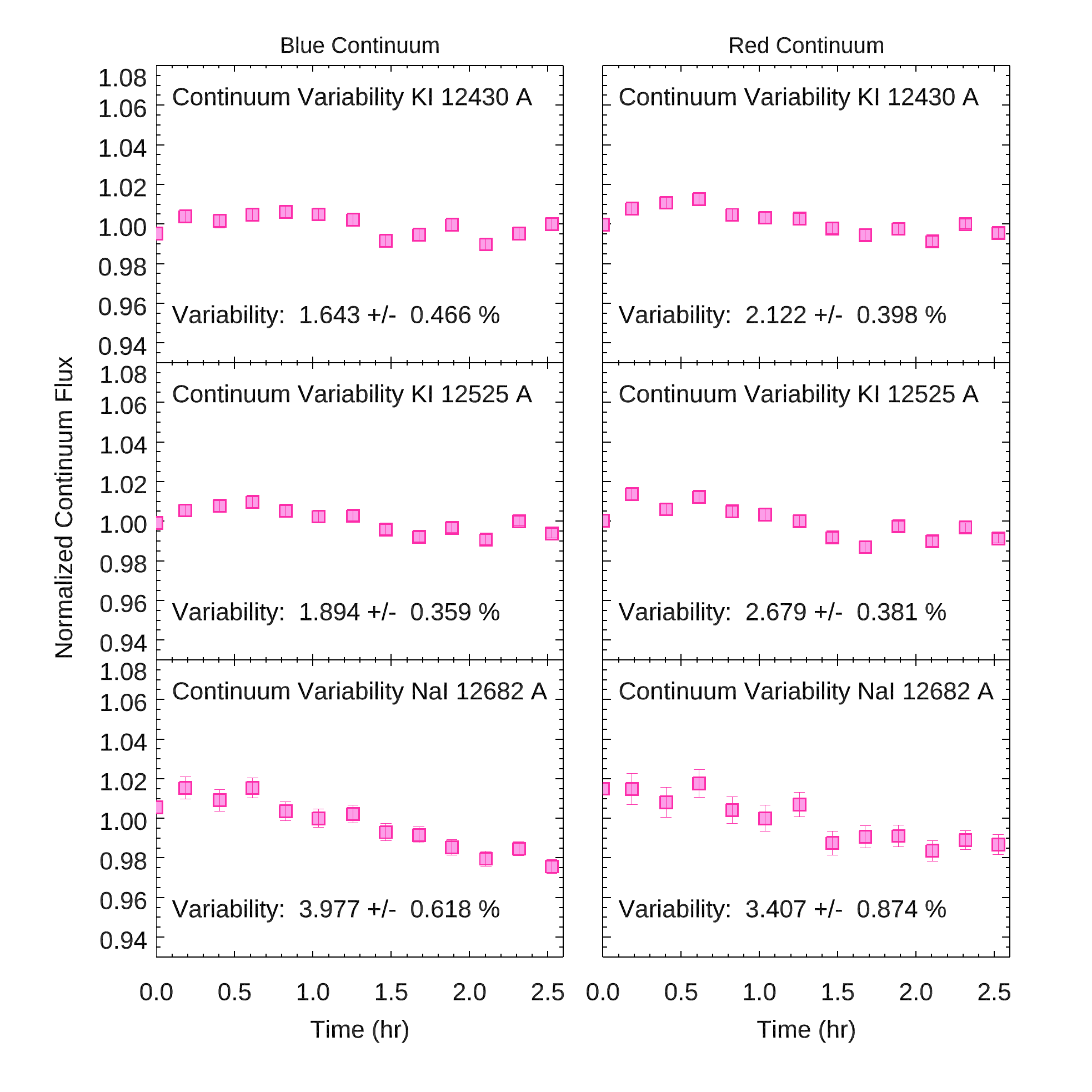}
    \includegraphics[width=0.41\textwidth]{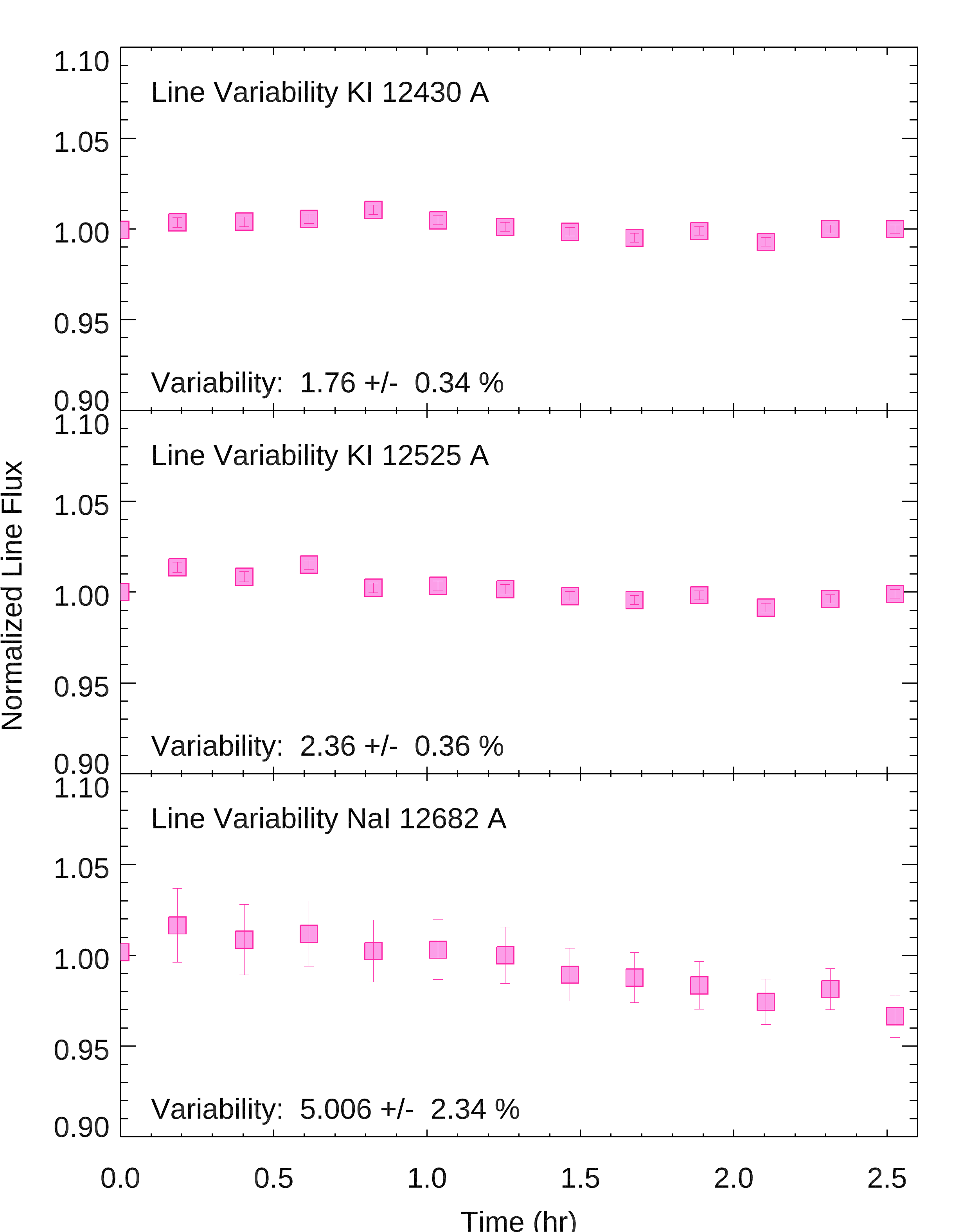}
    \caption{Variability inside the wavelength range of the blue and red continuum, and inside the alkali lines wavelength range for the calibration star 8 for spectra with R$\sim$100.}
    \label{corr_LCs_obj8_R100}
\end{figure*}

\bibliography{2M2208_variability}{}
\bibliographystyle{aasjournal}



\end{document}